%% file: emid_arxiv.tex
\documentclass[aps,prd,twocolumn,showpacs,superscriptaddress,groupedaddress,nofootinbib]{revtex4}
\usepackage{epsfig}
\usepackage{graphicx}
\usepackage{color}
\usepackage{subfigure}
\usepackage{setspace}
\usepackage{amssymb}
\usepackage{amsmath}
\usepackage{graphicx}
\usepackage{dcolumn}
\usepackage{bm}
\usepackage{epsfig}
\usepackage{psfrag}
\usepackage{xspace}
\usepackage{multirow}
\usepackage{lscape}
\usepackage{comment}
\usepackage{color}
\usepackage{lineno}
\usepackage{soul}

\newcommand{\dzero}     {D0}

\newcommand{\geant}     {\mbox{\textsc{geant}}\xspace}

\newcommand{\met}{\mbox{\ensuremath{E\kern-0.6em\slash_T}}\xspace}
\newcommand{\metx}{\mbox{\ensuremath{E\kern-0.6em\slash_x}}\xspace}
\newcommand{\mety}{\mbox{\ensuremath{E\kern-0.6em\slash_y}}\xspace}
\newcommand{\sigmet}{\ensuremath{\sigma_{\mbox{\ensuremath{E\kern-0.6em\slash_T}}}}\xspace}

\newcommand{\DO}{D0}

\newcommand{\phimod}{\mbox{$\phi_{\rm mod}$}}
\newcommand{\mee}{\mbox{$M_{ee}$}}

\newcommand{\ilum}{\mbox{$\mathcal{L}_{\rm inst}$}}
\newcommand{\wenu}{\mbox{$W \rightarrow e\nu$}}

\newcommand{\chisqr}{\mbox{$\chi^{2}$}}

\begin{document}

\hspace{5.2in}\mbox{FERMILAB-PUB-13-582-E}

\title{Electron and Photon Identification \\
in the D0 Experiment 
}

\input author_list.tex
\date{April 6, 2014}

\begin{abstract}
The electron and photon reconstruction and identification
algorithms used by the D0 Collaboration at the Fermilab
Tevatron collider are described. The determination of the electron energy scale and
resolution is presented. Studies of the performance of the electron and
photon reconstruction and identification are summarized. 
The results are based on measurements of $Z$ boson decay events of
$Z \to ee$ and 
$Z \to \gamma \ell\ell$ $(\ell = e, \mu)$
collected in $p\bar{p}$ collisions at a
center-of-mass energy of 1.96~TeV using an integrated luminosity of up
to 10~fb$^{-1}$. 
\end{abstract}

%\PACS{29.30.Aj \sep 29.85.-c }
\maketitle

\section{Introduction}
\label{sec::intro}
The precise and efficient reconstruction and identification of
electrons\footnote{In the following, if not
indicated otherwise, ``electron'' denotes both electrons and
positrons.} and photons at the D0 experiment
at the Fermilab Tevatron $p\bar{p}$ collider
is essential for a broad spectrum of physics
analyses, including high precision standard model (SM) measurements
and searches for new phenomena. To satisfy this requirement, the D0 detector was
designed to have excellent performance for the measurements of electrons
and photons of energies from a few GeV up to
${\cal{O}}$(100~GeV). Another design requirement
was to have good discrimination between jets and electrons or photons, since physics
measurements often suffer from large backgrounds induced by jets being
misidentified as electrons or photons.
In this paper, the reconstruction of electromagnetic (EM) objects
using D0 data is described.
The determination of the electron energy scale and
resolution and
the performance of
electron and photon identification using the Run~II dataset
recorded between 2002 and 2011 are presented.

\section{D0 detector}
\label{sec::det}
The D0 detector is described elsewhere \cite{d0det}.  The components
most relevant to electron and photon identification are the central
tracking system, composed of a silicon microstrip tracker (SMT) that
is located near the $p\bar{p}$ interaction point and a
central fiber tracker (CFT) embedded in a 2~T solenoidal magnetic
field, a central preshower (CPS), and a liquid-argon/uranium
sampling calorimeter.  The CPS is located before the inner
layer of the calorimeter, outside the calorimeter cryostat, 
and is formed of one radiation length of
absorber followed by three layers of scintillating strips.
The D0 coordinate system is right-handed. The $z$-axis points in
the direction of the Tevatron proton beam, and the $y$-axis points
upwards. Pseudorapidity is defined as $\eta=-\ln [\tan(\theta/2)]$, where
$\theta$ is the polar angle relative to the proton beam direction.
The azimuthal angle $\phi$ is defined in the plane transverse to
the proton beam direction. The SMT covers $|\eta| <3$, and the CFT
provides complete coverage out to $|\eta| \approx 1.7$.

\begin{figure}[h]
  \begin{center}
    \includegraphics[keepaspectratio,width=.48\textwidth]{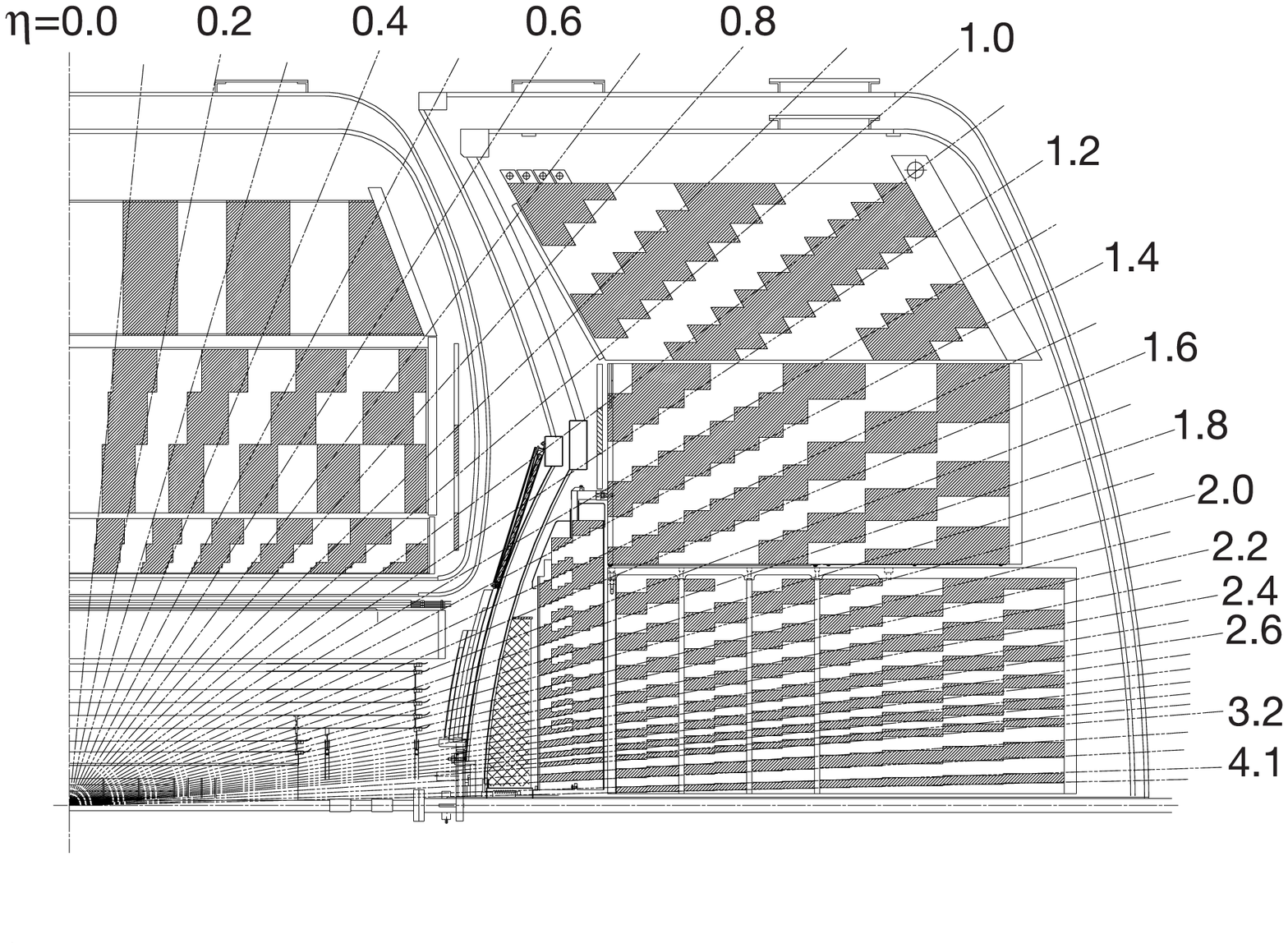}
    \end{center} \caption{Side view of a quadrant of the D0
    calorimeters
    showing the transverse and longitudinal segmentation. The
    alternating shading
    pattern indicates the cells for signal readout. The lines indicate
    the pseudorapidity intervals defined from the center of the
    detector. The CC covers the region $|\eta| < 1.1$ and the EC
    extends the coverage to $|\eta| \approx 4.2$.
}
  \label{fg:d0_cal_detetaview}
\end{figure}

The calorimeter consists of a central section (CC) with coverage in
pseudorapidity of $|\eta|<1.1$, and two endcap calorimeters (EC) covering
up to $|\eta|\approx 4.2$, as shown in Fig.~\ref{fg:d0_cal_detetaview}.
The region $1.1 < |\eta| < 1.5$ is not fully covered
by the calorimeter. Therefore, the reconstruction, identification,
energy scale and resolution estimation methods described in the paper
can not be used. In that particular region, the tracking system is mainly
used for reconstruction which is beyond the scope of this paper.
Each part of the calorimeter is contained in its
own cryostat and comprises an EM section, closest to the
interaction region, and a hadronic section.  The EM section of the
calorimeter is segmented into four longitudinal layers with transverse
segmentation of $\Delta\eta\times\Delta\phi = 0.1\times 0.1$, except
in the third layer (EM3), where the segmentation is $0.05\times 0.05$.
There are 32 azimuthal modules for EM layers in the CC. The hadronic
section is composed of fine (FH) and coarse (CH) layers. The FH
layers are closer to the interaction point, followed by the CH layers.

There is material varying between 3.4 and 5 radiation lengths ($X_0$)
between the beam line and the CC.
For EC, it varies between 1.8 and 4.8 $X_0$.
The amount of material depends on the incident angle of
the electron or photon \cite{MwPRD}.
At $\eta \approx 0$, the amount of material in front of the calorimeter
is 0.2 $X_0$ in the tracking detector, 
0.9 $X_0$ in the solenoid, 0.3 $X_0$ in the
preshower detector plus 1.0 $X_0$ in the associated lead, and
1.3 $X_0$ in the cryostat walls plus related support structures.

\section{Data and Monte Carlo samples}
\label{sec::datamc}
Data and Monte Carlo (MC) simulated events have been used to study
reconstruction and identification efficiencies, to measure the energy
scale and resolution, and to derive correction factors to compensate
for any residual mismodelling of the detector.
The electron candidates are selected from $Z \to ee$ data and MC
using the ``tag-and-probe method'' as described in Sect. \ref{tagandprobe}.
The photon candidates are selected from diphoton MC and
$Z \to \gamma \ell\ell \ (\ell = e,\mu)$ data and MC,
where the photons are radiated from charged leptons in $Z$ boson decays
by requiring the dilepton invariant mass to be less than 82 GeV while
the three-body mass of dilepton and photon $M_{\rm \ell\ell\gamma}$ is
required to be
$82 < M_{\rm \ell\ell\gamma} < 102$ GeV \cite{Zgam-ref}.
To evaluate misidentification of jets as electrons or photons,
dijet events are selected.
For dijet MC, an EM cluster passing the preselection as described in
Sect.~\ref{emclus} is selected as jets misidentified as electrons or photons.
For dijet data, a jet~\cite{jesNIM} with transverse momentum $p_T > 20$ GeV
and $|\eta| < 2.5$ is selected, then a preselected EM cluster is selected
in the opposite azimuthal plane with $|\Delta\phi (\rm jet, EM)| > 2.9$ radian
as the jets misidentified as electrons or photons.
To eliminate possible contamination from diboson,
$Z$ $+$ jets and $W$ $+$ jets
processes, events with at least one isolated high-$p_T$ muon~\cite{muonNIM},
events with an invariant mass of the EM cluster and an isolated track
between 60 and 120 GeV, and events with
missing transverse energy~\cite{MET-ref}
greater than 10 GeV are rejected.
For studies of jets misidentified as photons, the $\gamma$ $+$ jet
component containing a real photon is removed from the dijet sample by
requiring that the EM cluster be non-isolated by cutting on the
shower isolation fraction
(see Sect. \ref{emclus}) of $0.07 < f_{\rm iso} < 0.15$.
The $Z \to ee$ and $Z \to ee\gamma$ data events
are collected using single electron triggers 
as described in Sect.~\ref{ele_trigger}.
For $Z \to \mu\mu\gamma$ and dijet data events, 
single muon triggers~\cite{muonNIM}
and jet triggers~\cite{jesNIM} are used, respectively.

The data used in physics analyses were collected by the D0
detector during Tevatron Run~II between April 2002 and September 2011
and correspond to an integrated luminosity of approximately 10 fb$^{-1}$.

The $Z \to ee$ signal samples are generated
using the {\sc alpgen} generator~\cite{alpgen-ref} interfaced to {\sc
pythia}~\cite{pythia-ref} for parton showering and hadronization. The simulated
transverse momentum $p_T$ distribution of the $Z$ boson is weighted to match the distribution
observed in data~\cite{zpt-ref}.
Diphoton and $Z \to \gamma \ell\ell \ (\ell = e,\mu)$ signal events, and dijet
background samples are generated using {\sc pythia}~\cite{pythia-ref}.
All MC samples used here are generated using the CTEQ6L1 \cite{cteq-ref} parton
distribution functions,
followed by a {\sc geant}~\cite{geant-ref} simulation of the D0
detector. To accurately model
the effects of multiple $p\bar{p}$ interactions in a single bunch crossing
and detector noise, data
from random $p\bar{p}$ bunch crossings are overlaid on the MC events.
The instantaneous luminosity spectrum of these overlaid events is
matched to
that of the events used in the data analysis.
Simulated events are
processed using the same reconstruction code that is used for data.

\section{EM object reconstruction and identification}
\label{sec::recoid} 
EM objects -- electrons and photons -- are reconstructed by detecting localized
energy deposits in the EM calorimeter.
Confirmation of the existence of an electron track
is sought from the central tracking system since an
isolated high-$p_{T}$ track should originate from the
interaction vertex. The hadronic calorimeter, preshower, and
tracking systems can be used
to differentiate electrons and photons from jets.

\subsection{EM cluster reconstruction}\label{emclus}
EM objects in the \dzero\ detector are reconstructed using the nearly 55,000
calorimeter channels. Only channels with energies above
noise are read out~\cite{jesNIM}.
We use the same cluster reconstruction algorithm for
electrons and photons, since their showers both consist of
collimated clusters of energy deposited mainly in the EM layers of
the calorimeter. Calorimeter cells with the same $\eta$ and $\phi$
are grouped together to form towers.
For the calculation of the energy of EM clusters, 
we sum the energies measured in the four EM
layers and the first hadronic (FH1) layer which is included to account
for leakage of energy of EM objects into the hadronic part of the calorimeter.
Starting with the highest transverse energy tower
($E_{T} > 500$ MeV), energies of
adjacent towers
in a cone of radius ${\cal R}=\sqrt{(\Delta \eta )^2 + (\Delta \phi)^2}=0.4$
around the highest $E_T$ tower are added to form EM clusters
in the CC \footnote{
We use the terms ``CC EM cluster'' to denote EM clusters in the
pseudorapidity range $|\eta| < 1.1$, and ``EC EM cluster'' to denote EM clusters
in the pseudorapidity range $1.5 < |\eta| < 3.2$.}. 
In the EC, EM clusters are a set of adjacent cells with a
transverse distance of less than 10 cm from an initial cell with the
highest energy content in the EM3 layer.

To be selected as an EM candidate, EM clusters must satisfy the
following set of criteria:
\begin{list}{$\bullet$}{}
\item The cluster transverse energy must be $E_{T} >$~1.5~GeV.
\item
The fraction of energy in the EM layers is

\begin{equation}
\label{eq-emf}
f_{\rm EM}=\frac{E_{\rm EM}}{E_{\rm tot}},
\end{equation}

where $E_{\rm EM}$ is the cluster energy in the EM layers, and $E_{\rm tot}$ is
the total energy of the cluster
in all layers within the cone. At least 90\% of the energy 
should be deposited in the EM layers of the calorimeter.
\item
The isolation fraction is the ratio of the energy in an isolation cone
surrounding an EM cluster to the energy of the EM cluster,

\begin{equation}
\label{eq-caliso}
f_{\rm iso}=\frac{E_{\rm tot}({\cal R}<0.4)-E_{\rm EM}({\cal R}<0.2)}{E_{\rm EM}({\cal R}<0.2)},
\end{equation}

where $E_{\rm tot}({\cal R}<0.4)$ is the total energy within a cone
of radius ${\cal R}=0.4$ around 
the cluster, summed over the entire depth of
the calorimeter except the CH layers,
and $E_{\rm EM}({\cal R}<0.2)$ is the energy in the towers in a cone of radius
${\cal R}=0.2$ summed over the EM layers only.
To select isolated electron or photon clusters in the calorimeter,
we require an isolation fraction of less than 0.2.
\end{list}

For each EM candidate, the centroid of the EM cluster is
computed by weighting cell coordinates
with cell energies
in the EM3 layer of the calorimeter.
The shower centroid position
together with the location of the $p\bar{p}$ collision vertex
is used to calculate the direction of
the EM object momentum.

Since EM objects begin to shower in the preshower detector,
clusters are also formed in that detector.
Single layer clusters are formed from scintillating strips for each layer.
A preshower cluster is built by combining the single layer 
clusters from each of the three layers.
These preshower clusters are extensively used to help
identify the electrons and photons, and to build the 
multivariate identification methods, as well as to
find the right interaction vertex for the photon
as described in the following sections. 

Electron candidates are distinguished from photon candidates by the presence
of a track with $p_{T}>$ 1.5 GeV within a window of 
$\Delta \eta \times \Delta \phi = 0.05 \times 0.05 $ around the
coordinates of the EM cluster. 
The momentum of an electron candidate is recalculated using the
direction of the best spatially matched track
while the energy of the electron is
measured by the calorimeter due to limited momentum resolution of the
central tracking system.
An EM cluster is considered to be a photon candidate
if there is no associated track.

\subsection{EM object identification}\label{emid}
After applying the above criteria to EM clusters,
there remains a considerable fraction of jets misidentified as EM objects.
Further criteria must be
applied to reject these misidentified jets and increase the
purity of the selected electron and photon candidates. 
The following is a description of the quantities
employed for electron and photon identification. There are a
number of different selection criteria for these quantities to meet
the needs of different physics analyses.

\paragraph{EM energy fraction}\label{emf}

Because the development of EM and hadronic showers are
substantially different,
shower shape information can be used to differentiate between
electrons, photons, and hadrons.  Electrons and photons deposit
almost all their energy in the EM section of the calorimeter while
hadrons are typically much more penetrating.  EM clusters typically
have a large EM fraction, $f_{\rm EM}$ (Eq. \ref{eq-emf}).
The requirement of large values of
$f_{\rm EM}$ is very efficient for rejecting hadrons, but also removes
electrons pointing to
the module boundaries (in $\phi$) of the central EM calorimeter,
since they deposit a considerable fraction of their energy in the
hadronic calorimeter.

\paragraph{EM shower isolation}\label{iso}
Electrons and photons from a prompt decay of $W$ and $Z$ bosons tend to be
isolated in the calorimeter,
and therefore usually have a low isolation fraction $f_{\rm iso}$ (Eq. \ref{eq-caliso}).
In this case most of the energy of the EM
cluster is deposited in a narrow cone of radius
${\cal R}=0.2$ in the calorimeter.

\paragraph{EM shower width}\label{sigphi}
Showers induced by electrons and photons are usually narrower than those
from jets.
The EM3 layer of the calorimeter has a fine segmentation,
providing sensitive variables to separate
electrons and photons from misidentified jets.
The squared width, $\sigma_{\phi}^{2}$, of the shower shape in the transverse plane
is calculated as

\begin{equation}
\small
\sigma_{\phi}^{2} = \frac{ \sum (5.5 + \log\left(\frac{E_{\rm cell}^{i}}{E_{\rm EM3}}\right)) \cdot( r_{\rm cell}^{i} \times \sin(\phi_{\rm cell}^{i} - \phi_{\rm EM}))^{2}} 
       {\sum (5.5 + \log\left(\frac{E_{\rm cell}^{i}}{E_{\rm EM3}}\right))},
\end{equation}

where $E_{\rm cell}^{i}$, $r_{\rm cell}^{i}$ and $\phi_{\rm cell}^{i}$
are the energy, radius calculated from $z$-axis and 
azimuthal angle for cell~$i$ in the EM3 layer associated to the EM cluster,
and $E_{\rm EM3}$ and $\phi_{\rm EM}$ are the total energy
and centroid azimuthal angle of the EM cluster at the EM3 layer.
A value of 5.5 was chosen as a result of studies to eliminate effect of low energy cells.
Only the cells with positive weight $(5.5 + \log\left(\frac{E_{\rm cell}^{i}}{E_{\rm EM3}}\right)) > 0$
are used in the calculation.
The width $\sigma_{\eta}$ of the shower
in the pseudorapidity direction
is calculated as

\begin{equation}
\tiny
\sigma_{\eta} = \sqrt{ \frac{\sum (5.5 + \log\left(\frac{E_{\rm cell}^{i}}{E_{\rm EM3}}\right)) \cdot \eta_{i}^{2}}{\sum (5.5 + \log\left(\frac{E_{\rm cell}^{i}}{E_{\rm EM3}}\right))} - \left(\frac{\sum (5.5 + \log\left(\frac{E_{\rm cell}^{i}}{E_{\rm EM3}}\right)) \cdot \eta_{i}}{\sum (5.5 + \log\left(\frac{E_{\rm cell}^{i}}{E_{\rm EM3}}\right))}\right)^{2}},
\end{equation}

where $E_{\rm cell}^{i}$ and $\eta_{i}$ are the energy and
pseudorapidity of cell $i$.

\paragraph{H-matrix technique}\label{hmx}

The shower shape of an electron or a photon is distinct
from that of a jet.  Fluctuations cause the energy deposition to vary
from the average in a correlated fashion among the cells and layers.
Longitudinal and transverse shower shapes, and
the correlations between energy depositions in the calorimeter cells
are taken into account to obtain the best discrimination against hadrons,
using a covariance matrix (``H-matrix'')
technique~\cite{hmatrix-1, hmatrix-2}.  
A covariance matrix is formed from a set of
eight well-modeled variables describing shower shapes:
\begin{itemize}
\item The longitudinal development is described by the fractions
of shower energy in the four EM layers (EM1, EM2, EM3, EM4).
\item
To characterize the lateral development of the shower, we consider the
shower width in both dimensions in the third EM layer
($\sigma_{\phi}^{2}$ and $\sigma_{\eta}$), which is the layer with the
finest granularity.
The logarithm of the total shower energy and the
coordinate of the $p\bar{p}$ collision vertex along the beam axis are included,
so that the dependence of the H-matrix on these quantities 
is properly parametrized.
\end{itemize}
In the EC the matrix is of dimension $8 \times 8$, while in the CC
$\sigma_{\eta}$ is not used and therefore the matrix has the dimension
$7 \times 7$.
A separate matrix is built for each ring of
calorimeter cells with the same $|\eta|$ coordinate.
To measure how closely the shower
shape of an electron candidate matches expectations from MC
simulations, a $\chi^{2}$ value is calculated ($\chi^2_{\rm Cal}$).
Since the electron and photon candidates tend
to have smaller $\chi^2_{\rm Cal}$ than jets, this variable can be used to
discriminate between EM and hadronic showers.

\paragraph{Track isolation}\label{trkiso}
For electrons and photons that are isolated,
the scalar sum of the $p_T$ of all charged particle tracks with $p_T>0.5$ GeV,
excluding the associated track for the EM cluster, originating from the $p\bar{p}$ collision vertex
in an annular cone of $0.05<{\cal R}<0.4$ around the
electron and photon candidates, $\Sigma p_{T}^{\rm trk}$, is expected to
be small. It is therefore a sensitive variable for discriminating between
EM objects and jets.

\paragraph{Track match}\label{trk}
For electron identification, to suppress photons and jets
misidentified as electrons,
the cluster is required to be associated with a track
in the central tracking system in a road between the EM calorimeter
cluster and the $p\bar{p}$ collision vertex satisfying the conditions
$|\Delta\eta_{\rm EM,trk}|<0.05$ and $|\Delta\phi_{\rm EM,trk}|<0.05$
for the differences between $\eta$ and $\phi$ of the EM cluster and
the associated track.
To quantify the quality of the cluster-track matching,
a matching probability 
P($\chi^2_{\rm spatial}$) is defined using $\chi^2_{\rm spatial}$, which is given by

\begin{equation}\label{spchi}
\chi^2_{\rm spatial} = \left(\frac{\Delta\phi}{\delta_{\phi}}\right)^2 +
\left(\frac{\Delta\eta}{\delta_{\eta}}\right)^2.
\end{equation}

The probability is computed for each matched track. In these expressions,
$\Delta\eta$ and $\Delta\phi$ are the differences between the track
position and the EM cluster position in the EM3 layer of the calorimeter.  The
variables $\delta_{\phi}$ and $\delta_{\eta}$ are the resolutions
of the associated quantities.  The track with the highest P($\chi^2_{\rm spatial}$) is
taken to be the track matched to the EM object.
If there is no matched track, P($\chi^2_{\rm spatial}$) is set to $-1$.

\paragraph{Hits on road}\label{hor}
Due to tracking inefficiencies, the cluster-track matching probability
method is not fully efficient in separating electron from photon candidates,
in particular in events with high instantaneous luminosity.
To improve the separation between electron and photon candidates,
a ``hits on road'' discriminant, $D_{\rm hor}$, is used in the CC.
For each EM object, a ``road'' is defined
between the $p\bar{p}$ collision vertex
position and the CPS cluster position, if it is matched to
the EM object, or else to the EM cluster position.
To account for the different sign of the electric charge
of electrons and positrons, two roads (positive-charge and negative-charge roads) are
defined. The number of
hits from CFT fibers and SMT strips along the EM cluster's trajectory,
$N_{\rm hits}$, is counted.
The discriminant $D_{\rm hor}$ is defined by

\begin{equation}
D_{\rm hor} = \frac{P_{e}(N_{\rm hits})}{P_{e}(N_{\rm hits})+P_{\gamma}(N_{\rm hits})}
,
\end{equation}

where $P_{e}$ and $P_{\gamma}$ are the probabilities
in the bin of $N_{\rm hits}$, given by

 \begin{equation}
P_{e}(N_{\rm hits}) = \frac{\sum_{i=0}^{N_{\rm hits}} N_{e}^{i}}
{\sum_{i=0}^{24} N_{e}^{i}} ,
\end{equation}
\begin{equation}
P_{\gamma}(N_{\rm hits}) =
\frac{\sum_{i=N_{\rm hits}}^{24} N_{f}^{i}}{\sum_{i=0}^{24} N_{f}^{i}}
,
\end{equation}

where $N_{e}^{i}$ and $N_{f}^{i}$ are the number of
electrons and fake electrons in the bin $N_{\rm hits} = i$
from $Z \to ee$ and multijet data events, respectively.
The maximum number of hits is
24, as the maximum of CFT hits is 16 and the maximum of SMT hits is 8.
Electrons tend to have $D_{\rm hor} \approx 1$, while
photons tend to have values close to 0.

\begin{figure*}[htbp]
\centering
\includegraphics[width=0.49\textwidth]{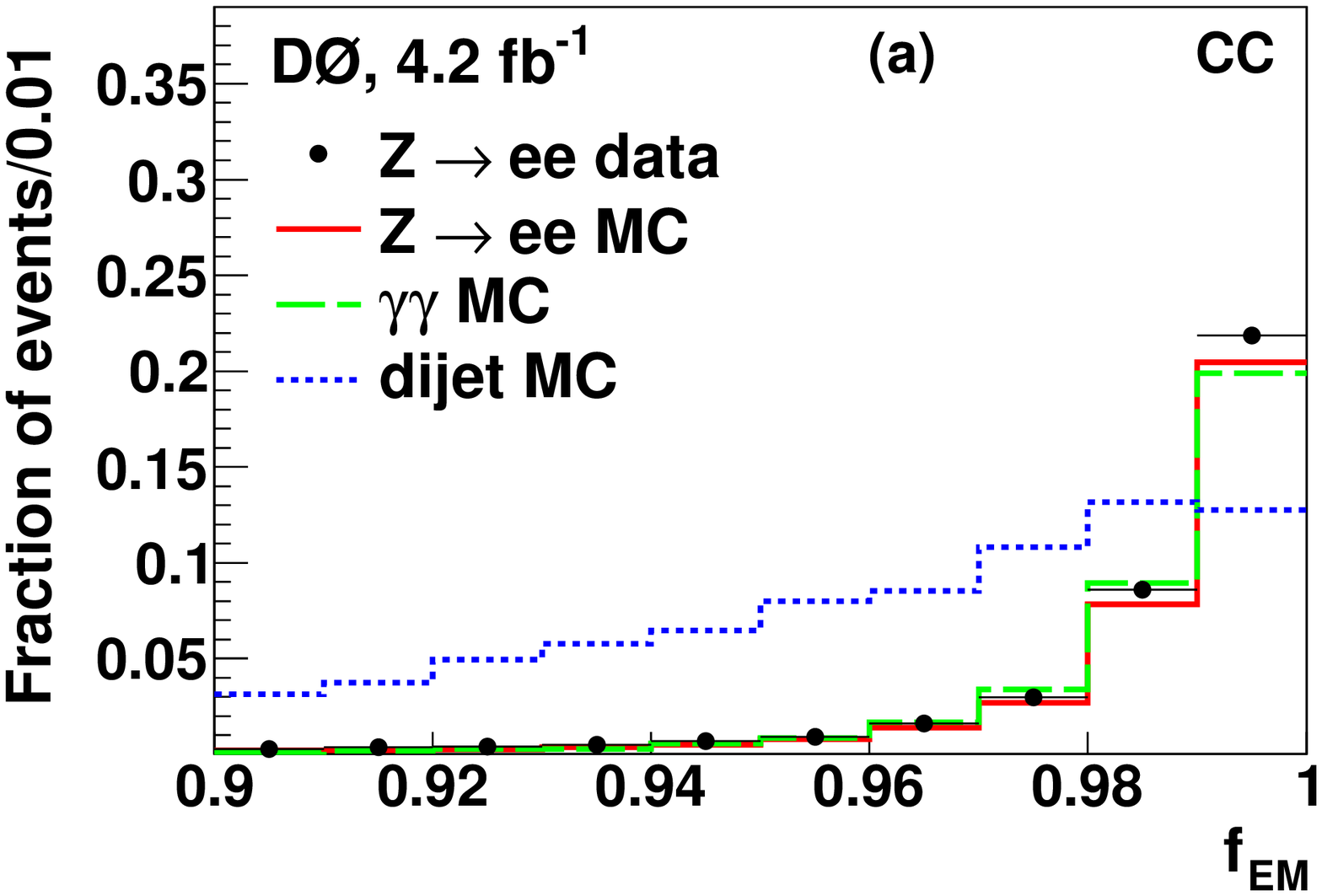}
\includegraphics[width=0.49\textwidth]{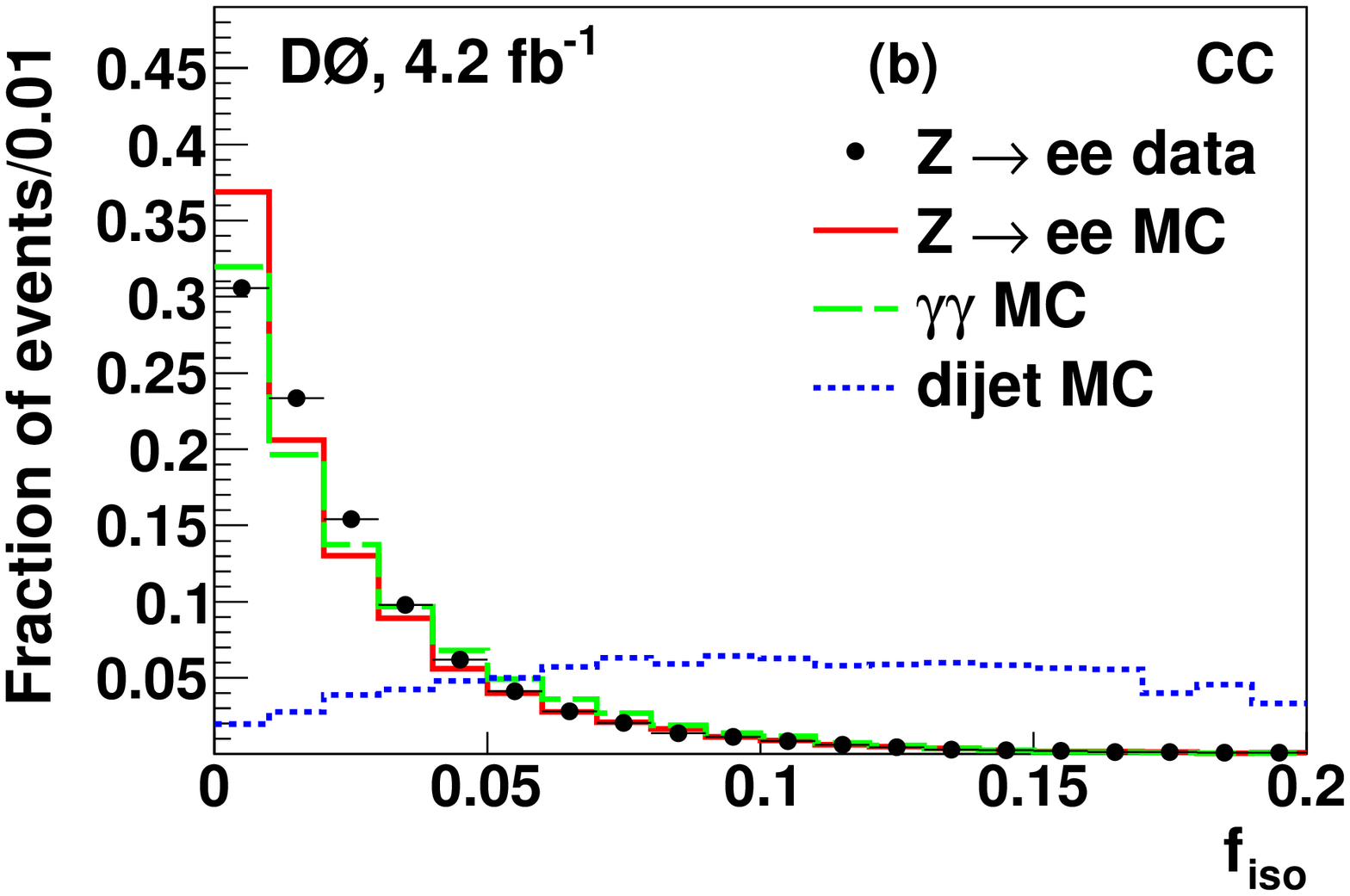}
~\\
\includegraphics[width=0.49\textwidth]{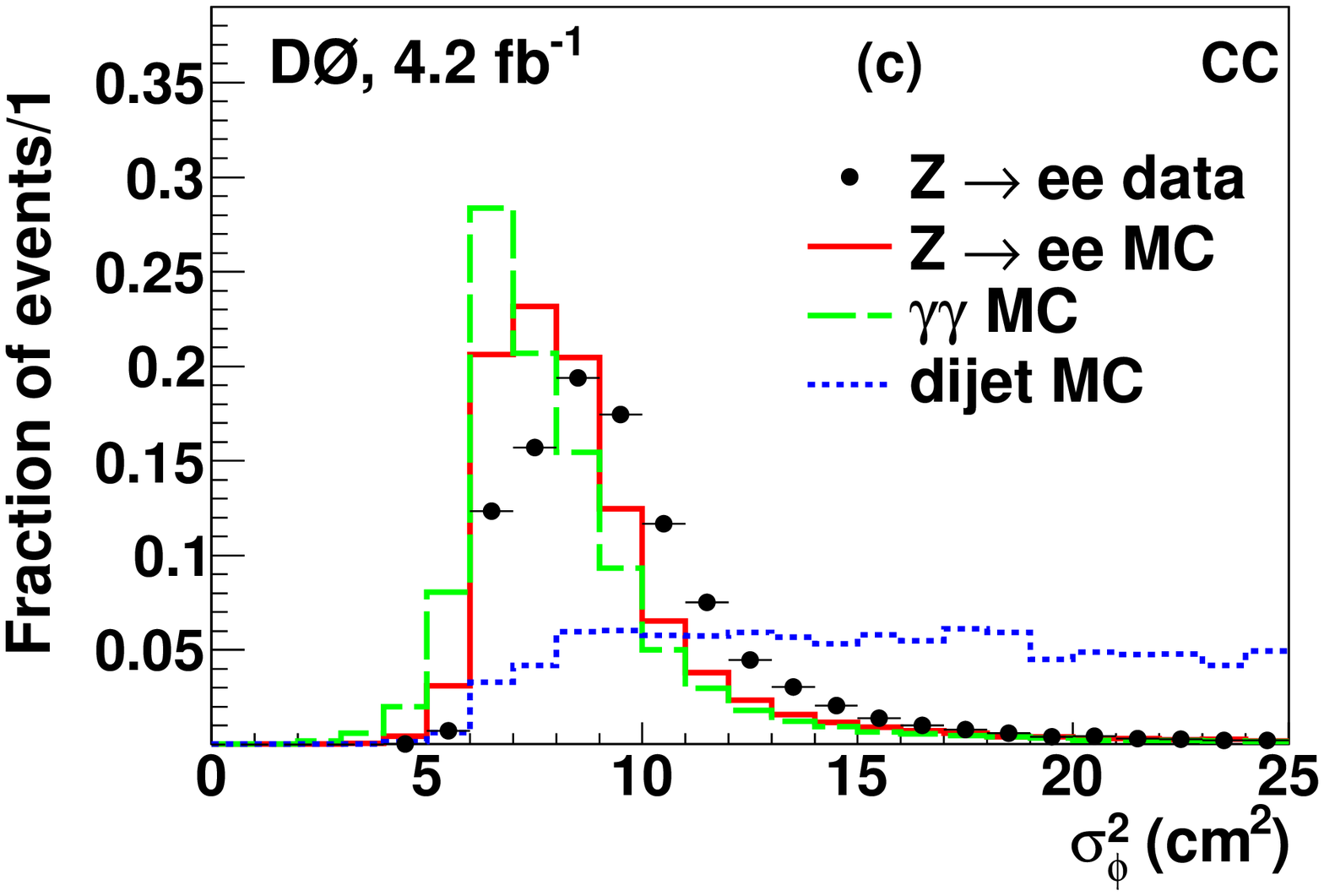}
\includegraphics[width=0.49\textwidth]{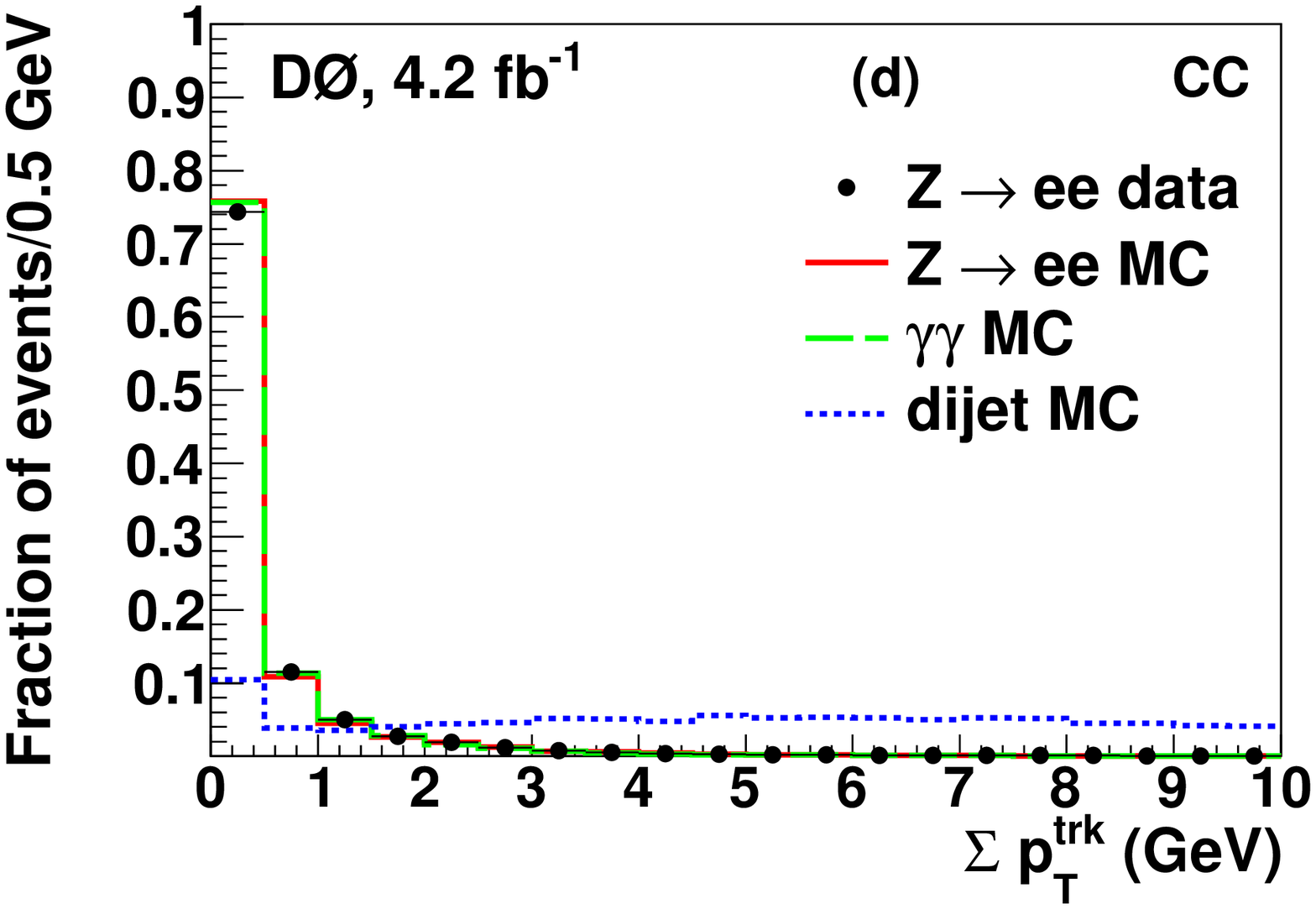}
~\\
\includegraphics[width=0.49\textwidth]{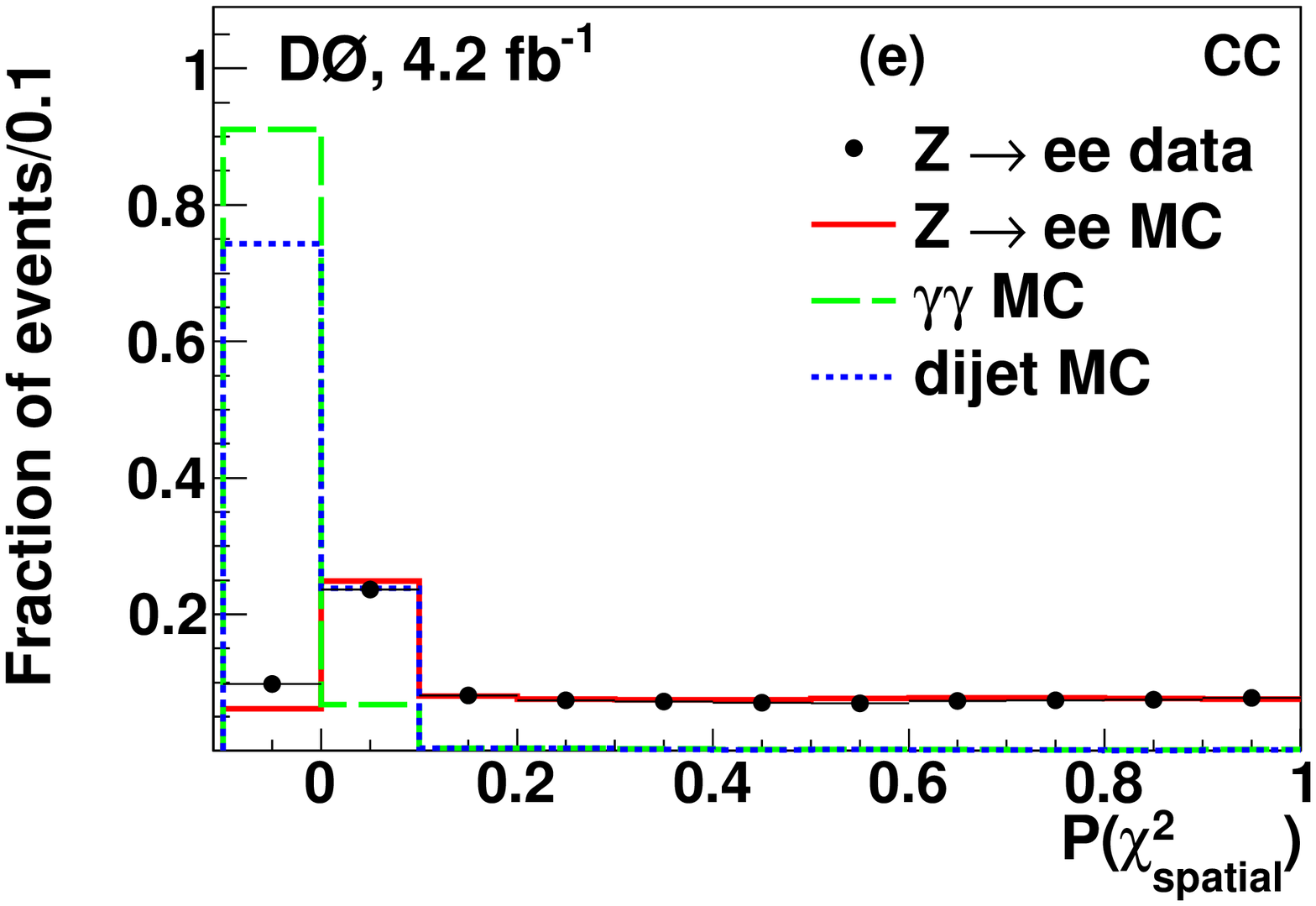}
\includegraphics[width=0.49\textwidth]{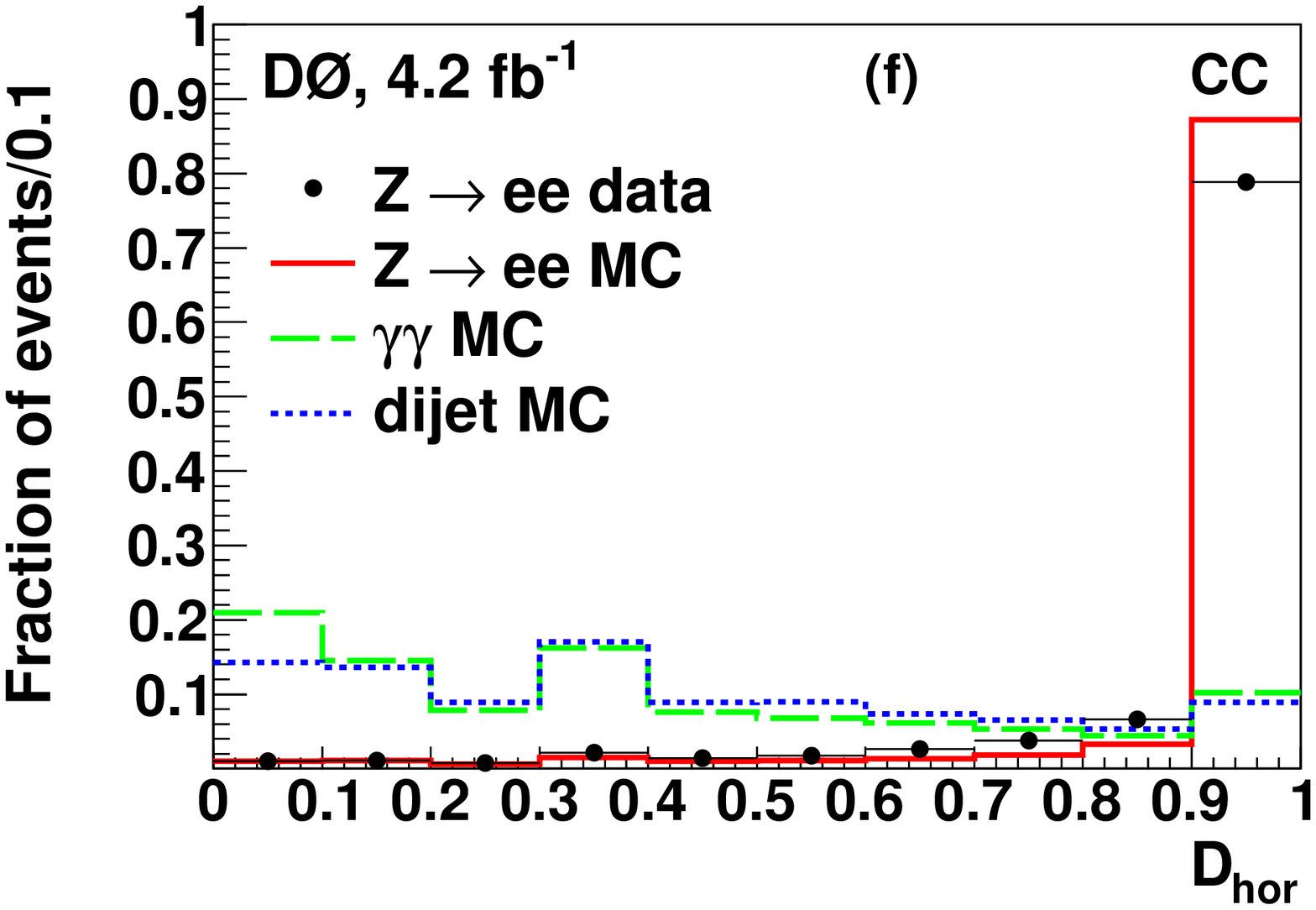}
\caption{
Normalized distributions of EM object identification variables
as defined in Sect.~\ref{emid} for $Z \to ee$ data and MC events,
and for diphoton and dijet MC events in the CC.
Presented are (a) the EM energy fraction, (b) the EM shower isolation,
(c) the width of the EM shower in the transverse plane,
(d) the track isolation, (e) the track matching probability,
and (f) the hits on road discriminant.
The first bin of the track matching probability distribution indicates no track match.
}
\label{fig:emid-varible-CC}
\end{figure*}

\begin{figure*}[htbp]
\centering
\includegraphics[width=0.49\textwidth]{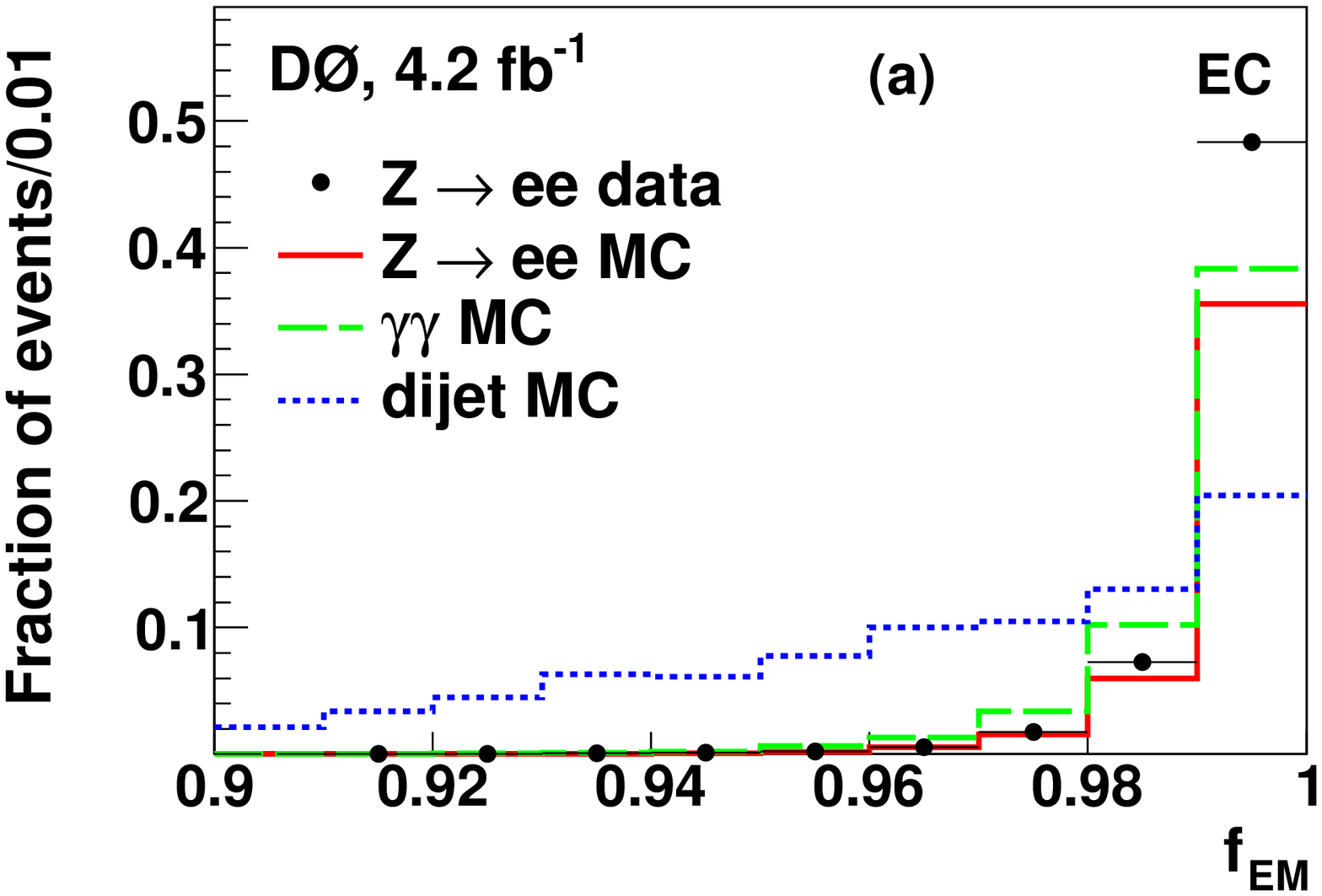}
\includegraphics[width=0.49\textwidth]{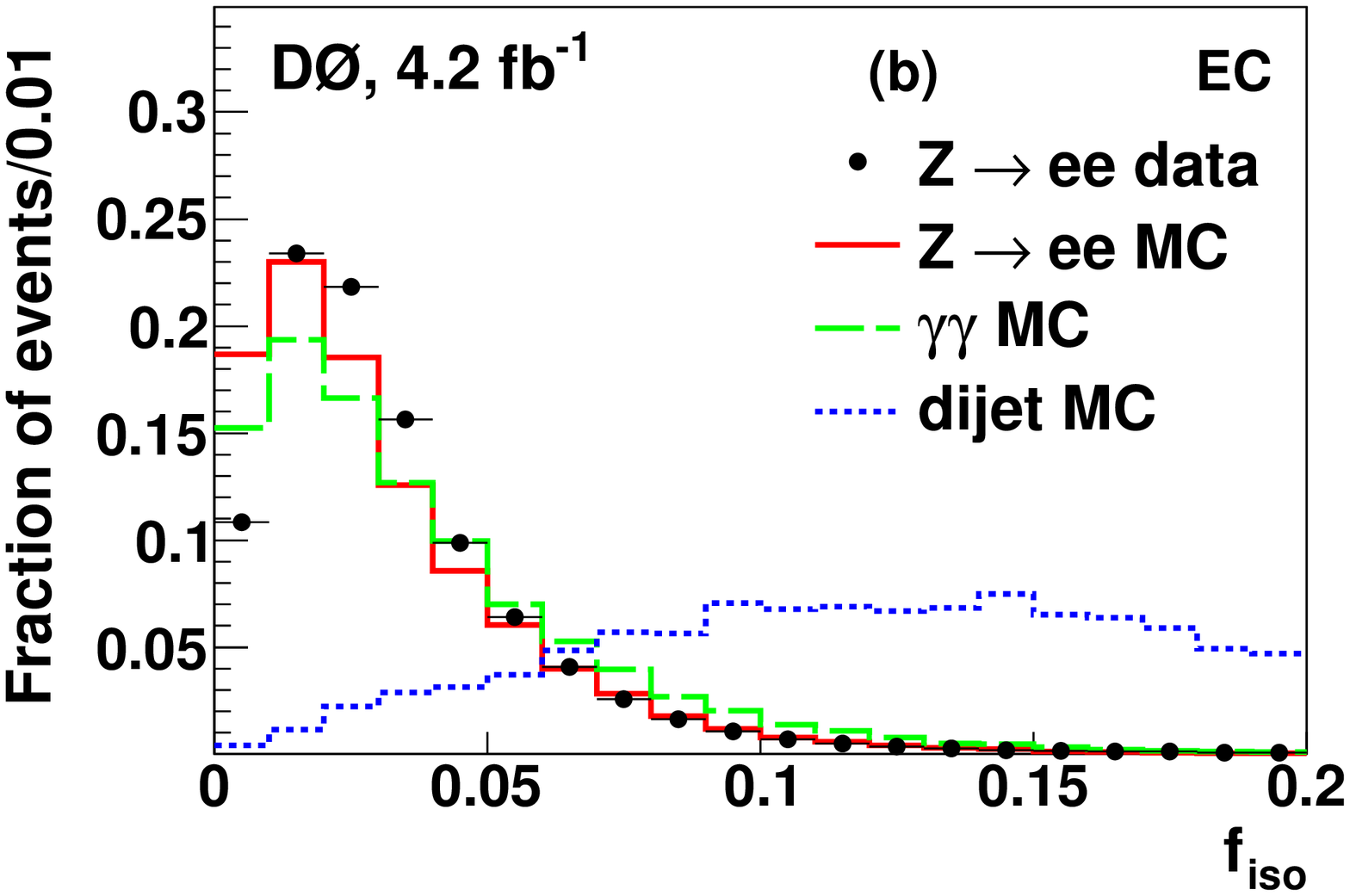}
~\\
\includegraphics[width=0.49\textwidth]{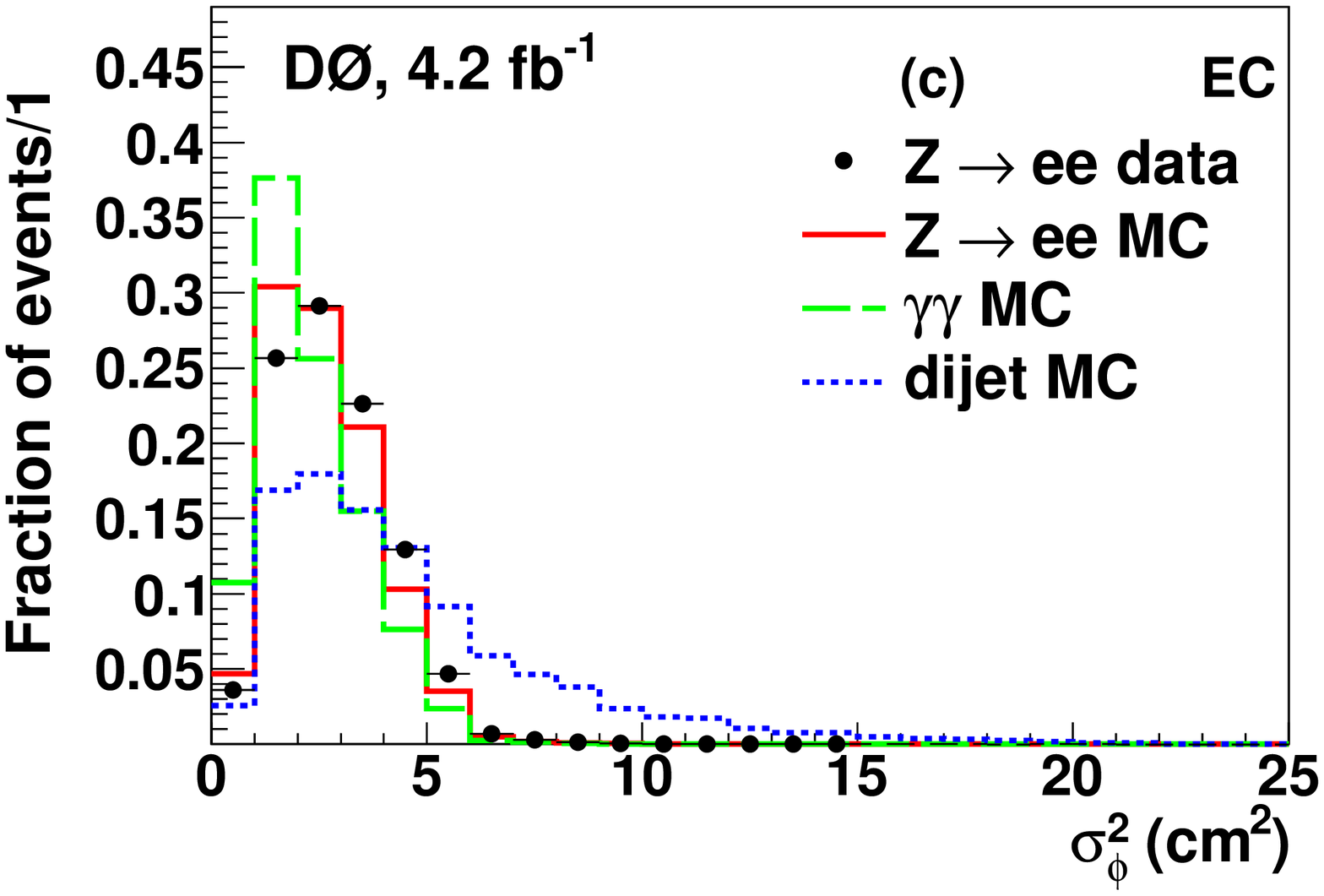}
\includegraphics[width=0.49\textwidth]{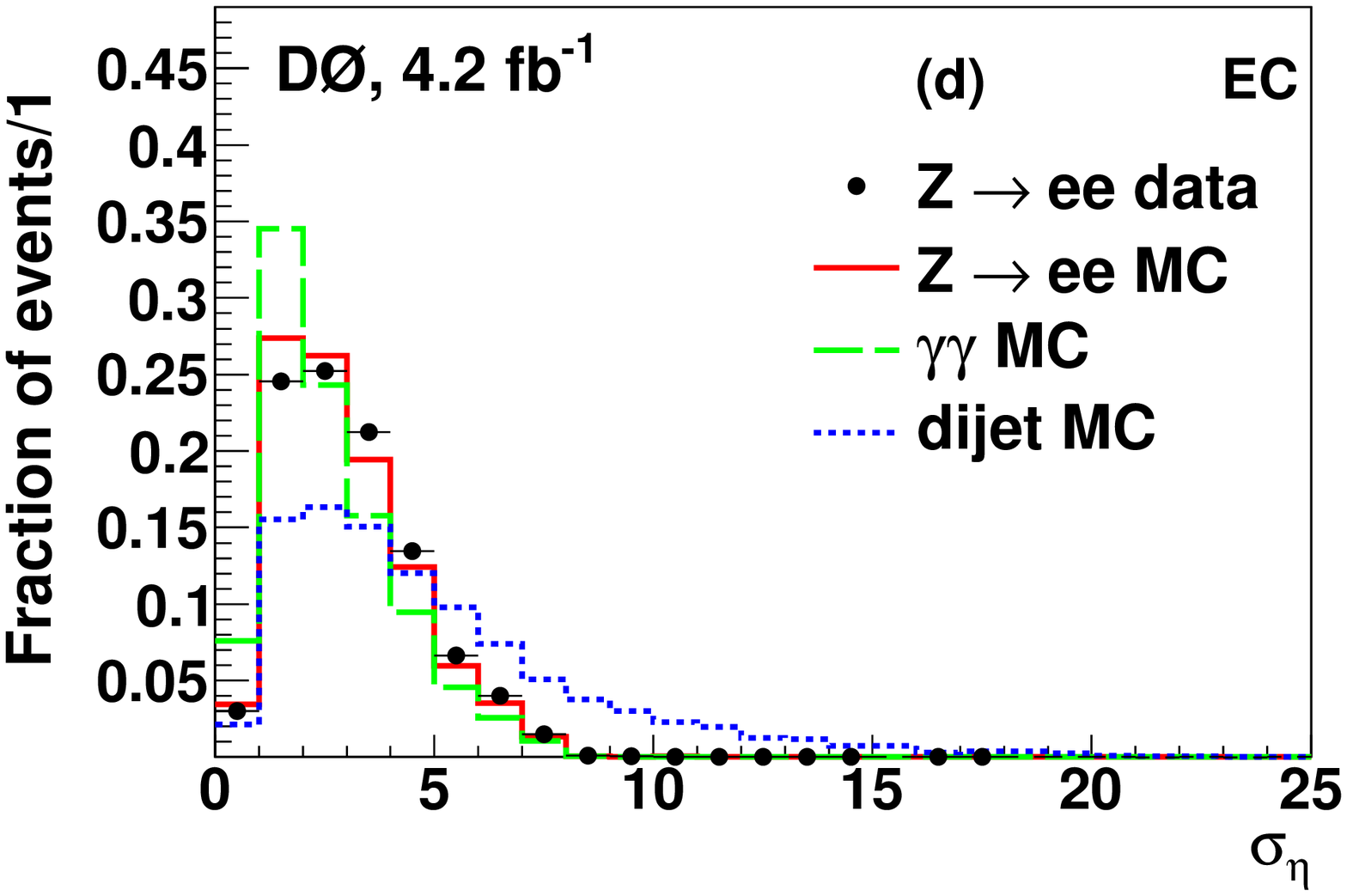}
~\\
\includegraphics[width=0.49\textwidth]{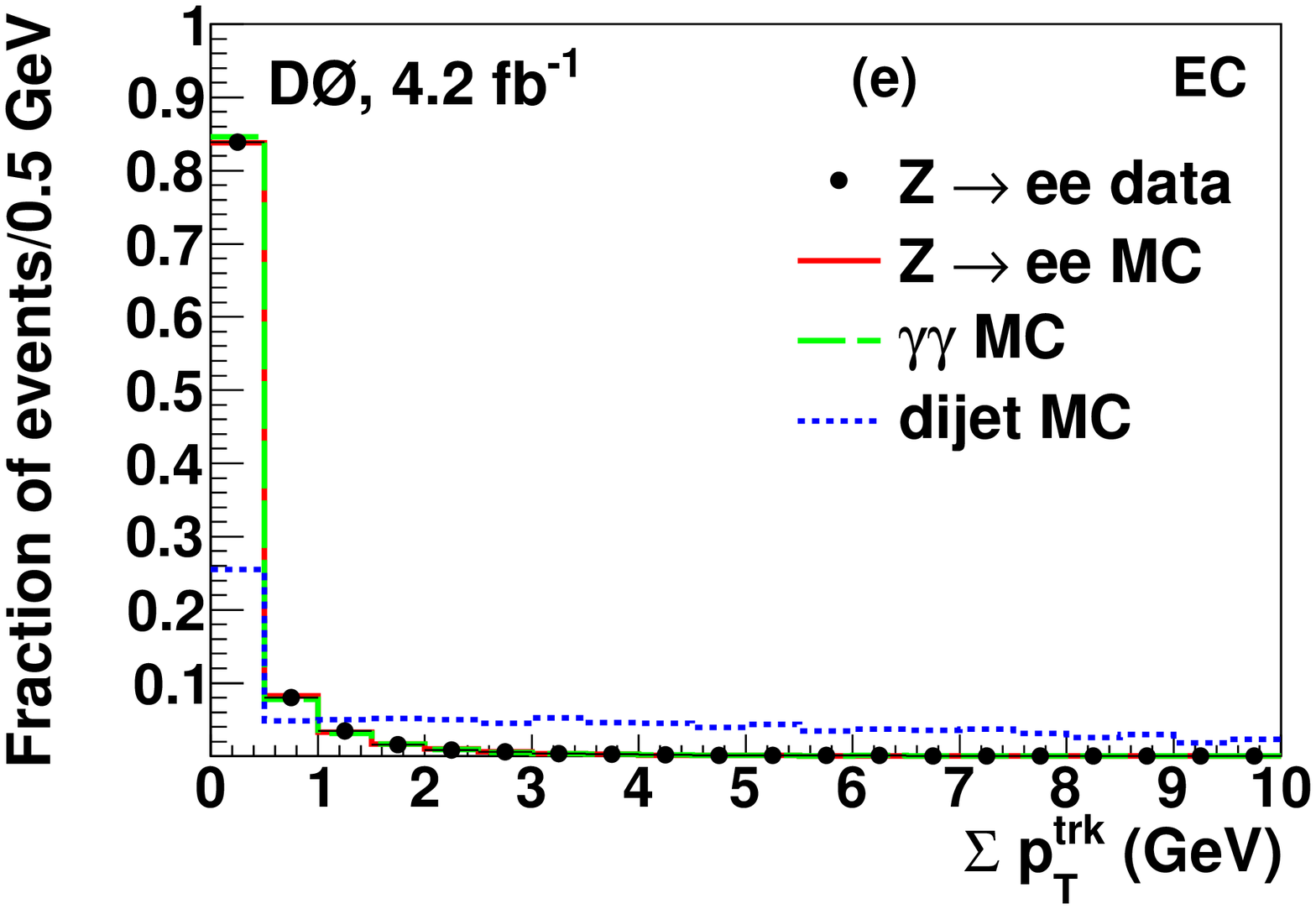}
\includegraphics[width=0.49\textwidth]{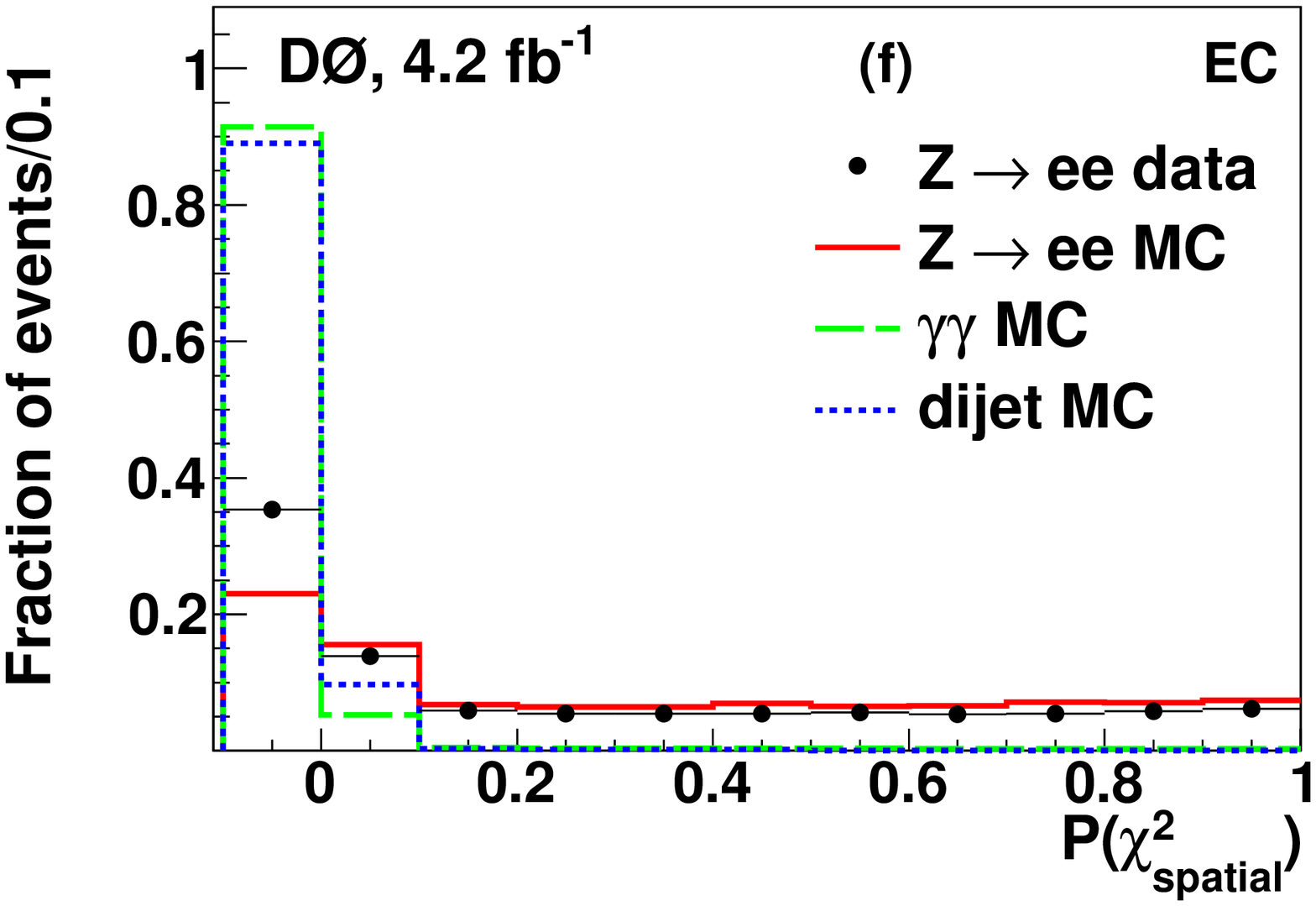}
\caption{
Normalized distributions of EM object identification variables
as defined in Sect.~\ref{emid} for $Z \to ee$ data and MC events,
and for diphoton and dijet MC events in the EC.
Presented are (a) the EM energy fraction, (b) the EM shower isolation,
(c) the width of the EM shower in the transverse plane,
(d) the width of the EM shower in the pseudorapidity direction,
(e) the track isolation, and (f) the track matching probability.
The first bin of the track matching probability distribution indicates no track match.
}
\label{fig:emid-varible-EC}
\end{figure*}

\begin{figure*}
\centering
\includegraphics[width=0.49\textwidth]{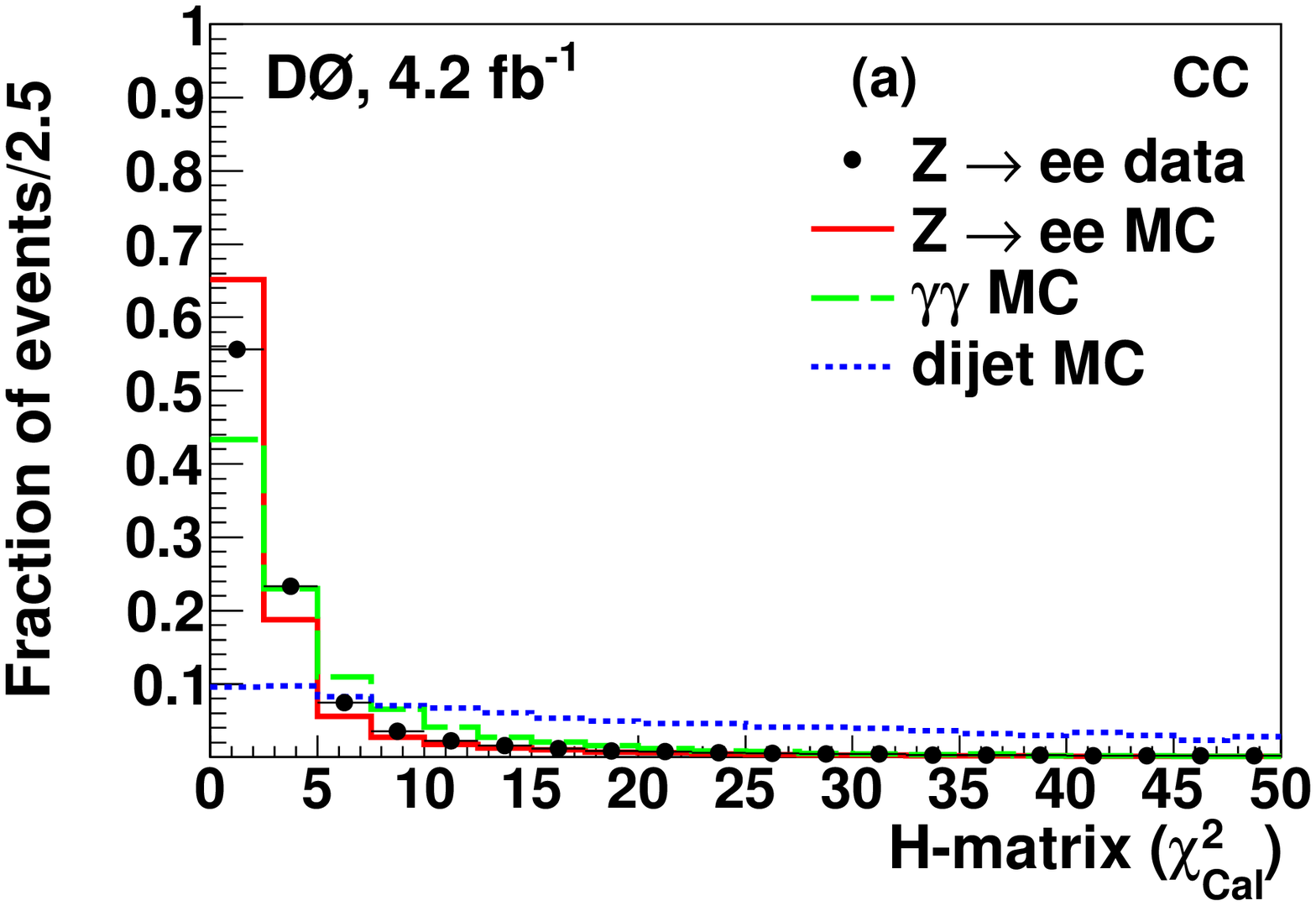}
\includegraphics[width=0.49\textwidth]{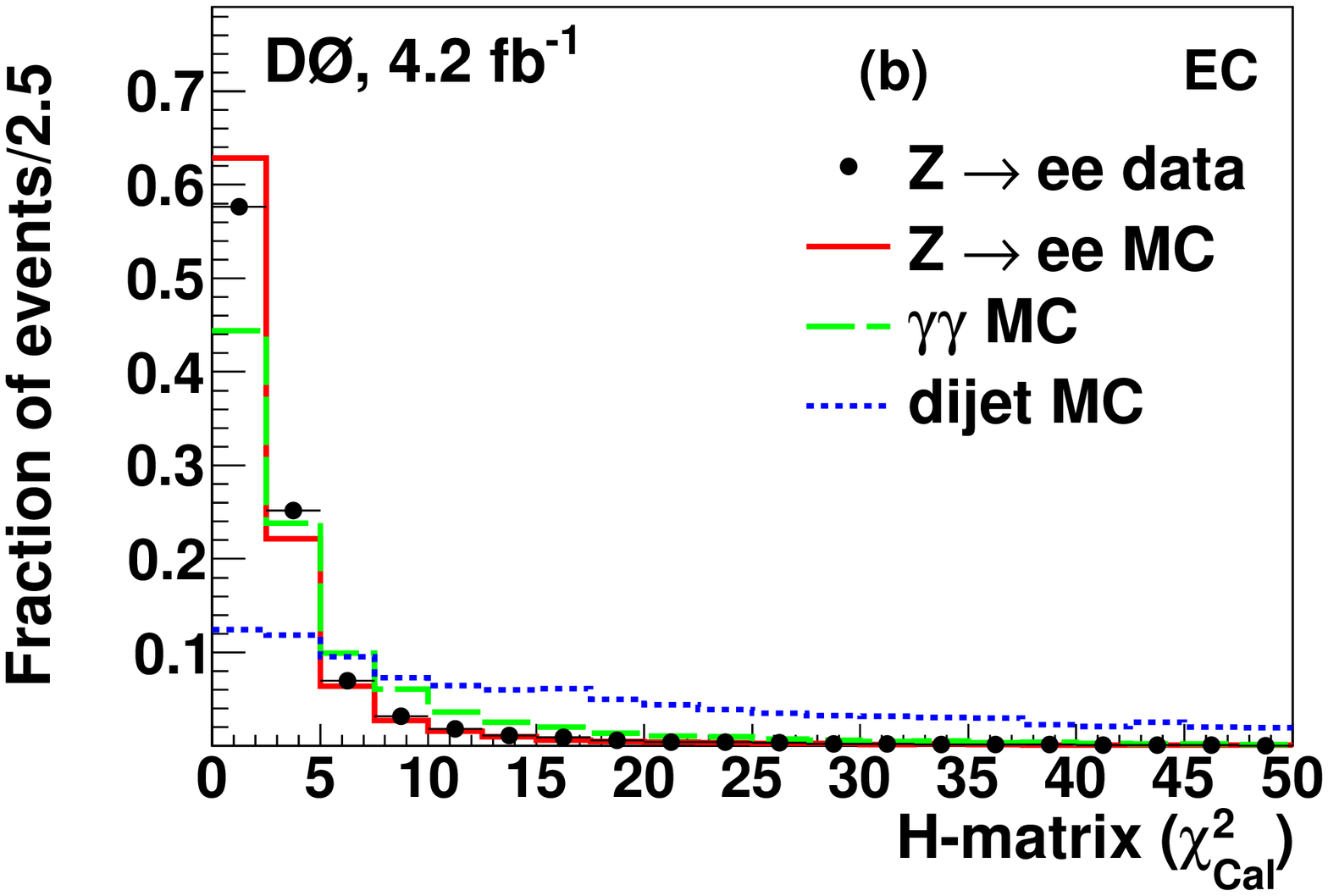}
\caption{
The distributions of $\chi^{2}_{\rm Cal}$
for EM candidates for
$Z \to ee$ data and MC events, and for diphoton and dijet MC events
in the CC (a) and EC (b).
}
\label{fig:emid-varible-hmx}
\end{figure*}

Figs.~\ref{fig:emid-varible-CC}-\ref{fig:emid-varible-hmx}
show distributions of identification variables
for EM candidates from
$Z \to ee$ data and MC events, as well as from diphoton and dijet MC events\footnote{
The distributions shown in this paper are generally derived
from subsets of the Run II data sample.  The small variations observed between
different periods of Run II are treated as systematic uncertainties in
physics analyses.}.
As can be inferred from the distributions, the simulation has some
imperfections in modeling the shower shapes mainly caused by an
insufficient description of uninstrumented material \cite{MwPRD}. 
This is accounted for
when correcting simulated electron and photon identification
efficiencies utilizing data as described in
Sects.~\ref{sec::results_ele} and~\ref{sec::results_gam}, respectively.

\section{Multivariate identification methods} 
\label{sec::recoid_ann}
The variables described in Sect.~\ref{emid} allow efficient
identification of electron and photon candidates.  However, to
maximize the identification efficiencies
of electrons and photons
and to minimize the misidentification rate from jets in
physics analyses, various
multivariate analysis (MVA) techniques are explored.
One MVA technique, the H-matrix method, has already been discussed in
Sect.~\ref{hmx}.
Two more types of MVAs that are used in physics analyses are
a Likelihood method
for electrons and a neural network (NN) method for electrons and
photons.
H-matrix, Likelihood, and NN achieve an improved
background rejection.
However, the H-matrix is mainly based on the calorimeter
information, while the Likelihood method 
includes the tracking information in addition, 
while the advantage of the NN is that it includes CPS information.
The electron identification efficiency
and purity are therefore found to be improved when
these MVA output variables are utilized together with other electron
reconstruction variables as input to a Boosted Decision Tree (BDT)~\cite{BDT}.
All MVAs except the H-matrix are described in this section.

\subsection{Electron Likelihood}

Likelihood-based identification of electron candidates is
an efficient technique for separating electrons from background
by combining information from various detector components into a single
discriminant.

There are several mechanisms by which particles,
either isolated or in jets, may produce electron signatures.
Photon conversions may be marked
by the presence of a track very close to the track matched to
the EM cluster, or a
large $E_T/p_T$ when the closely situated $ee$ pair is
reconstructed as a single EM cluster and only one track is
identified. Here, $E_T$ is the transverse energy of the cluster
measured by the calorimeter and $p_T$ is the transverse momentum of
the associated track measured by the tracker. The calorimeter
quantities describing the shower shape, however, are nearly identical
to that of an electron,
though photon calorimeter clusters may be slightly wider than an
electron shower. Neutral pions $(\pi^{0})$ may also have nearby tracks, as they
are generally produced in association with other charged hadrons. Since the
$\pi^0 \to \gamma\gamma$ decay would have to overlap with a charged
hadron track in
order to fake an electron, the track matching quantity could be poor, and the track
would not necessarily be isolated.
The H-matrix $\chi^2_{\rm Cal}$ and
$f_{\rm EM}$ of the EM object may be influenced by the surrounding hadrons.
The following eight variables are used to calculate the electron
likelihood\footnote{For
definitions see Sect.~\ref{sec::recoid}.}:
\begin{itemize}
\item EM energy fraction $f_{\rm EM}$;
\item EM shower isolation $f_{\rm iso}$;
\item H-matrix $\chi^2_{\rm Cal}$;
\item $E_T/p_T$;
\item transverse impact
parameter of the selected track with respect to the $p\bar{p}$ collision vertex;
\item number of tracks with $p_{T} > 0.5$ GeV in a cone of radius ${\cal R} = 0.4$ around and
including the matched track;
\item cluster-track matching probability P($\chi^{2}_{\rm spatial}$);
\item track isolation variable $\Sigma p_{T}^{\rm trk}$.
\end{itemize}

\begin{figure*}
\centering
\includegraphics[width=0.49\textwidth]{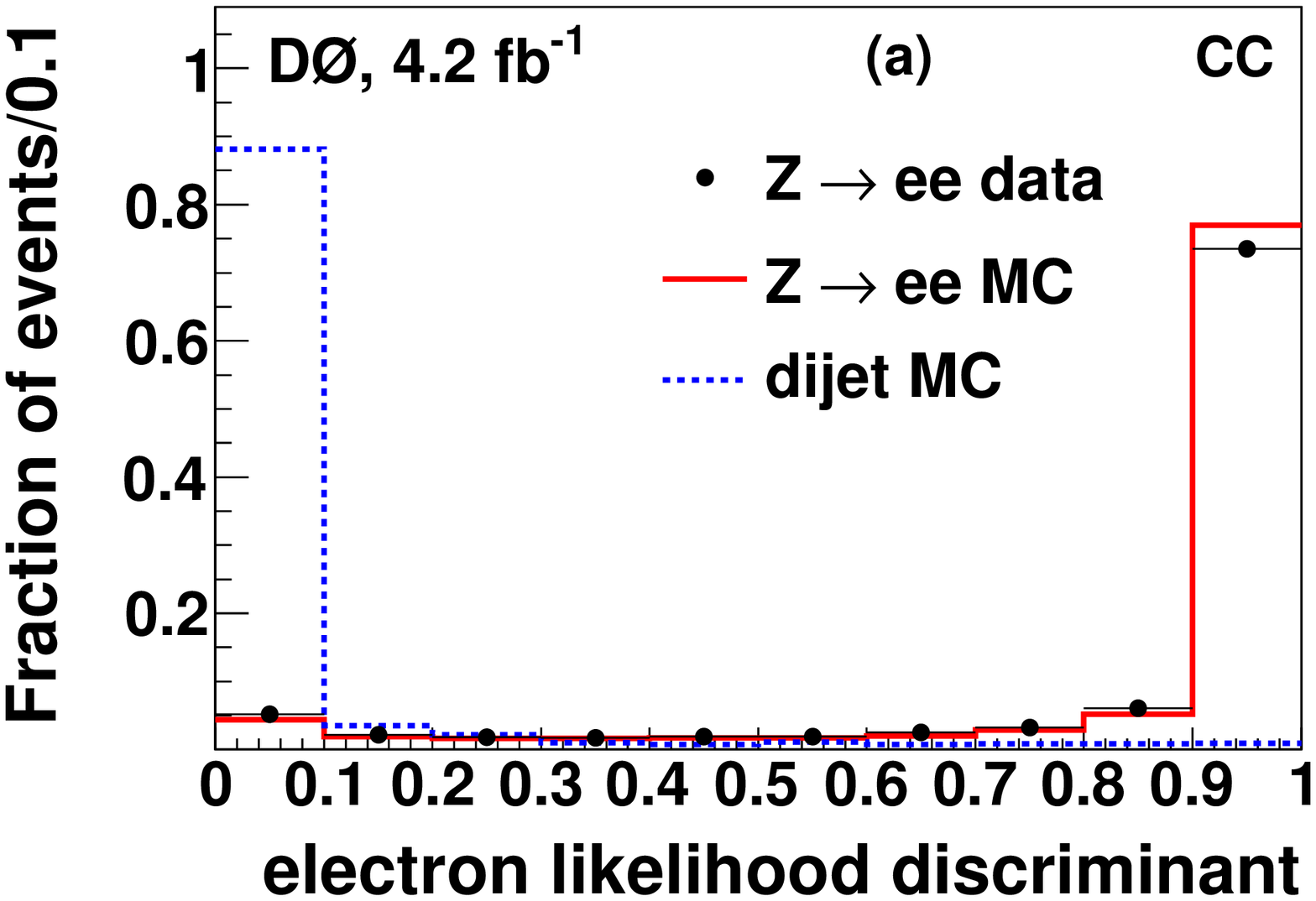}
\includegraphics[width=0.49\textwidth]{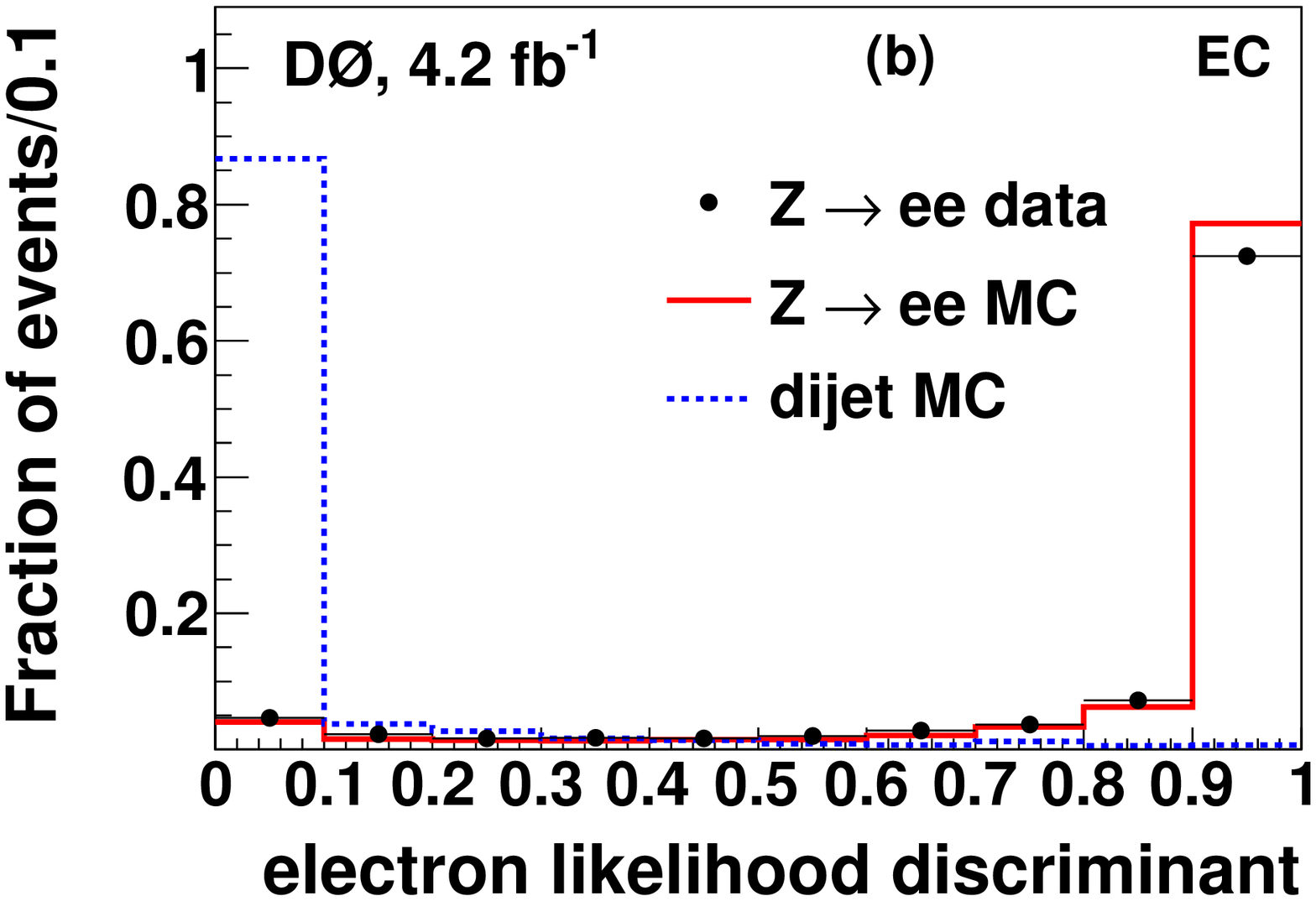}
\caption{
Distribution of the electron likelihood discriminant of electron
candidates in $Z \to ee$ data and MC events,
and in dijet MC events in the CC (a) and EC (b).
}
\label{fig:emid-varible-EC2}
\end{figure*}

The distributions of these eight variables are
normalized to unit area to generate probability density distributions for
each variable 
from $Z \to ee$ and dijet data for signal and background, respectively.
These distributions are used to assign a probability for a given
EM object to be signal or background.
To quantify the degree of correlation between the input variables, we
calculate the correlation coefficients. We find that most of the
combinations have correlation coefficients close to zero and hence are
mutually uncorrelated. Others do not exceed 55\% for signal or fake electrons.
The product of individual probabilities from all variables 
is correlated with the overall probability 
for the EM object to be an electron.
To differentiate between signal-like and background-like
electron candidates, a likelihood discriminant is calculated:

\begin{equation}
\mathcal{L} = \frac{P_{\rm sig}}{P_{\rm sig}+P_{\rm bkg}},
\end{equation}

where $P_{\rm sig}$ and $P_{\rm bkg}$ are the overall probabilities
for signal and background, respectively.
Distributions of this discriminant for electron candidates in the CC
and EC are presented
in Fig.~\ref{fig:emid-varible-EC2}.
This demonstrates the enhanced
power to separate between genuine electrons, which peak at
large values of the discriminant, and jets,
which peak at low values.

\subsection{Neural Network for electron and photon
identification}
To further suppress jets misidentified as electrons and photons, we
train an NN~\cite{NN} using a set of variables that are
sensitive to differences between electrons (photons) and jets.
The variables, selected to explore both the
tracker activity and the energy distribution in the calorimeter and
CPS, are listed below.
\begin{itemize}
\item fraction of the EM cluster energy deposited in the first EM
calorimeter layer ($f_{\rm EM1}$);
\item number of cells above an EM cluster $E_{T}$-dependent
threshold, given by $0.004 \times E_{T}$ (in GeV) $+$ 0.25~GeV in the first
EM calorimeter layer within ${\cal R} < 0.2$ ($N_{\rm cells}^{{\cal R} < 0.2}$)
and $0.2 < {\cal R} < 0.4$ ($N_{\rm cells}^{0.2 < {\cal R} < 0.4}$) of
the EM cluster;
\item track isolation variable $\Sigma p_{T}^{trk}$;
\item number of charged particle tracks with $p_T > 0.5$ GeV originating
from the $p\bar{p}$ collision vertex within ${\cal R} < 0.05$ of the EM cluster
($N_{\rm trks}^{{\cal R} < 0.05}$);
\item number of CPS clusters within ${\cal R}< 0.1$ of the EM
cluster ($N_{\rm cps}$);
\item squared width of the energy deposit in the CPS:

\begin{equation}
\sigma_{\rm CPS}^{2}=\frac{\sum_{i} E_{i}^{2} \times (\phi_{\rm EM} -
\phi_{i})^{2}}{\sum_{i}E^{2}_{i}},
\end{equation}

where $E_{i}$ and $\phi_{i}$ are the energy and azimuthal angle of
the $i^{th}$ strip in CPS in the direction of the EM cluster
and $\phi_{\rm EM}$ is the azimuthal angle of the EM cluster at the EM3 layer;
\item $\chi^2_{\rm Cal}$ calculated from the H-matrix.
\end{itemize}

Separate NNs are built for electrons and photons in the CC,
whereas a single NN is used for electrons and photons in the EC.
Table~\ref{ANN-input} lists the input variables utilized in each NN.

For the construction of the NN for electrons in the CC, the seven variables above
are used as inputs ($e$NN7).
Here, $Z \to ee$ data events define the signal,
and dijet data events define the background.
Performance checks have been performed using $Z \to ee$
and dijet MC events.

The NN for CC photons ($\gamma$NN5) is built from the same variables as $e$NN7
but excluding the tracker-based variable $N_{\rm trks}^{{\cal R} < 0.05}$, and $f_{\rm EM1}$
since its distribution varies significantly with the $E_T$
of the EM cluster.
The direct diphoton MC defines the
signal, and dijet MC events are used as background in training
the NN.  For testing, the reconstructed radiated photon from
$Z \to \ell\ell\gamma$ ($\ell = e, \mu$) events in data and MC events, and dijet MC
events are used.

A photon NN ($\gamma$NN4) is built with four input variables as
listed in Table~\ref{ANN-input} for the EC region.
The training is based on direct diphoton and dijet MC events.
The same types of events used to test $\gamma$NN5
are used to test $\gamma$NN4.
Considering the similar performance of the input variables
of electrons and photons in the EC, $\gamma$NN4 is found to work well, 
and is used, for electron identification in the EC.

\begin{figure*}
\centering
\includegraphics[width=0.49\textwidth]{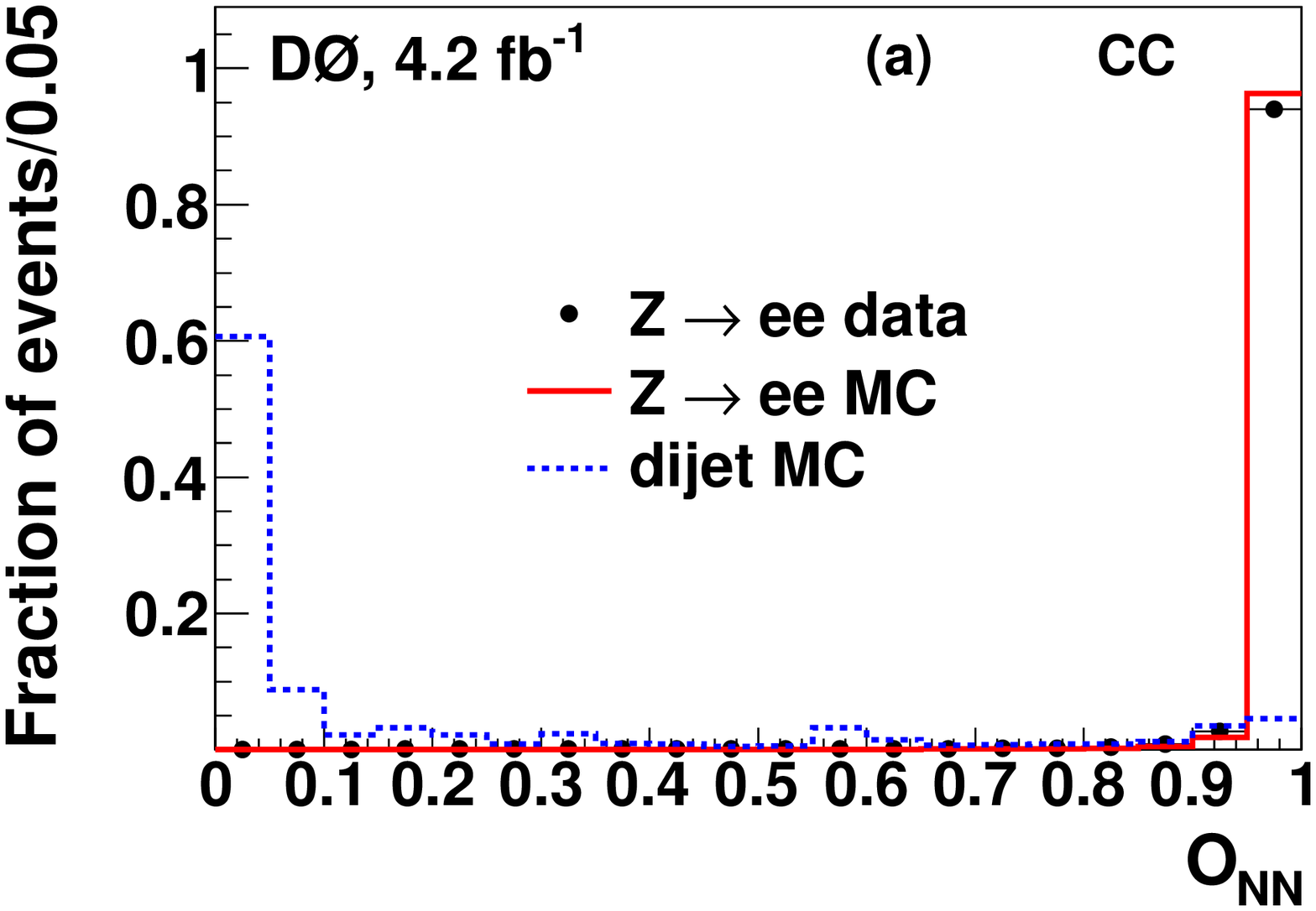}
\includegraphics[width=0.49\textwidth]{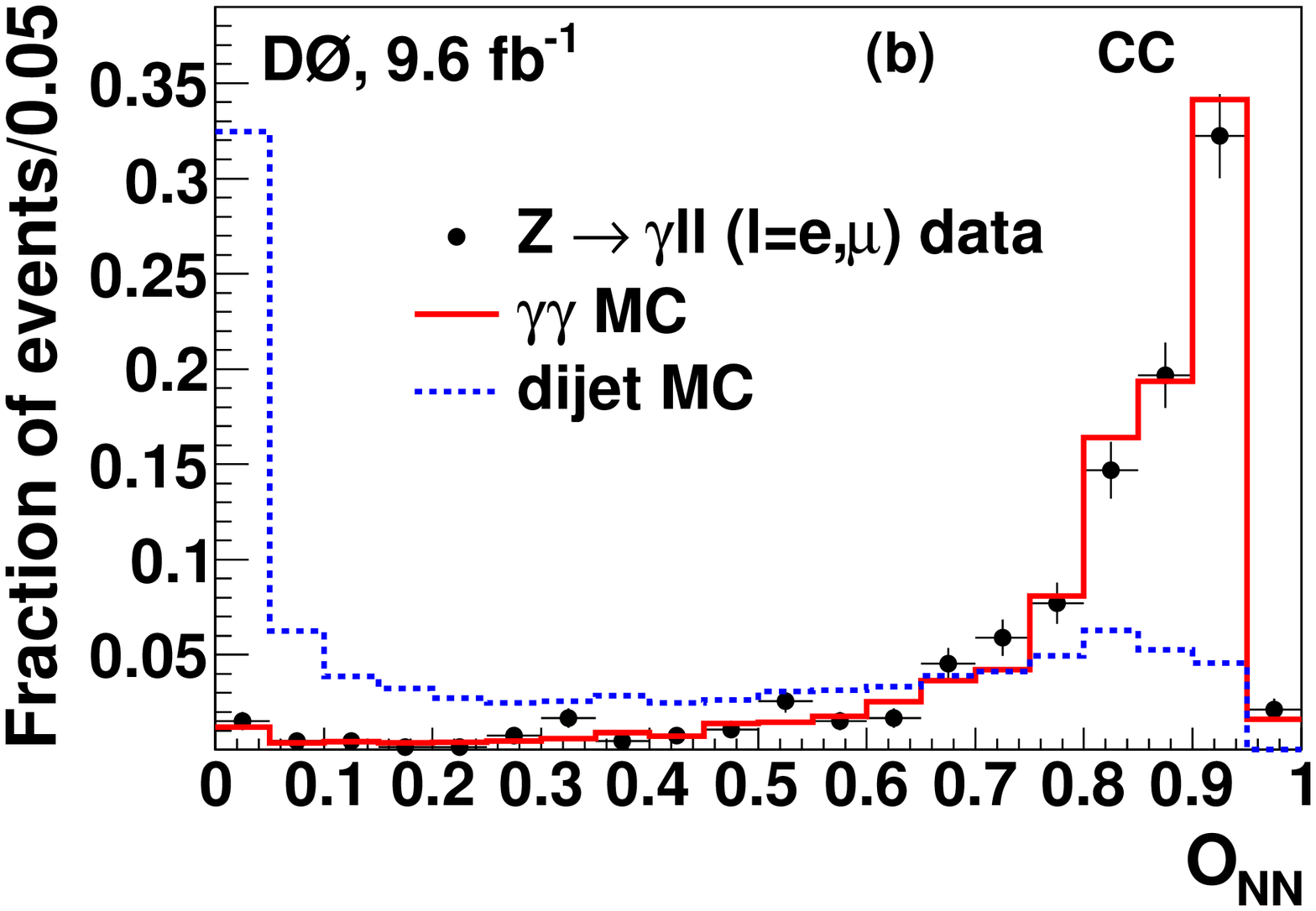}
\includegraphics[width=0.49\textwidth]{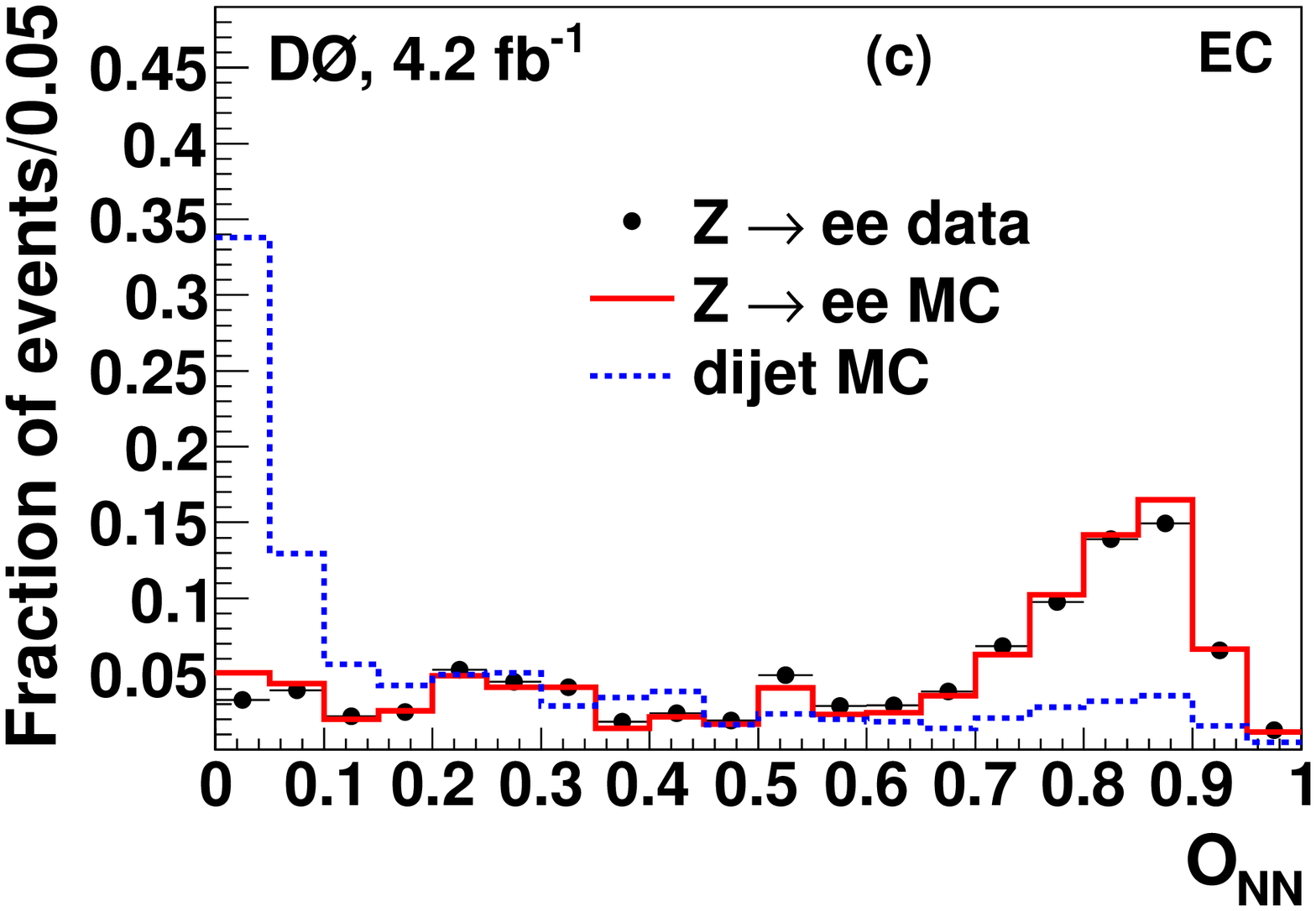}
\includegraphics[width=0.49\textwidth]{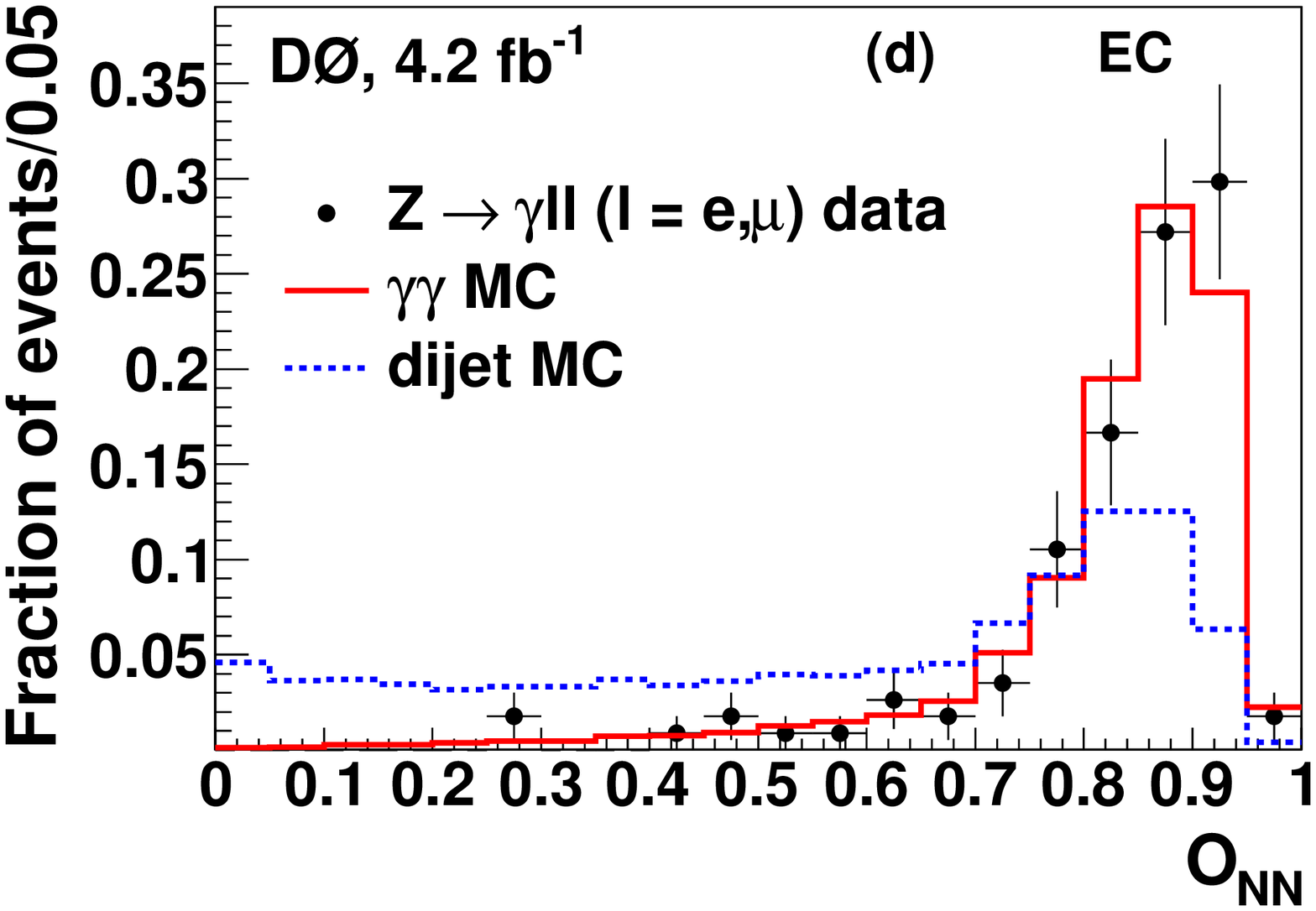}
\caption{
The output distributions of eNN7 for CC electrons (a),
$\gamma$NN5 for CC photons (b), $\gamma$NN4
for EC electrons (c) and EC photons (d).
}
\label{fig:ANN}
\end{figure*}

Figure~\ref{fig:ANN} shows the NN output distributions for
reconstructed EM clusters with P($\chi^2_{\rm spatial}$) $>$ 0.001
(electron candidates) and without track match (photon
candidates) for $Z \to ee$ data and MC events, and for dijet background MC events. The
distributions show good agreement between data and MC simulation and demonstrate
good separation between signal and background.

\begin{figure*}
\centering
\includegraphics[width=0.49\textwidth]{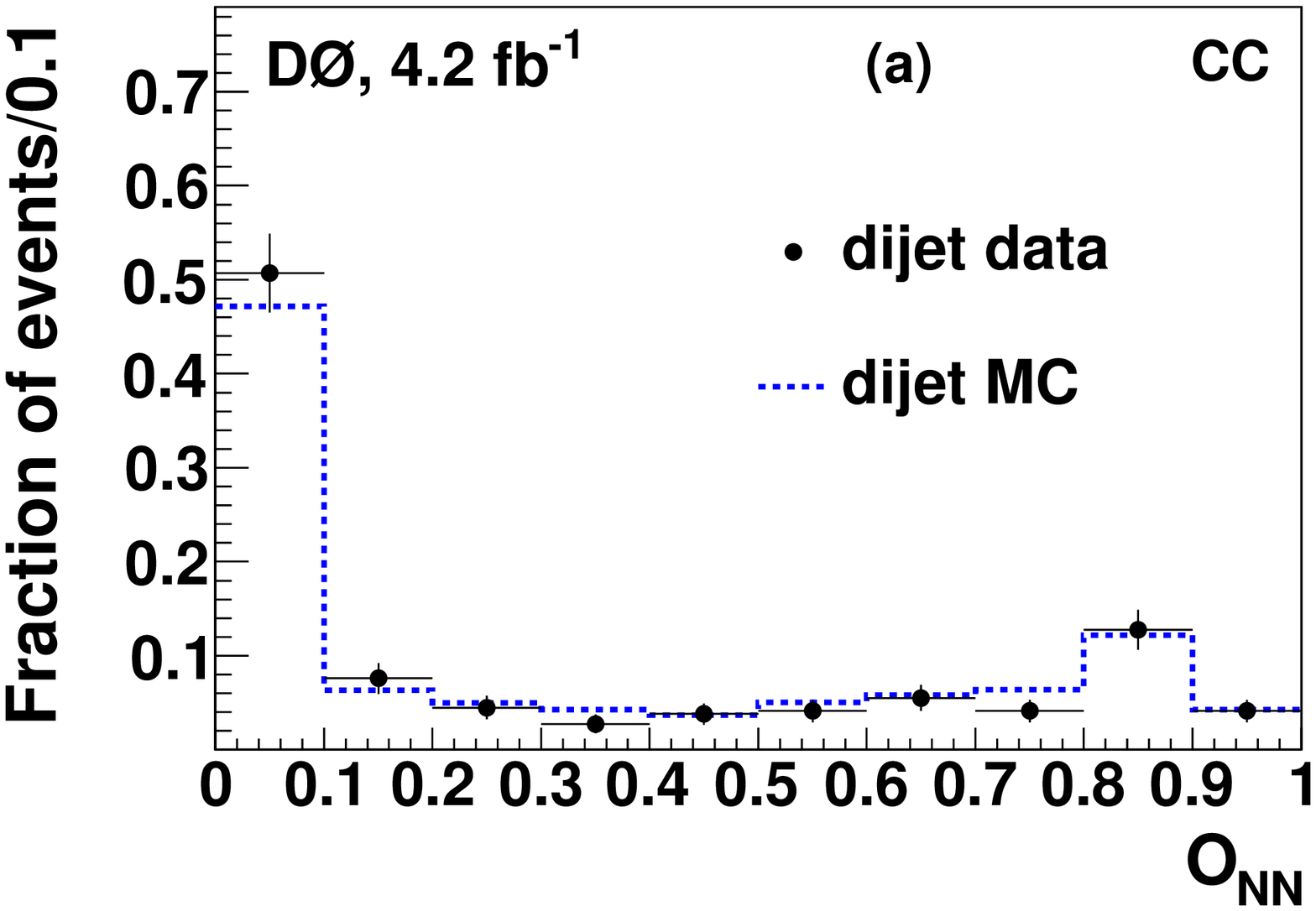}
\includegraphics[width=0.49\textwidth]{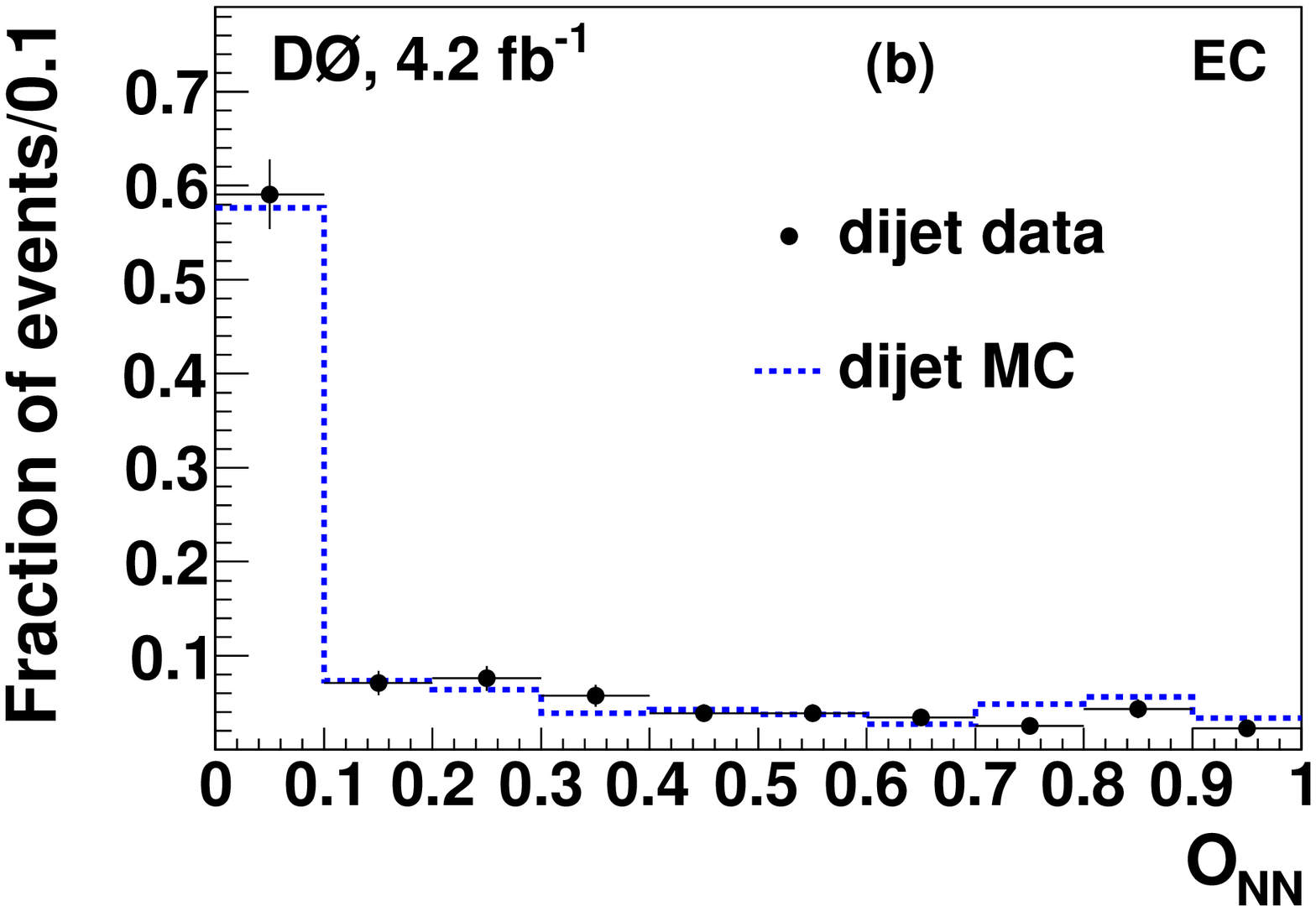}
\caption{
 The output distributions of $\gamma$NN5 in CC (a)
and $\gamma$NN4 in EC (b)
for jet-like EM cluster candidates
from dijet data and MC events.
}
\label{fig:ANNej}
\end{figure*}

To validate the photon NNs for jets, dijet data events in the jet-enriched
calorimeter isolation region $0.07 < f_{\rm iso} < 0.15$ are
compared to MC simulation.
As shown in
Fig.~\ref{fig:ANNej}, good agreement between data and MC is observed.

\begin{table}[htb]
  \centering
  \begin{tabular}{|c|c|c|c|}
    \hline
Input variables & $e$NN7 in CC & $\gamma$NN5 in CC &
  $\gamma$NN4 in EC\\
\hline
$f_{\rm EM1}$& $\surd$  & $-$ & $-$  \\
\hline
$N_{\rm cells}^{{\cal R} < 0.2}$& $\surd$  &$\surd$ & $\surd$ \\
\hline
$N_{\rm cells}^{0.2 < {\cal R} < 0.4}$& $\surd$  & $\surd$ & $\surd$\\
\hline
$\Sigma p_{T}^{\rm trk}$& $\surd$  & $\surd$ & $\surd$ \\
\hline
$N_{\rm trks}^{{\cal R} < 0.05}$& $\surd$  & $-$ & $-$ \\
\hline
$N_{\rm cps}$& $\surd$ & $\surd$ & $-$ \\
\hline
$\sigma_{\rm CPS}^{2}$& $\surd$  & $\surd$ & $-$ \\
\hline
H-matrix $\chi^2_{\rm Cal}$ &$-$ &  $-$ & $\surd$  \\
\hline
 \end{tabular} \caption{\label{ANN-input} \small Input variables used
  in the NNs for electrons and photons in the CC and EC. For electrons
  in the EC, $\gamma$NN4 is used.}
\end{table}

\subsection{Boosted Decision Trees for electron identification}
To enhance the efficiency and purity in electron identification, a
BDT is constructed utilizing variables that are significantly
different for signal and background leading to a strong discrimination
power of the BDT output distribution.
The following variables are used to construct the BDT:
%
\begin{comment}
The $Z \to ee$ and dijet data events are
used for training of a BDT.
The shapes of all utilized variables are significantly different for
signal and background leading to a strong discrimination power
for the BDT output distribution.
Four separate BDTs are
trained, corresponding to the four combinations of CC and EC with high
and low instantaneous luminosity (\ilum).
The motivations for training separate BDTs for the CC
and EC are the differences in signal-to-background ratio in the two
regions, and the fact that the CC region has better tracking coverage. Similarly,
differences in signal-to-background ratio and in the resolution of some
variables motivate the training of separate BDTs for high
(\ilum\ $> 1.6 \times 10^{32}$ cm$^{-2}$s$^{-1}$)
and low instantaneous luminosity events.
The following variables are used to build the BDTs:
\end{comment}
%
\begin{itemize}
\item EM energy fraction $f_{\rm EM}$;
\item EM shower isolation $f_{\rm iso}$;
\item energy fraction in EM1, EM2, EM3, EM4 and FH1;
\item $\sigma_{\phi}^{2}$ in EM1, EM2, EM3, EM4 and FH1;
\item $\sigma_{\eta}$ in EM1, EM2, EM3, EM4 and FH1;
\item H-matrix $\chi^{2}_{\rm Cal}$;
\item $\Sigma p_{T}^{\rm trk}$;
\item cluster-track matching probability P($\chi^2_{\rm spatial}$);
\item ``hits on road'' discriminant $D_{\rm hor}$ in CC;
\item ratio $E_T/p_T$;
\item number of hits from CFT fibers $N_{\rm CFT}$;
\item number of hits from SMT strips $N_{\rm SMT}$;
\item ratio $N_{\rm CFT}/N_{\rm SMT}$;
\item number of hits in the first layer of the SMT;
\item number of charged particle tracks with $p_T > 0.5$ GeV originating
from the $p\bar{p}$ collision vertex within ${\cal R} < 0.05$ of the EM cluster;
\item electron likelihood discriminant $\mathcal L$;
\item output distribution of $e$NN7 in CC;
\item output distribution of $\gamma$NN4 in EC;
\item squared width of the energy deposit in the CPS $\sigma^{2}_{\rm CPS}$ in CC;
\item $\chi^{2}$ for matching the spatial positions between CPS cluster and EM cluster in CC.
\end{itemize}

For the training of the BDT  $Z
\to ee$ and dijet data are used. The BDT is trained separately for the
CC and EC and for high (\ilum\ $> 1.6 \times 10^{32}$
cm$^{-2}$s$^{-1}$) and low instantaneous luminosities (\ilum\ $< 1.6 \times 10^{32}$
cm$^{-2}$s$^{-1}$) leading to a different ranking of the utilized
input variables. The training of separate BDTs for CC and EC is of
advantage since the signal-to-background ratio is different in the two
calorimeter regions, and the CC has a better coverage 
by the tracking
devices. Similarly, differences in the signal-to-background ratio and
in the resolution of various variables motivate the training
of separate BDTs for high and low instantaneous luminosities.

\begin{figure*}
\centering
\includegraphics[width=0.49\textwidth]{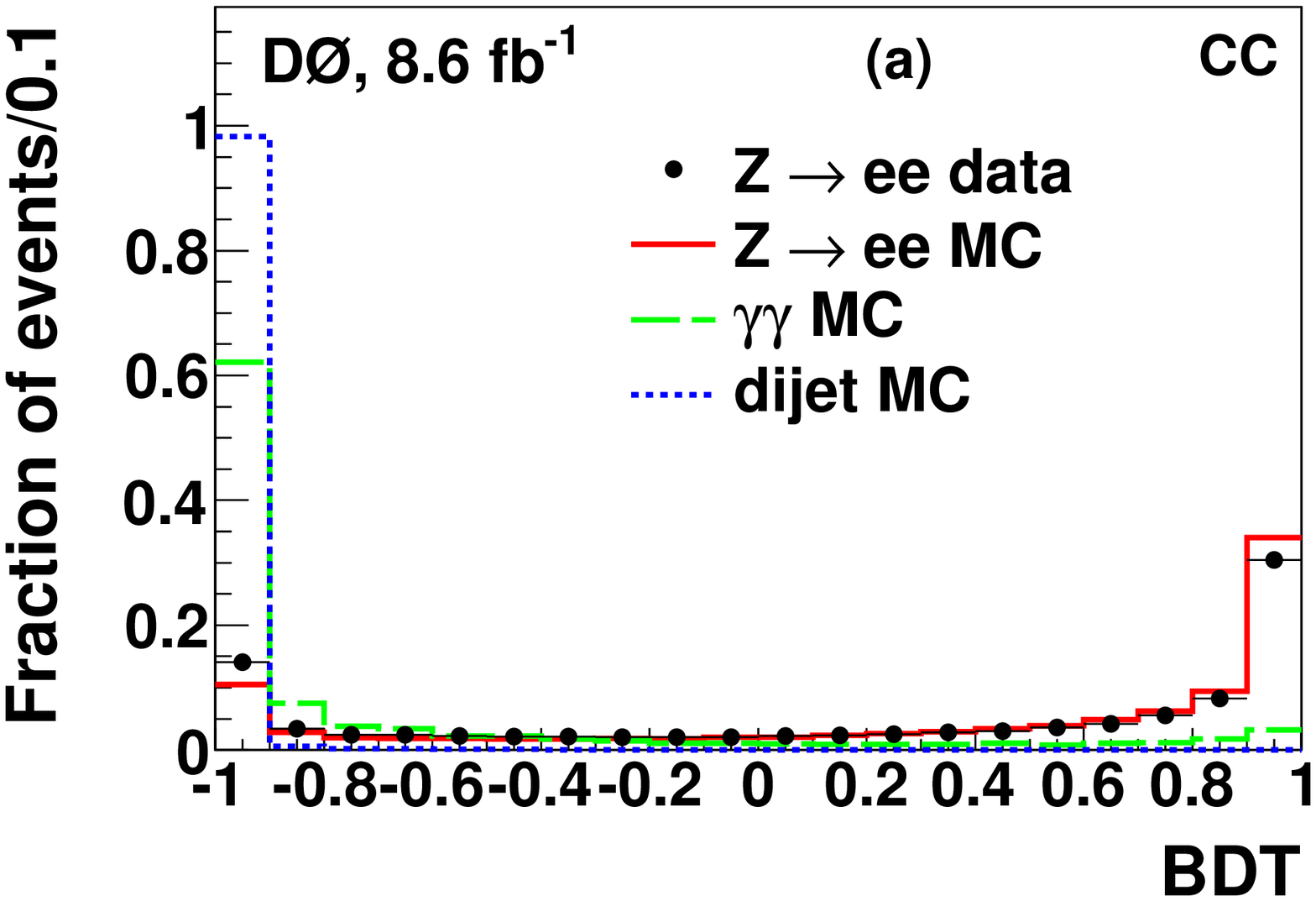}
\includegraphics[width=0.49\textwidth]{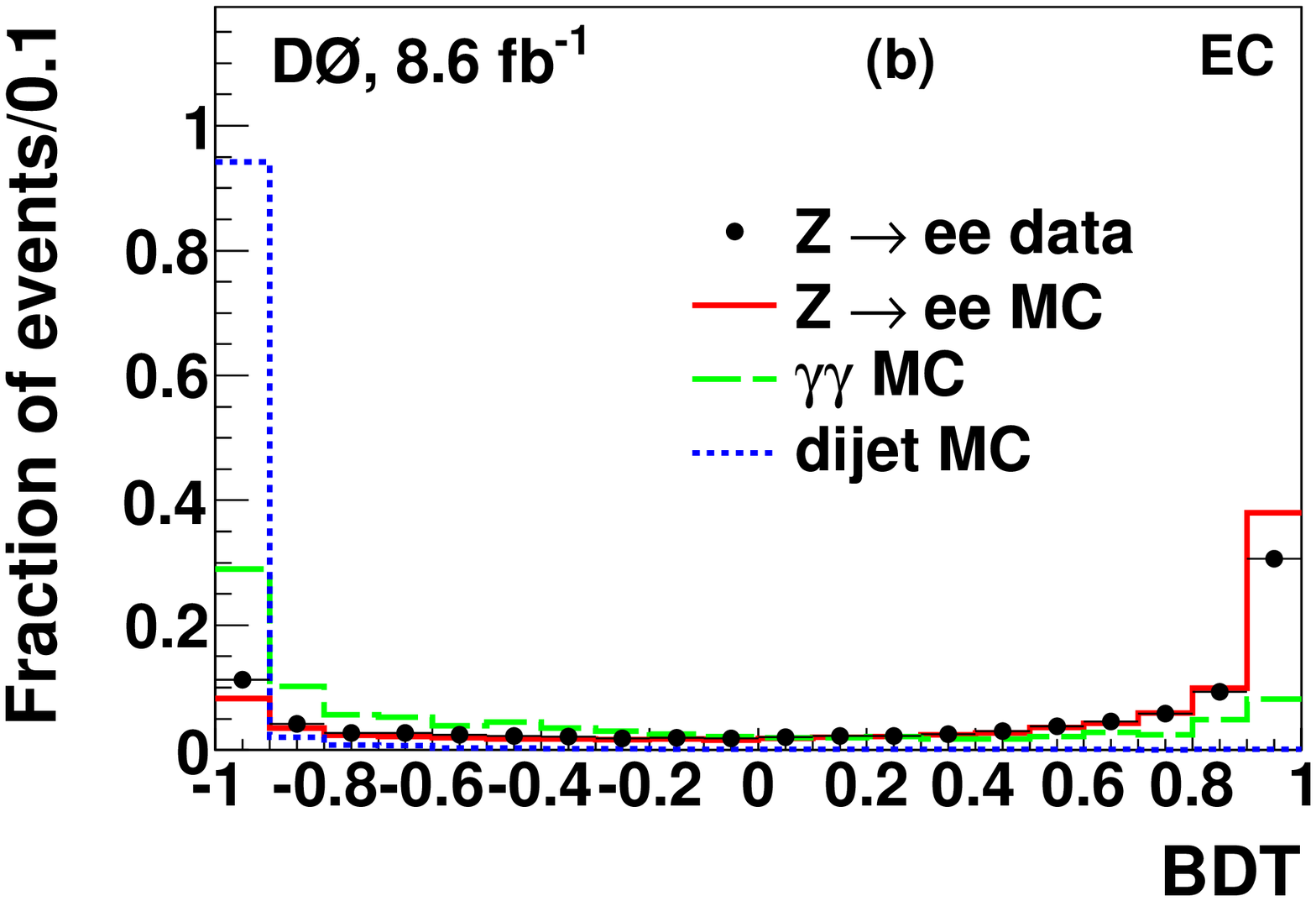}
\caption{
 BDT output distributions of electron candidates in the CC (a) and
 EC (b) region for $Z \to ee$ data and MC events, and for dijet and diphoton MC events.
}
\label{fig:BDT}
\end{figure*}

The BDT output distributions are shown combined for all instantaneous
luminosities but separately for CC and EC in Fig.~\ref{fig:BDT}. They
represent the most powerful
identification variables among the methods presented here.
Typically, the signal efficiency is increased by 4\%--8\% while
maintaining a similar fake rate as other methods.
Due to the insufficient description of uninstrumented material in the
MC simulation, discrepancy between data and MC exists.
This has been studied and taken into account by applying
corrections to the simulation.

\section{Energy scale and resolution calibration}
\label{sec::calib}
After EM objects are identified as described in Sects.~\ref{sec::recoid} and
\ref{sec::recoid_ann},
the detector response to the energy of electrons and
photons is calibrated.
The electron energy scale and resolution are determined from
$Z \to ee$ data events. EM showers induced by
electrons and photons have similar distributions in the D0
calorimeter.
However, EM clusters deposit energy in
the passive material such as the inner detector and solenoid
before reaching the calorimeter. On average, electrons lose more
energy in this material than photons~\cite{em-calorimeter}.  To
account for this difference, MC simulations tuned to reproduce the
response for electrons in data are used to derive the response
difference between electrons and photons.  In this section, the
electron energy scale and resolution, and the energy scale
difference between electrons and photons, are described.

\subsection{Energy scale}

The amount of material in front of the calorimeter
varies between 3.4 and 5 $X_0$ in the CC
and between 1.8 and 4.8 $X_0$ in the EC \cite{MwPRD}.
The fraction of energy deposited in each longitudinal
layer of the calorimeter depends on the amount of that passive material.
The energy loss in passive material is studied taking
into account the energy profile dependence on the incident angle
\cite{rafael-thesis}.
The differences of the energy response between data 
and the MC simulation are determined using $Z \to ee$ events 
and the corrections are applied to the MC simulation.

\begin{figure}[htbp]
\centering
\includegraphics[width=0.48\textwidth]{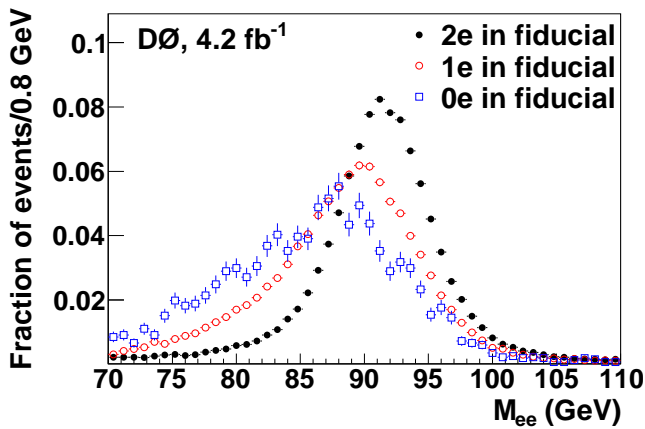}
\caption{
The $M_{ee}$ distribution in a sample of $Z \to ee$ data events, where
both electrons are in the CC, and separating events with 0, 1 and 2
fiducial
electrons. All three distributions are normalized to unit area.
}
\label{fig:Mfiducial210}
\end{figure}

The energy response is degraded near the module $\phi$ boundaries for
the EM layers of the CC.  In addition to a degradation of
energy response, the centroid position of the EM cluster
is shifted. To study these effects,
the following variable is defined:

\begin{equation}
\phimod\ = {\rm mod}\left(\frac{16 \cdot \phi_{\rm EM}}{\pi},1\right),
\end{equation}

where $\phi_{\rm EM}$ is the azimuthal angle of the EM cluster.  For track-matched electrons,
\phimod\ is determined by extrapolating the associated track through
the known magnetic field towards the calorimeter. For photons and
non-track-matched electrons, an average correction of the $\phimod$
is applied which was determined from
track-matched electrons.
Regions of 0.1 $<$ $\phimod$ $<$ 0.9 are referred to as ``in-fiducial'', the
values outside this range are defined as ``non-fiducial''.
Figure \ref{fig:Mfiducial210} shows the dielectron invariant mass (\mee)
distribution for $Z \to ee$ data events with two CC electrons.  The
distribution is shown separately for events with 0, 1, and 2 electrons
located in fiducial regions. Electrons in or close
to module boundaries suffer significant energy losses.  To correct for such energy
loss, the \phimod\ dependent energy scale corrections are derived for both
data and MC simulation using $Z \to ee$ events.
Due to the different
amount of material traversed by the electrons before reaching the
calorimeter, the events are split into five $\eta$ regions to derive the
correction parameters.  In addition, the energy loss near $\phi$ boundaries is larger
for electrons with a poorly measured shower shape corresponding to a
large H-matrix $\chi^2_{\rm Cal}$.  The energy scale
corrections are therefore derived as a function of \phimod\ and H-matrix
$\chi^2_{\rm Cal}$.

With increasing \ilum\ during Run~II,
the uncalibrated $Z$ boson mass is shifted to lower values in data events.
The cause of this effect is discussed in Ref.~\cite{MwPRD}.
The MC simulation, however, predicts
an increase in the average EM energy with \ilum\ due to extra energy
from additional $p\bar{p}$ interactions.
In the data, calibration of the calorimeter largely corrects
for this energy scale dependence on \ilum.
Residual offline corrections are derived by fitting the distribution of
$E_{T}/p_{T}$ for electrons in \wenu\ events, taking
advantage of the fact that the $p_{T}$ scale is independent of \ilum.

Individual cells in the EM calorimeter are known to saturate at
energies varying from about 60 to 260 GeV, depending on the cell
position.
As a result, an EM cluster loses on average about 0.5\% (6\%) of
its nominal energy at 300 (500) GeV. A simple correction truncates the
energy of any cells in the MC that exceed the saturation value for
that cell.

Due to the different amount of energy loss between electrons and photons in the passive material, 
the photon energy is over-corrected by
applying the electron energy scale correction.
The correction is about 3\% at $p_T = 20$~GeV, and it
decreases at higher energies.  The correction required for forward
photons is slightly smaller.
The reconstructed photon energy is corrected accordingly
to compensate for the over-correction.

The systematic uncertainty for the electron energy scale
correction is $\approx$0.5\%, which is mainly caused by the limited
statistics of $Z \to ee$ data events.
For photons, additional 0.5\% systematic uncertainty is added in quadrature
from electron-to-photon energy scale correction.

\subsection{Energy resolution}
The energy resolution of the calorimeter
as a function of the electron/photon energy, E,
can be written as

\begin{equation}
  \frac{\sigma_{\rm EM}} {E} = \sqrt{C^{2}_{\rm EM} + \frac{S^{2}_{\rm EM}}{E} + \frac{N^{2}_{\rm EM}}{E^{2}} }
\end{equation}

with $C_{\rm EM}$, $S_{\rm EM}$ and $N_{\rm EM}$
as the constant, sampling and noise terms, respectively. 
The constant term accounts for the
non-uniformity of the calorimeter response.
Its effect on the fractional resolution is independent
of the energy, and
therefore it is the dominant effect at high energies. The sampling term
is due to the fluctuations related to the physical development of
the shower, especially in sampling calorimeters where the energy
deposited in the active medium fluctuates event by event because the
active layers are interleaved with absorber layers. The noise term
comes from the electronic noise of the readout system, radioactivity
from the Uranium, and underlying events. Since the noise
contribution is proportional to 1/E it is basically negligible
for high energy electrons/photons.
Due to the large amount of material in front of the calorimeter,
$S_{\rm EM}$ is not a constant and is parametrized
as a function of electron energy and incident angle \cite{MwPRD}.
The constant term $C_{\rm EM}$ is derived
by a fit to the measured width of the $Z \to ee$ peak \cite{MwPRD}.
These terms are measured and applied to the true
energy of electron for the fast simulation.

The electron and photon energy resolution predicted by the \geant-based~\cite{geant-ref}
simulation of the \DO\ detector is better than observed
in data.
Furthermore, there are non-Gaussian tails in the
resolution distribution that are poorly modeled by the fully simulated MC
described in Sect.~\ref{sec::datamc}, partly
because the finite charge collection time
of the readout system of the calorimeter
is neglected in the simulation.
To account for both effects,
an ad-hoc smearing is applied to the 
reconstructed energy of EM clusters 
following the {\sc geant} simulation
according to the following
function, which was introduced by the Crystal Ball Collaboration
\cite{crystalball-ref}:

\begin{equation}
\tiny
\label{eq-smear}
  f(x;\alpha,n,\bar{x},\sigma) =
  \begin{cases}
    \exp(-\frac{(x-\bar{x})^2}{2\sigma^2}), & \text{for }
  \frac{x-\bar{x}}{\sigma} > -\alpha \\
    (\frac{n}{\alpha})^n \exp(-\frac{\alpha^2}{2})
  (\frac{n}{\alpha} - \alpha - \frac{x-\bar{x}}{\sigma})^{-n},   &
  \text{for } \frac{x-\bar{x}}{\sigma} \leq -\alpha \\
  \end{cases}
\end{equation}

Here, the $\sigma$ parameter determines the width of the Gaussian core
part of the resolution. The $\alpha$ parameter
controls the energy below which the power law is used,
and the $n$ parameter governs the exponent of the power law.
The $\bar{x}$ parameter is the mean of the Gaussian core part of the
resolution.  Typically, an increase in the width of the non-Gaussian tail
needs to be compensated by an increase in the mean.
The mean of $f(x)$ is around 0,
and the simulated energy is scaled by $1+x$, where $x$ is
sampled from the probability distribution function according to Eq. \ref{eq-smear}.

To determine the parameters of Eq.~\ref{eq-smear}, a fit is performed
by varying parameters applied to the MC, and minimizing the \chisqr\
between the data and fully simulated MC in the $M_{ee}$ distribution.
The $n$ parameter is fixed since there is enough
freedom in the other three parameters to adequately describe the data.
A value of $n=7$ is found to be appropriate.

The parameters are fitted separately for the following three
categories of EM clusters \cite{mika-thesis}:
\begin{itemize}
\item {\bf Category 1: CC in-fiducial} \\
CC in-fiducial clusters are defined as $|\eta| < 1.1$ and 0.1 $<$
$\phimod$ $<$ 0.9.
The parameters are fitted using events in which both electrons are CC
in-fiducial.
\item {\bf Category 2: CC non-fiducial}\\
CC non-fiducial clusters are defined as $|\eta| < 1.1$ and
$\phimod$ $<$ 0.1 or $\phimod$ $>$ 0.9.
The parameters are fitted using events containing two CC electrons,
where at least one is non-fiducial.
Any CC in-fiducial electrons are smeared using their already tuned
parameters.
\item {\bf Category 3: EC}\\
EC clusters are defined as having $|\eta| > 1.5$.
The parameters are fitted using events containing two EC electrons, or
one CC in-fiducial or non-fiducial plus one EC electron.
For EC clusters, a simple Gaussian smearing is used where the fit has
only two parameters ($\bar{x}$, $\sigma$).
\end{itemize}

\begin{figure*}
\centering
\includegraphics[width=0.49\textwidth]{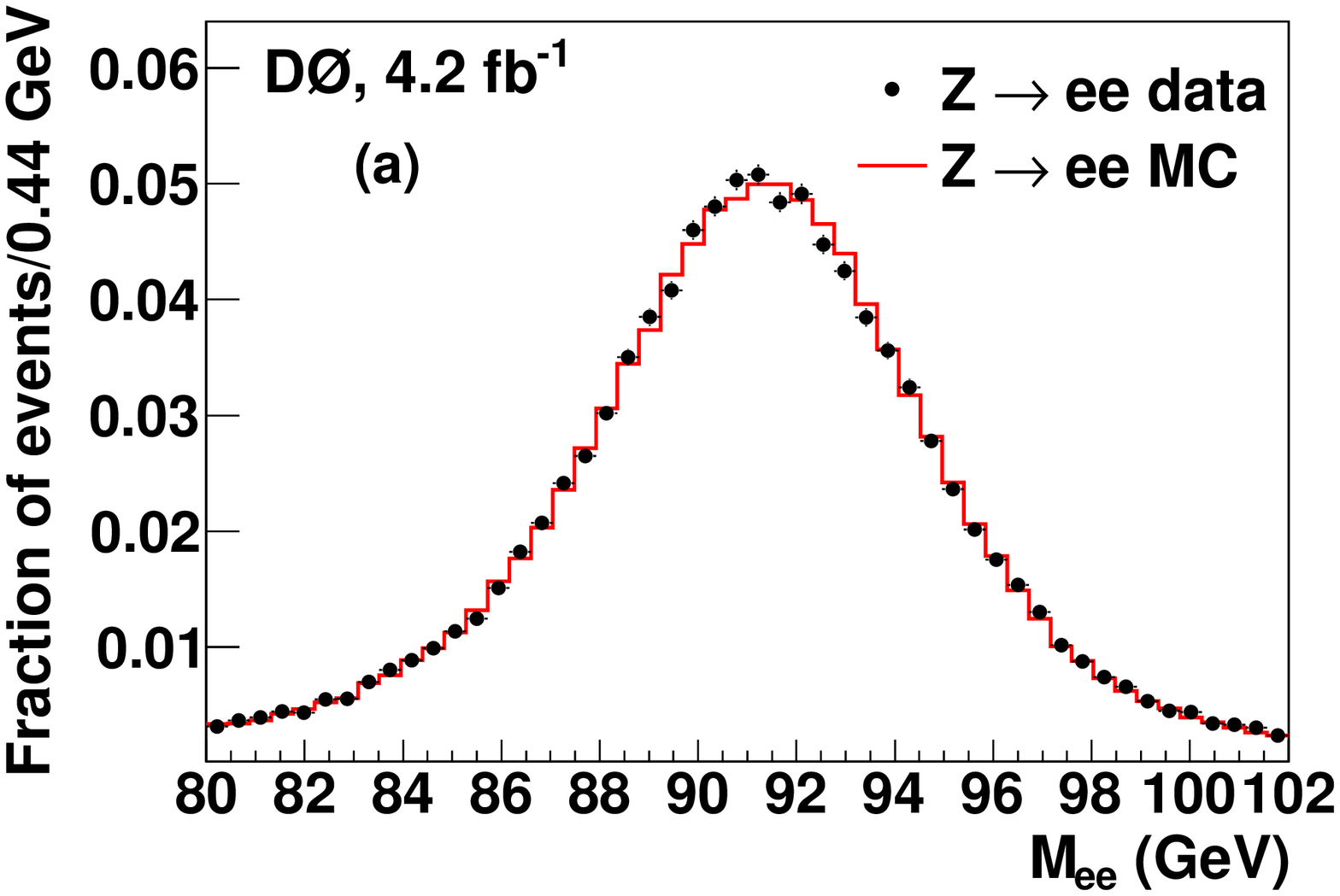}
\includegraphics[width=0.49\textwidth]{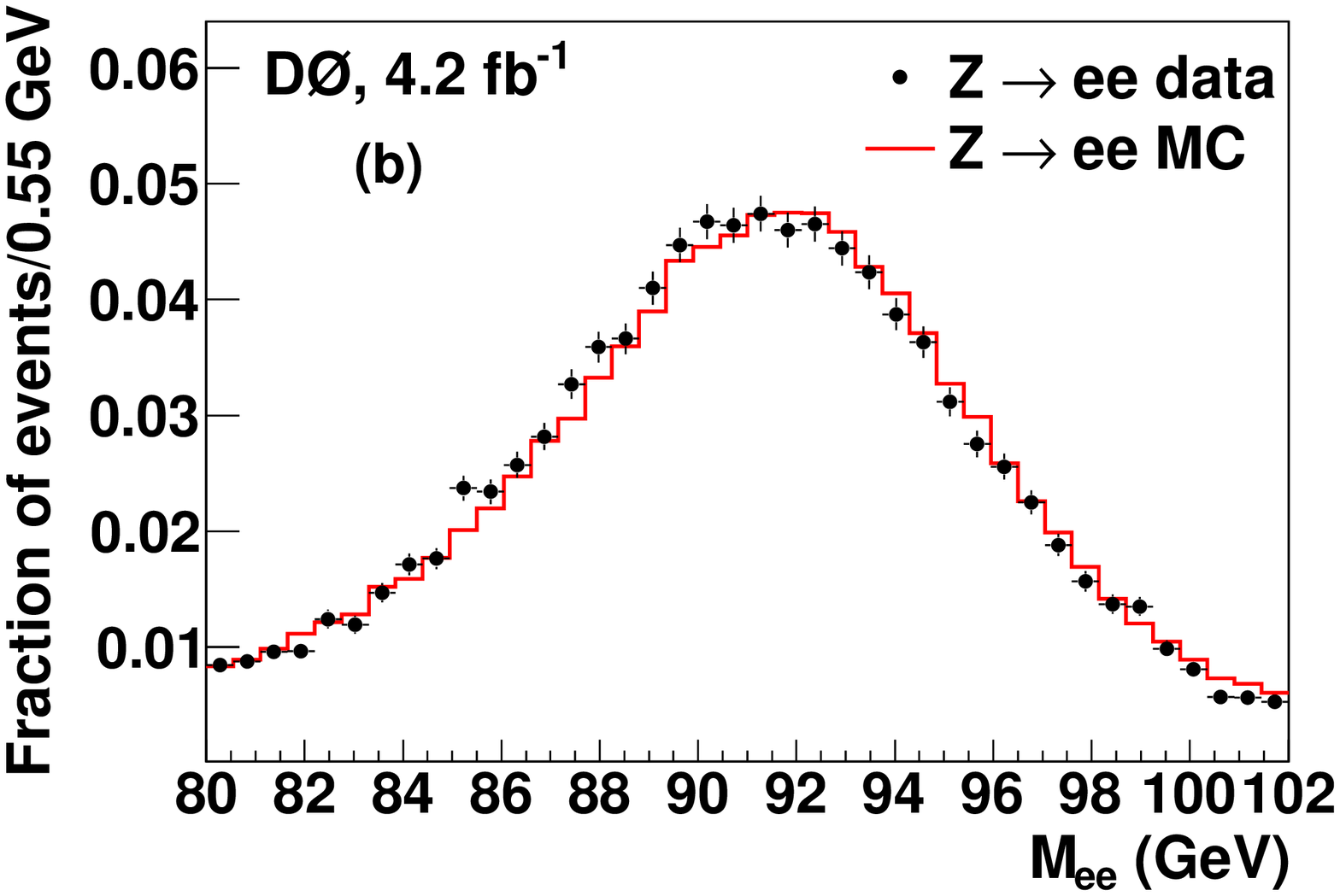}
\includegraphics[width=0.49\textwidth]{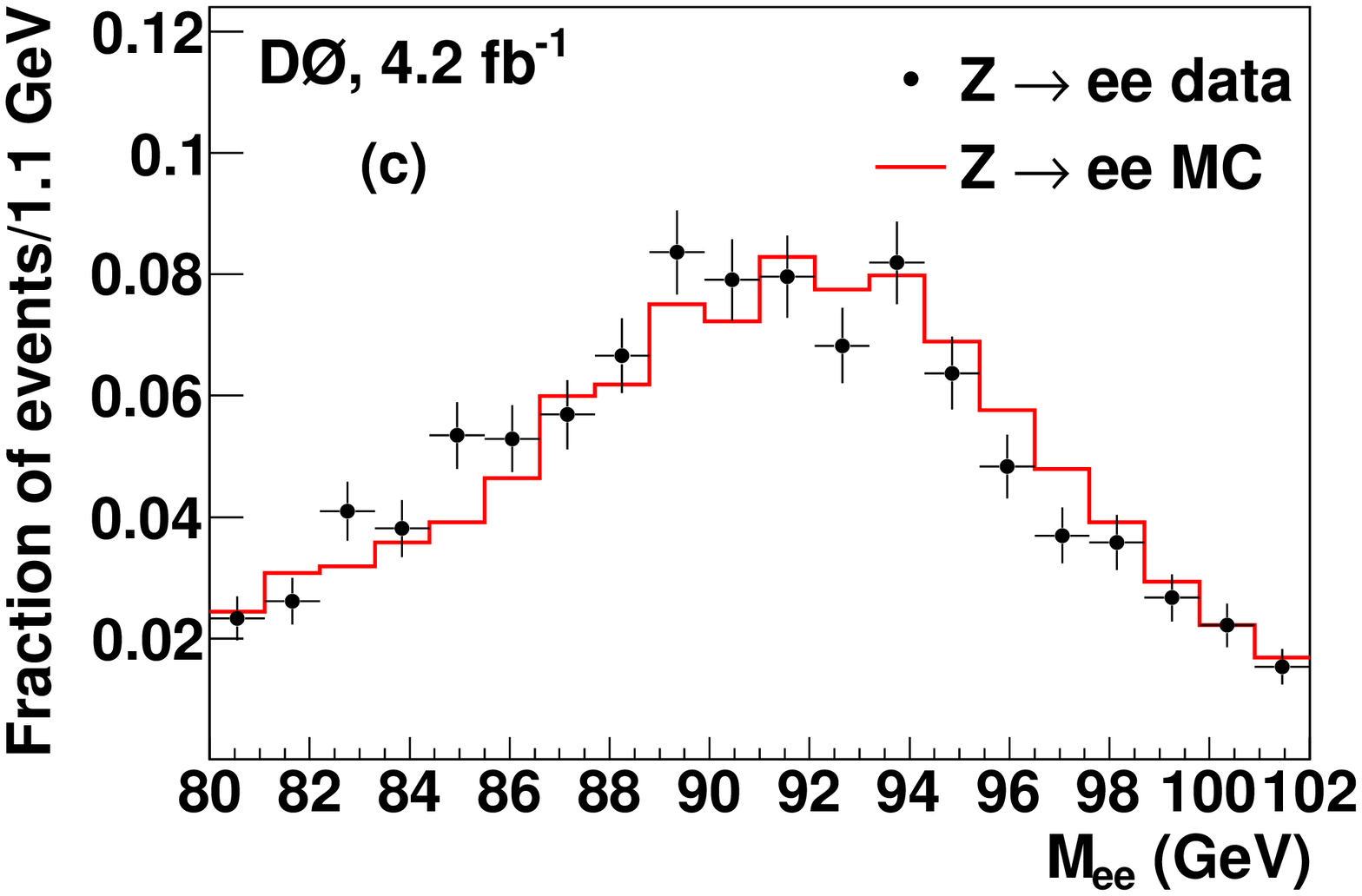}
\includegraphics[width=0.49\textwidth]{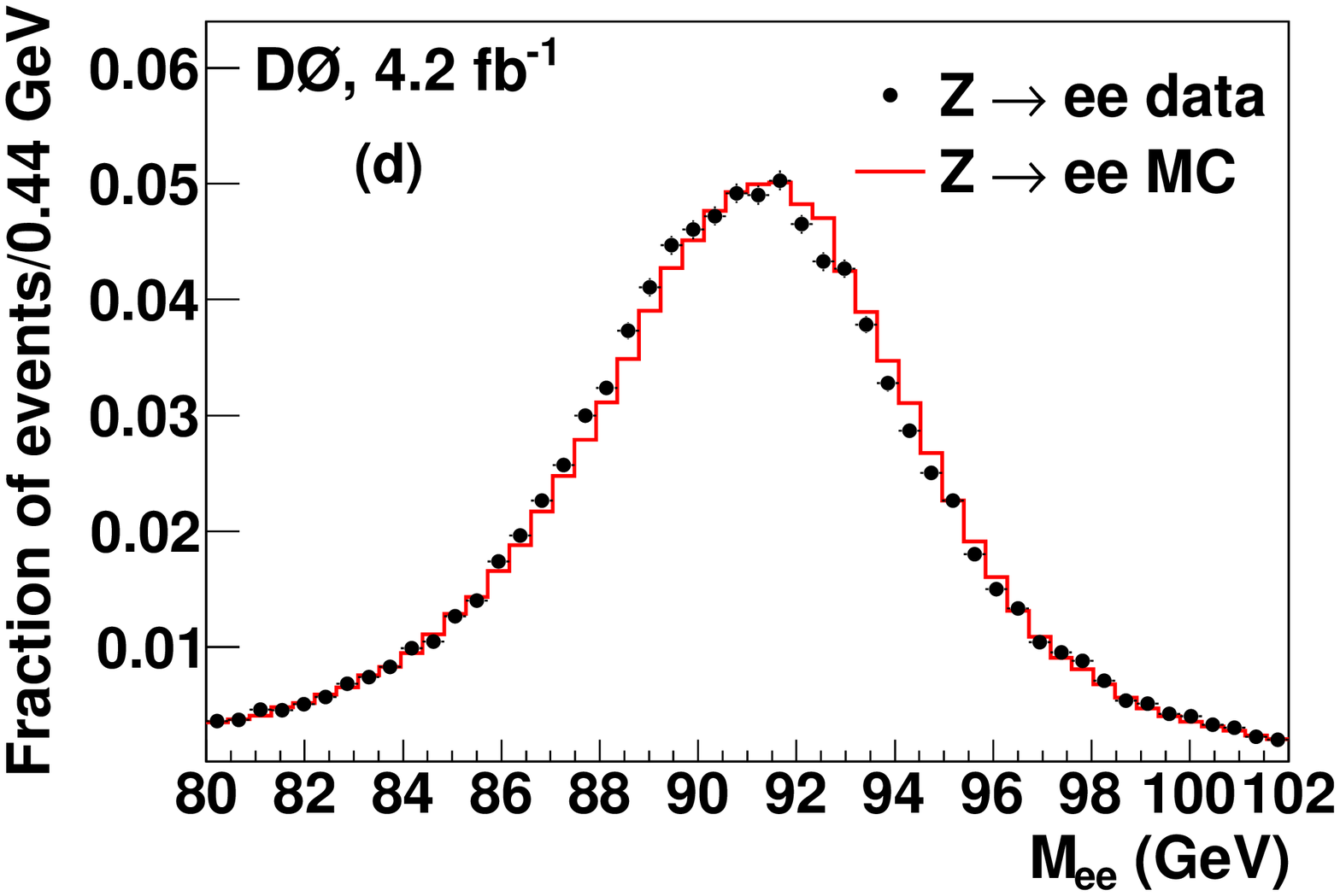}
\caption{
Dielectron invariant mass ($M_{ee}$) distributions for $Z \to ee$ data and MC events,
with two electrons in the CC fiducial regions (a), one electron in
the CC fiducial region and the other in the CC non-fiducial region (b), two
electrons in the CC non-fiducial regions (c), and at least one electron in
the EC region (d).}
\label{fig:Mee}
\end{figure*}

Figure \ref{fig:Mee} shows a comparison
of \mee\ distributions for $Z \to ee$ data and MC after applying the
energy scale and smearing corrections.
Good agreement between data and MC simulation is observed.

\section{Efficiencies of electron identification}
\label{sec::results_ele}
Electron trigger, preselection and identification
efficiencies are measured in $Z \to ee$ data and MC events by
selecting two high-$E_{T}$ electron candidates that have an
invariant mass close to the $Z$ boson mass peak.
To obtain an improved simulation,
differences between the efficiencies measured
in data and MC simulation are used to derive correction factors to be applied
to MC events taking into account kinematic dependences.

\subsection{Tag-and-probe method}\label{tagandprobe}
To measure the efficiencies,
a ``tag-and-probe method'' is used.
In $Z \to ee$ decays, a $E_T > 30$ GeV electron candidate
in CC fiducial is selected as the ``tag'' with the following requirements:
\begin{itemize}
\item $f_{\rm EM} > 0.96$;
\item $f_{\rm iso} < 0.07$;
\item $\Sigma p_{T}^{\rm trk} < 2$ GeV;
\item associated track $p_T > 15$ GeV;
\item $\mathcal{L} > 0.8$;
\item $e$NN7 $>$ 0.7.
\end{itemize}
The ``probe'' -- used to perform the measurement of the identification
efficiency -- is either an EM cluster or a track. The invariant mass
of the tag and probe electrons, $M_{tp}$, is required to be close to the
$Z$ boson mass. If the probe is an EM cluster,  $M_{tp}$ is required to
be greater than 80 GeV but less than 100 GeV.
The energy resolution for high-$p_{T}$ tracks is worse, and
the $M_{tp}$ is required to be greater than 70 GeV but less than 110 GeV
when the probe is a track.
If the probe passes the tag selection criteria, it will also be used
as a tag, resulting in the event being counted twice.  To avoid
bias, the same tag-and-probe method is used for both $Z \to ee$ data and MC events.

To remove the residual background from jet production in
data events, a template fit is applied to the $M_{tp}$
distributions. The
signal shape is obtained from $Z\to ee$ MC simulation, and the
background shape is
derived from dijet data.
To take into account
dependencies on the electron position in the detector,
the template fit is performed in various $\eta$ and $\phi$ regions.
The systematic uncertainty for the tag-and-probe method is 
dominated by the statistics of $Z \to ee$ data events.
It is $\approx$10\% for low probe $E_T$ ($<$ 20 GeV) region,
and $\approx$3\% for high probe $E_T$ region.

\subsection{Trigger efficiencies}\label{ele_trigger}
There are two types of single electron triggers \cite{MwPRD, trigger-ref}.
One class of triggers is solely based on calorimeter information and the other
class includes tracking information. Calorimeter-based triggers are used
for both electrons and photons. To have higher trigger efficiencies
for electrons, we combine both types of triggers by taking their logical OR.
The tag-and-probe method is used to measure the trigger efficiencies in data.
To be consistent with offline electron identification requirements
(described in Sect. \ref{ele_sel}), the
trigger efficiencies are measured with respect to each set of electron
identification requirements. To account for
dependencies on the EM cluster position in the detector, the trigger efficiencies
are parametrized as a function of $E_{T}$ and $\eta$ of the electron
candidate.
Single electrons are triggered with an efficiency $\approx$100\%
for transverse momenta above 30~GeV in the fiducial regions of the
calorimeter up to $|\eta| < 2.5$.

\subsection{Preselection efficiencies}
\label{ele_presel}
\begin{figure*}
\centering
\includegraphics[width=0.38\textwidth]{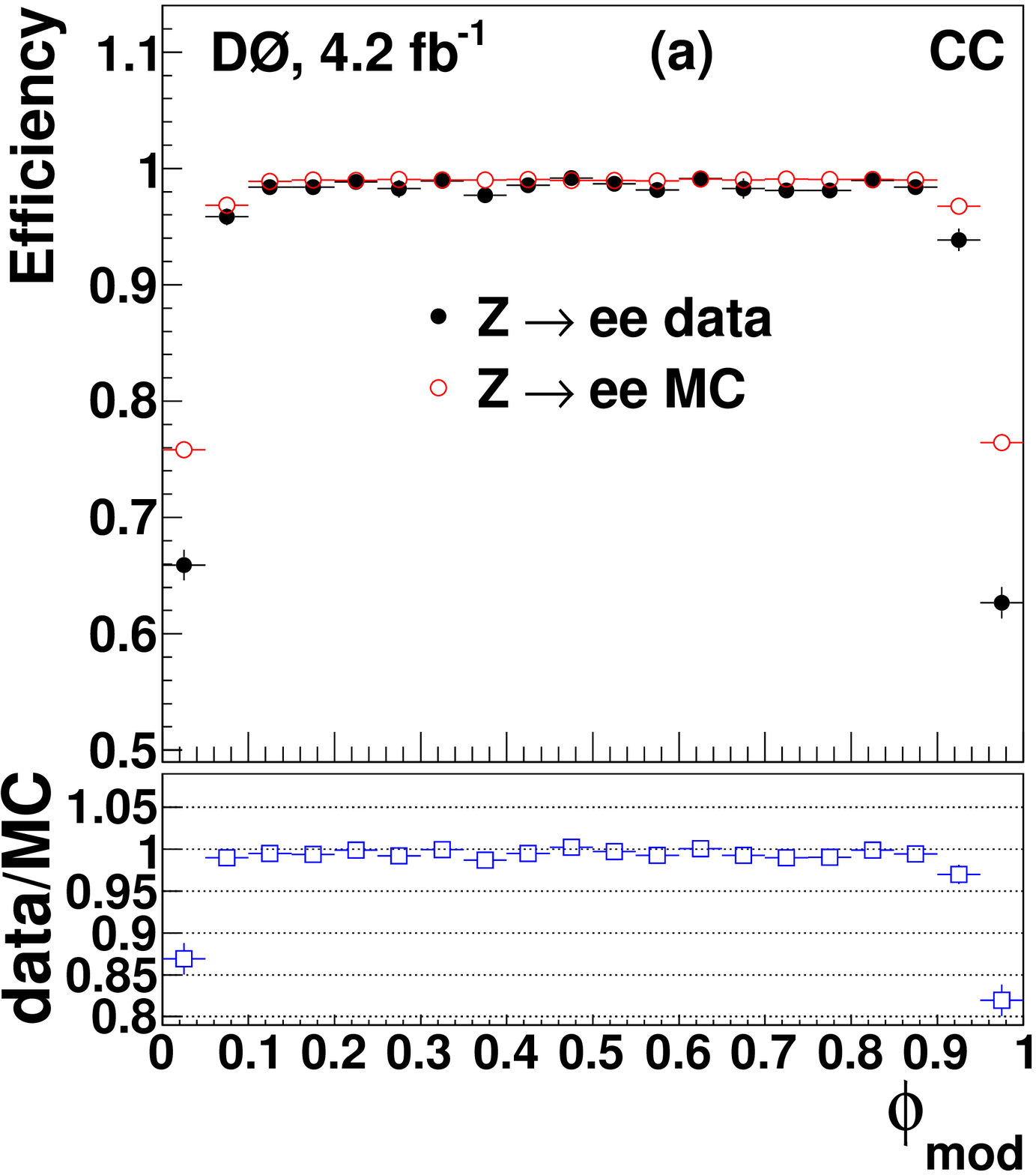}
\includegraphics[width=0.38\textwidth]{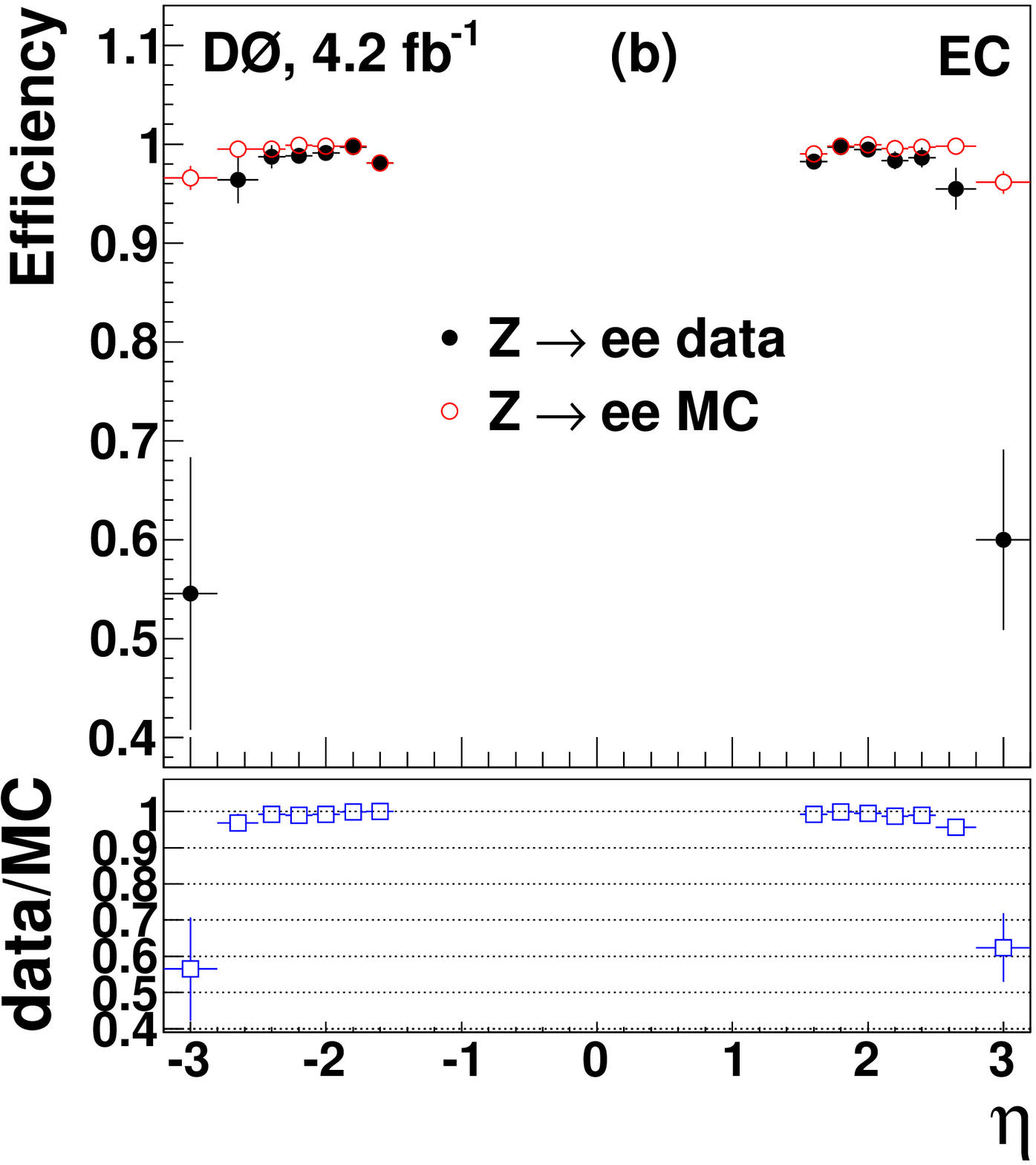}
\caption{
Preselection efficiencies of probe tracks as a function of
$\phi_{\rm mod}$ and $\eta$ for electrons in the CC (a) and EC (b).
$Z \to ee$ data is compared to the MC prediction, and the ratio
data/MC is presented.}
\label{fig:preeff}
\end{figure*}

Preselected electrons and photons are
EM clusters that satisfy
the criteria described in Sect.~\ref{emclus}.
The preselection
efficiency is given by the fraction of tracks that match an
EM cluster passing the preselection requirements for the probe
electron candidate.
In Fig.~\ref{fig:preeff}a the preselection efficiencies are presented
for probe tracks in the CC as a function of $\phi_{\rm mod}$ for
data and the MC simulation.  The average efficiency is
$\approx 98\%$. Data and MC simulation show good agreement, except in
non-fiducial regions. Therefore, the $\phi_{\rm mod}$-dependent correction
factors as shown in Fig.~\ref{fig:preeff}a are
applied to MC to improve the simulation.
Figure~\ref{fig:preeff}b shows the preselection efficiencies
as a function of $\eta$ for EC electrons.
Efficiency losses are observed in the
region $|\eta| > 2.5$ due to partial detector coverage
for increasing $\eta$. To correct for data versus MC differences in the EC
region, $\eta$-dependent factors are applied to the
simulation. No significant differences between data and MC
in other variables are
observed for either electrons or photons in the CC and EC regions.

\subsection{Electron identification efficiencies}\label{ele_sel}

Many sets of requirements for electron identification are
provided for use in physics analyses, each with
different electron selection efficiencies and
misidentification rates. As examples, the electron
identification efficiencies for two sets of requirements are presented here.
These sets are called ``loose'' and ``tight''.
Table~\ref{tab:sel} lists the specific requirements of these
two operating points.
\begin{table*}[htb]
  \centering
  \begin{tabular}{|c|c|c|c|c|}
    \hline
Variable & loose CC & loose EC & tight CC & tight EC\\
\hline
$f_{\rm EM} > $ & 0.9 & 0.9 & 0.9 & 0.9 \\
\hline
$f_{\rm iso} < $  & 0.09 & 0.1 & 0.08 & 0.06 \\
\hline
$\Sigma p_{T}^{\rm trk} < $  & 4 GeV & $(*)$ & 2.5 GeV & $(*)$ \\
\hline
H-matrix $\chi^2_{\rm Cal} <$  & -- & 40 & 35 & 40 \\
\hline
$\sigma_{\phi}^{2} < $  & -- & $(+)$ & -- & $(+)$ \\
\hline
$e$NN7(CC), $\gamma$NN4(EC) $>$  & 0.4 & 0.05 & 0.9 & 0.1 \\
\hline
P(${\chi^2_{\rm spatial}}$) $\neq$  & -1  & -- & -1 & -1 \\
or $D_{\rm hor} >$  & 0.6 & -- & -- & -- \\
\hline
$\mathcal{L} >$  & -- & -- & 0.6 & 0.65 \\
\hline
$E_T/p_T <$  & -- & --  & 3 & 6 \\
\hline
 \end{tabular}
  \caption{\label{tab:sel} \small
    Sets of requirements to identify electrons with loose and tight
    quality.\newline
   $(*)$: $\Sigma p_{T}^{\rm trk} < 0.01$ GeV or  $\Sigma p_{T}^{\rm trk} <
    (-2.5\, |\eta|+7.0 )$ GeV\newline
    $(+)$: $(6.5 \times (|\eta|-0.82)^{-1} - 2.8)$ cm$^2$ for $|\eta|<2.6$;
           $(6.5 \times (|\eta|-1.35)^{-1} - 2.8)$ cm$^2$ for $|\eta|>2.6$
}
\end{table*}

\begin{figure*}
\centering
 \includegraphics[width=0.42\textwidth]{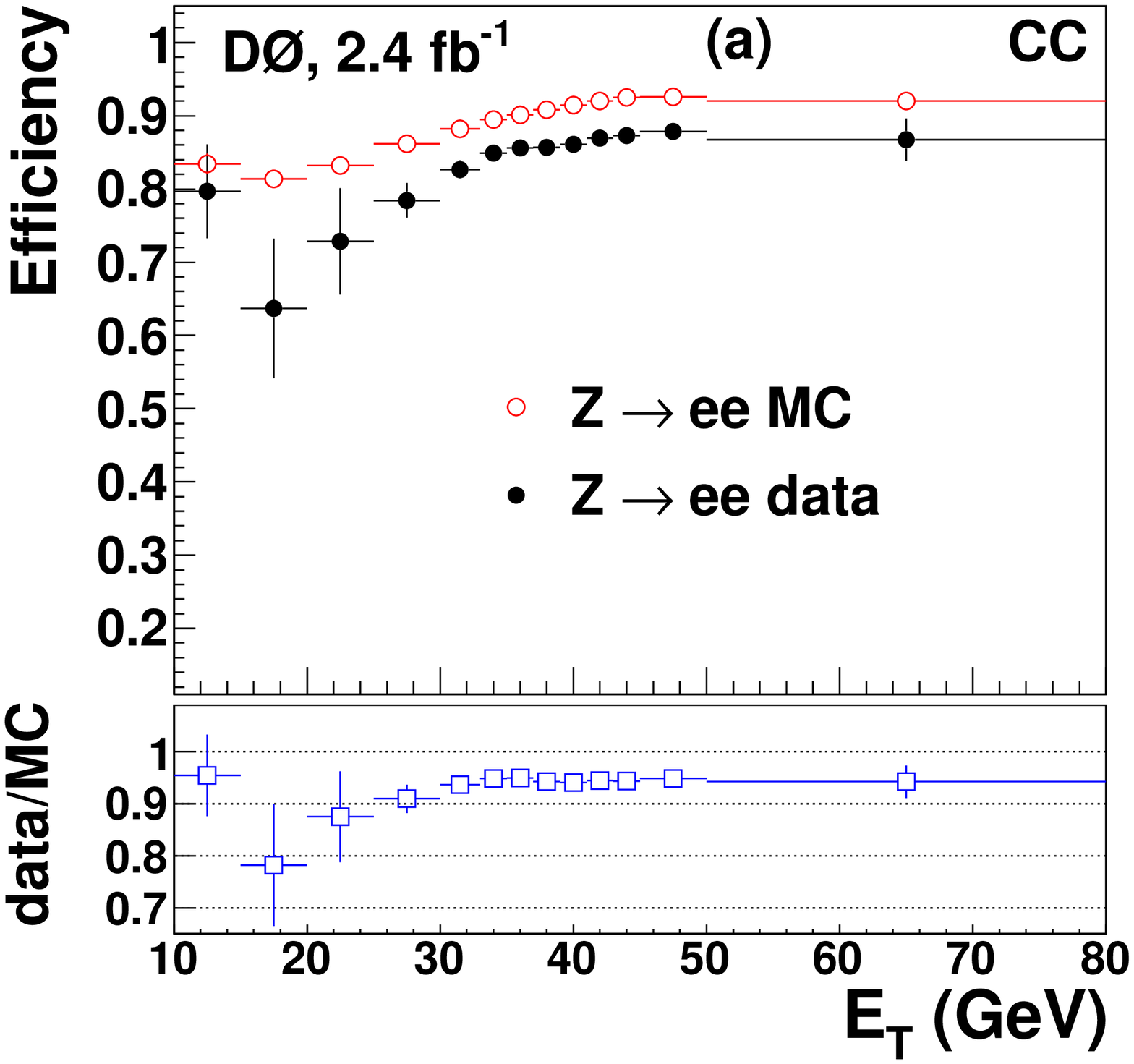}
 \includegraphics[width=0.42\textwidth]{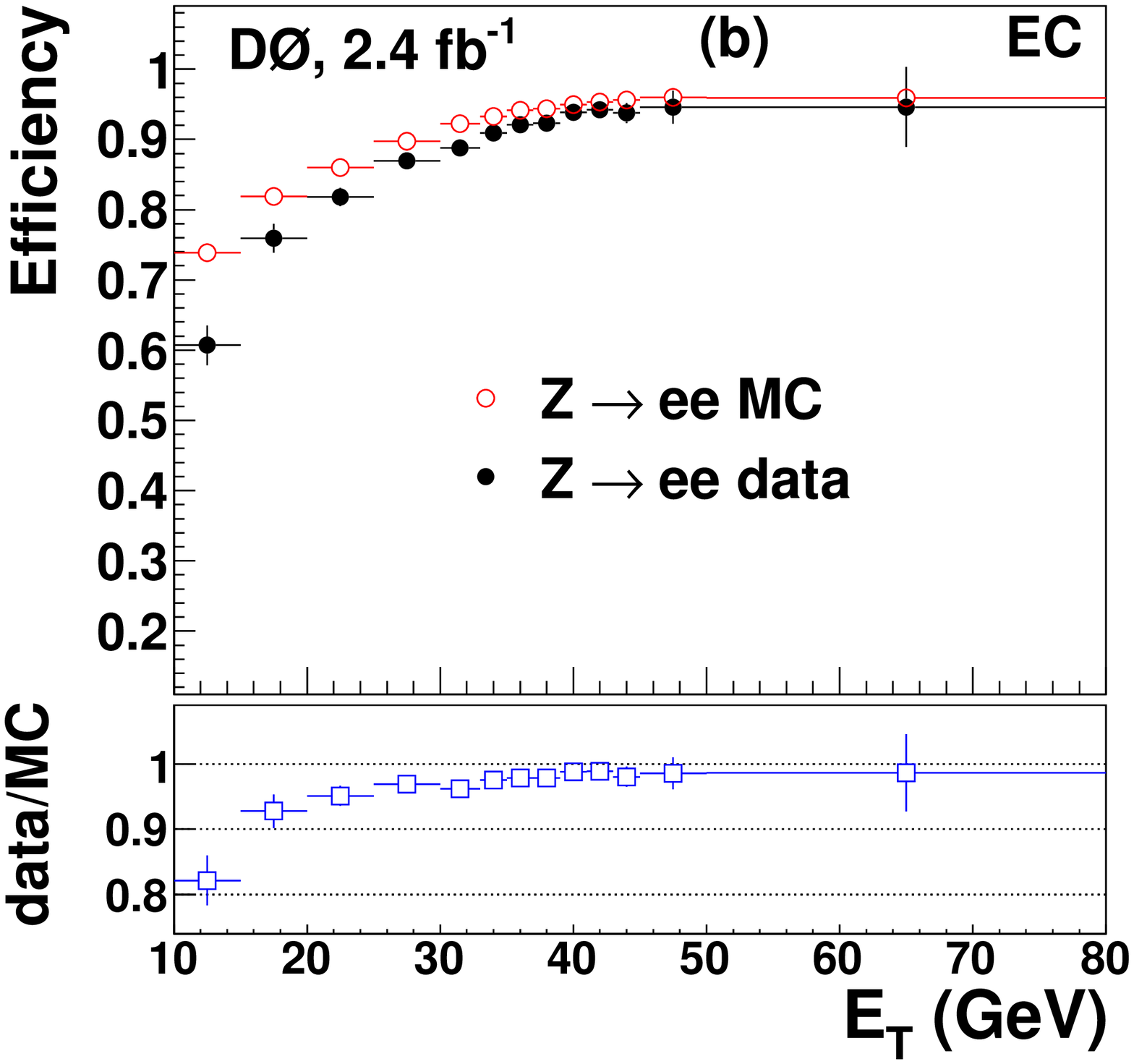}
 \includegraphics[width=0.42\textwidth]{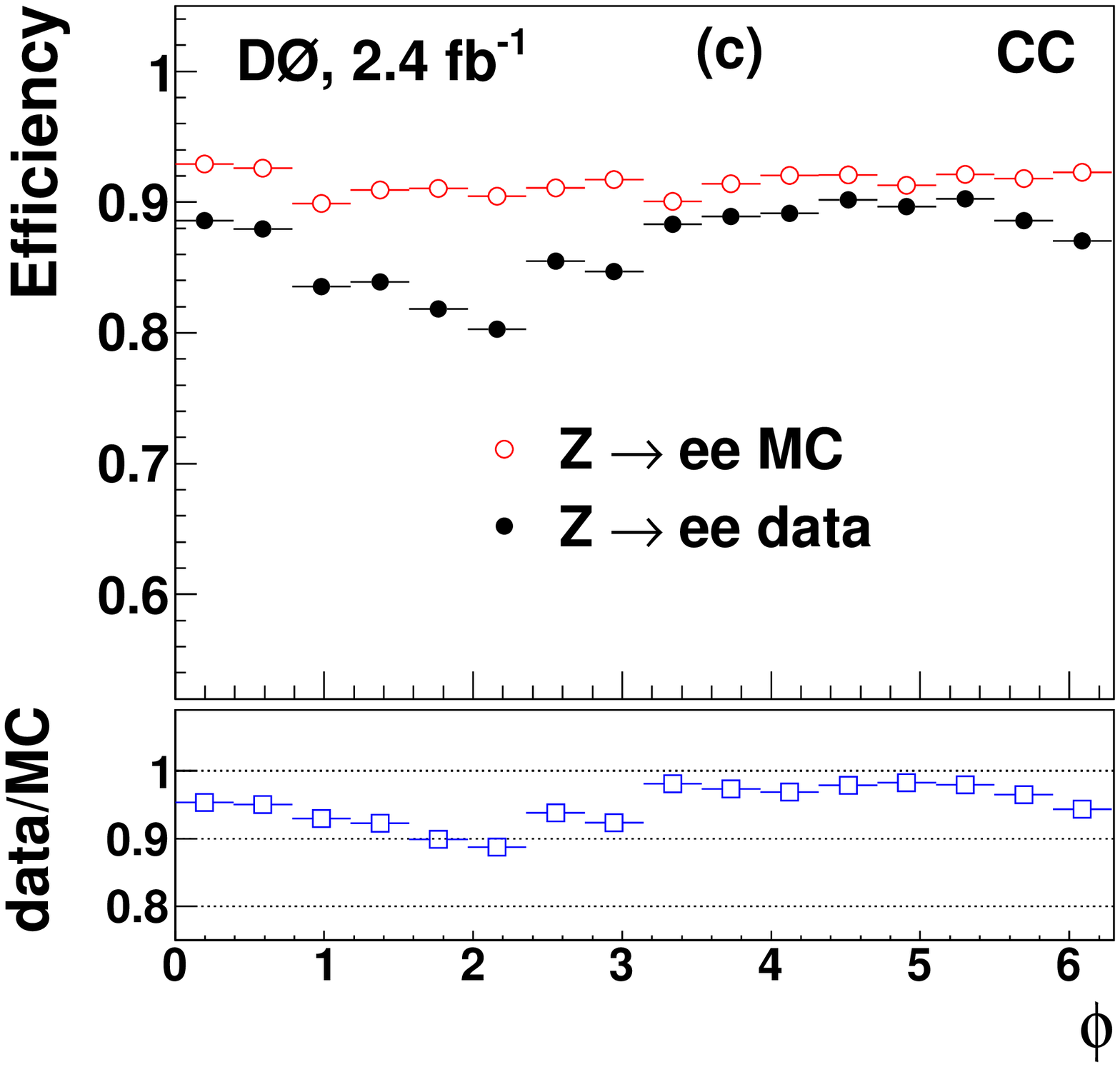}
 \includegraphics[width=0.42\textwidth]{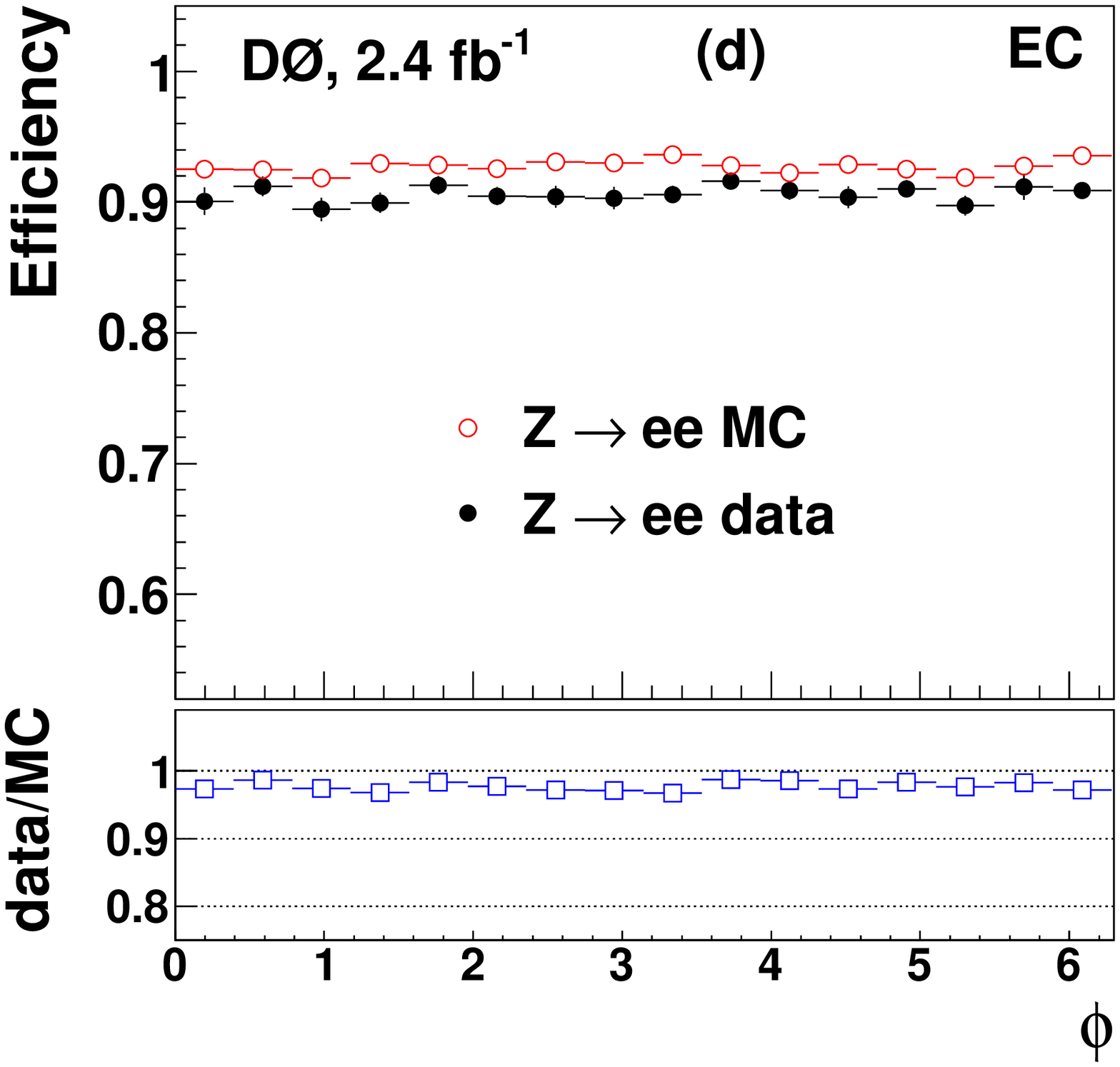}
 \includegraphics[width=0.42\textwidth]{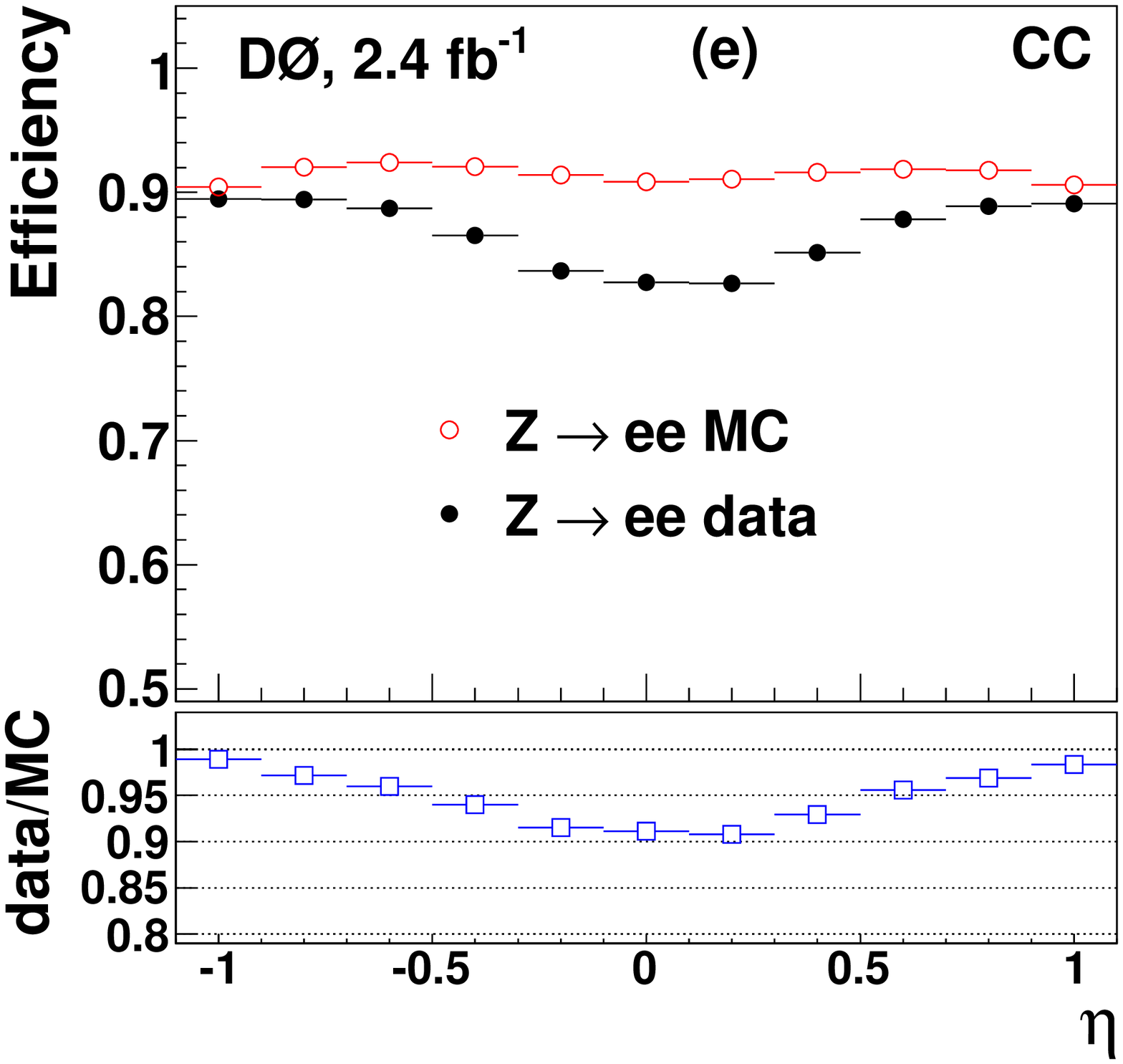}
 \includegraphics[width=0.42\textwidth]{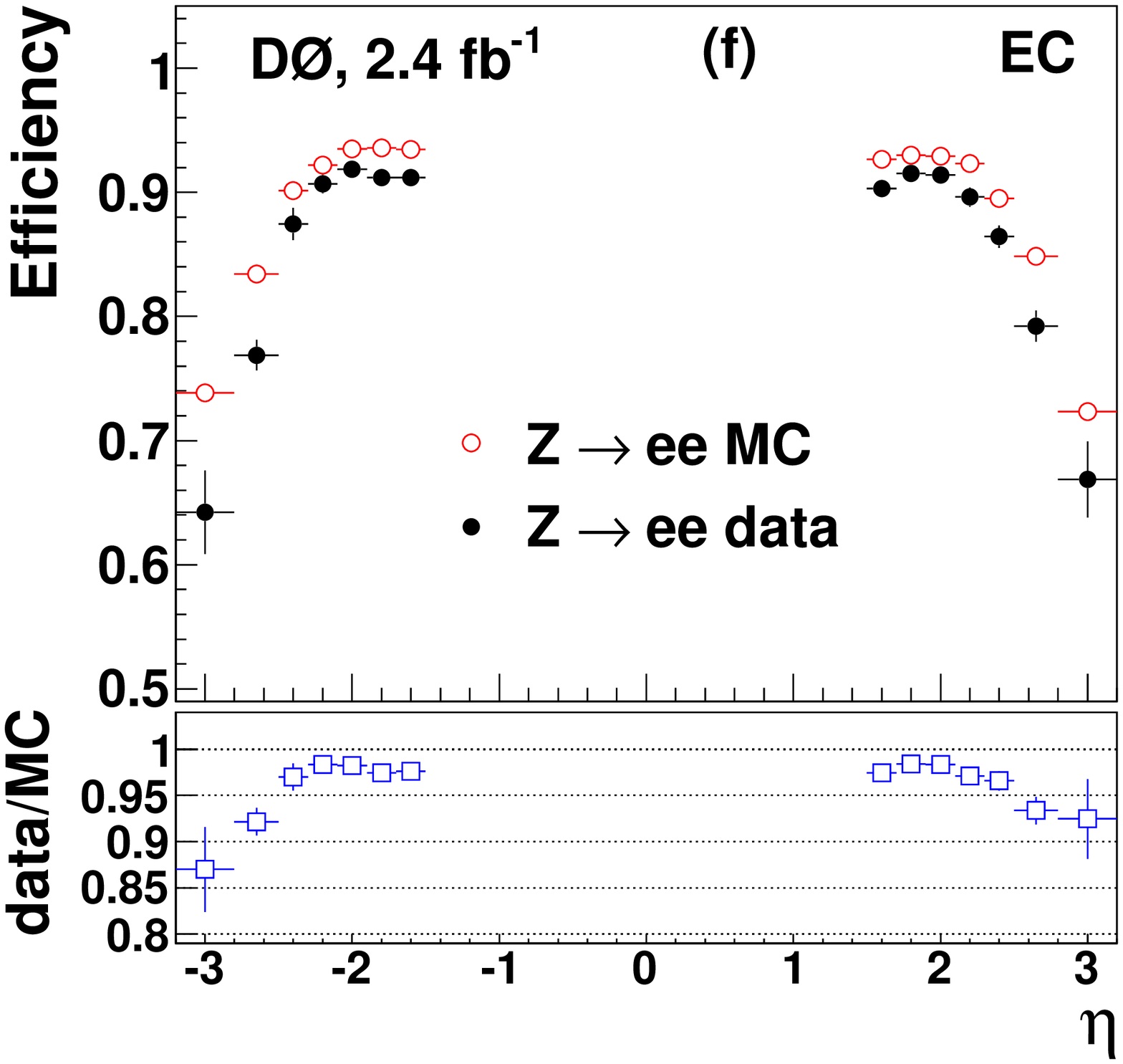}
\caption{
Electron identification efficiencies as a function of (a,b) $E_{T}$,
(c,d) $\phi$ and (e,f) $\eta$ for loose electron requirements in
CC and EC.
Efficiencies for data and MC simulated $Z \to ee$ events are shown,
as is the ratio of the data and MC efficiencies.}
\label{fig:eLooseEff}
\end{figure*}

\begin{figure*}
\centering
\includegraphics[width=0.42\textwidth]{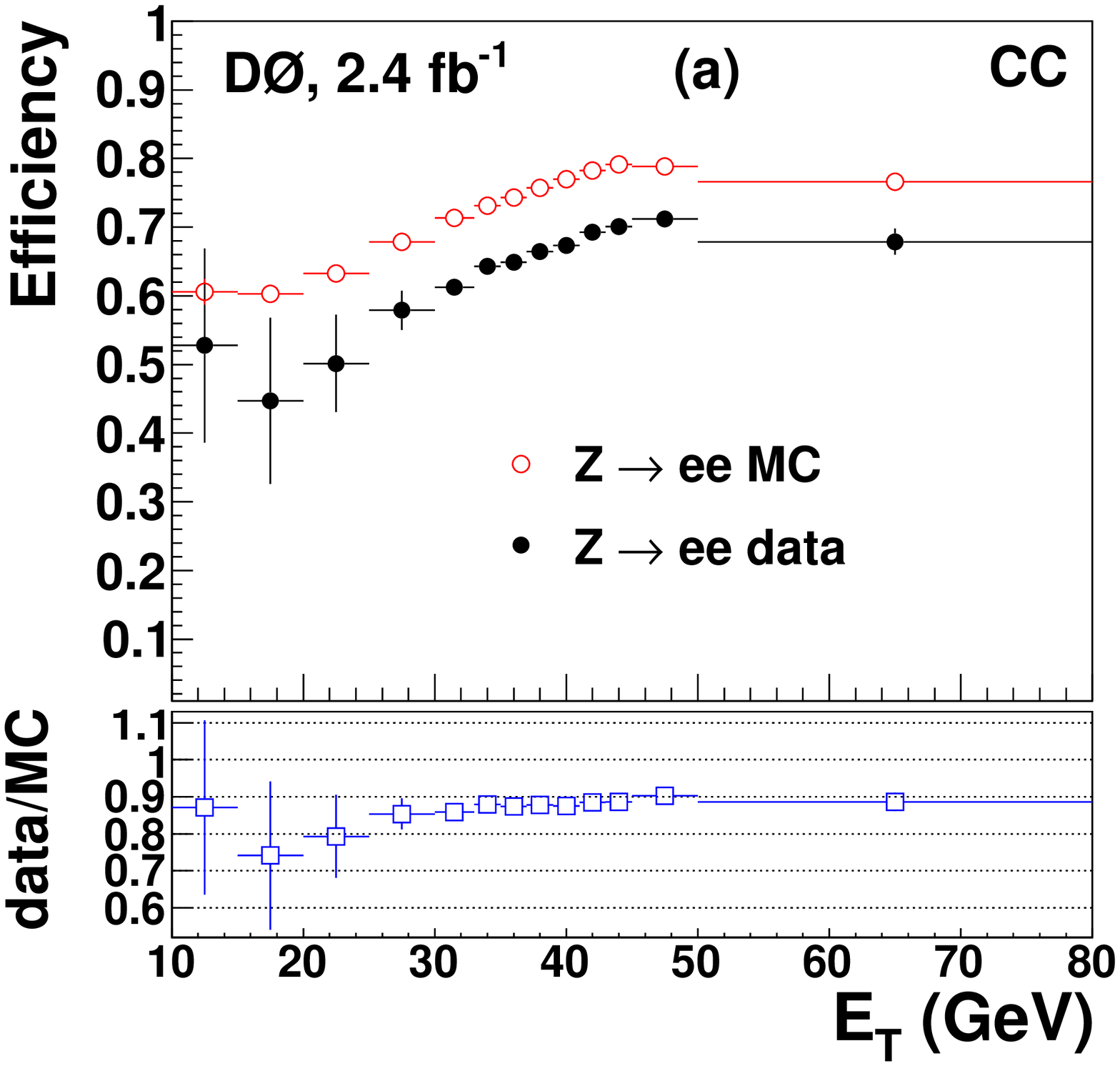}
\includegraphics[width=0.42\textwidth]{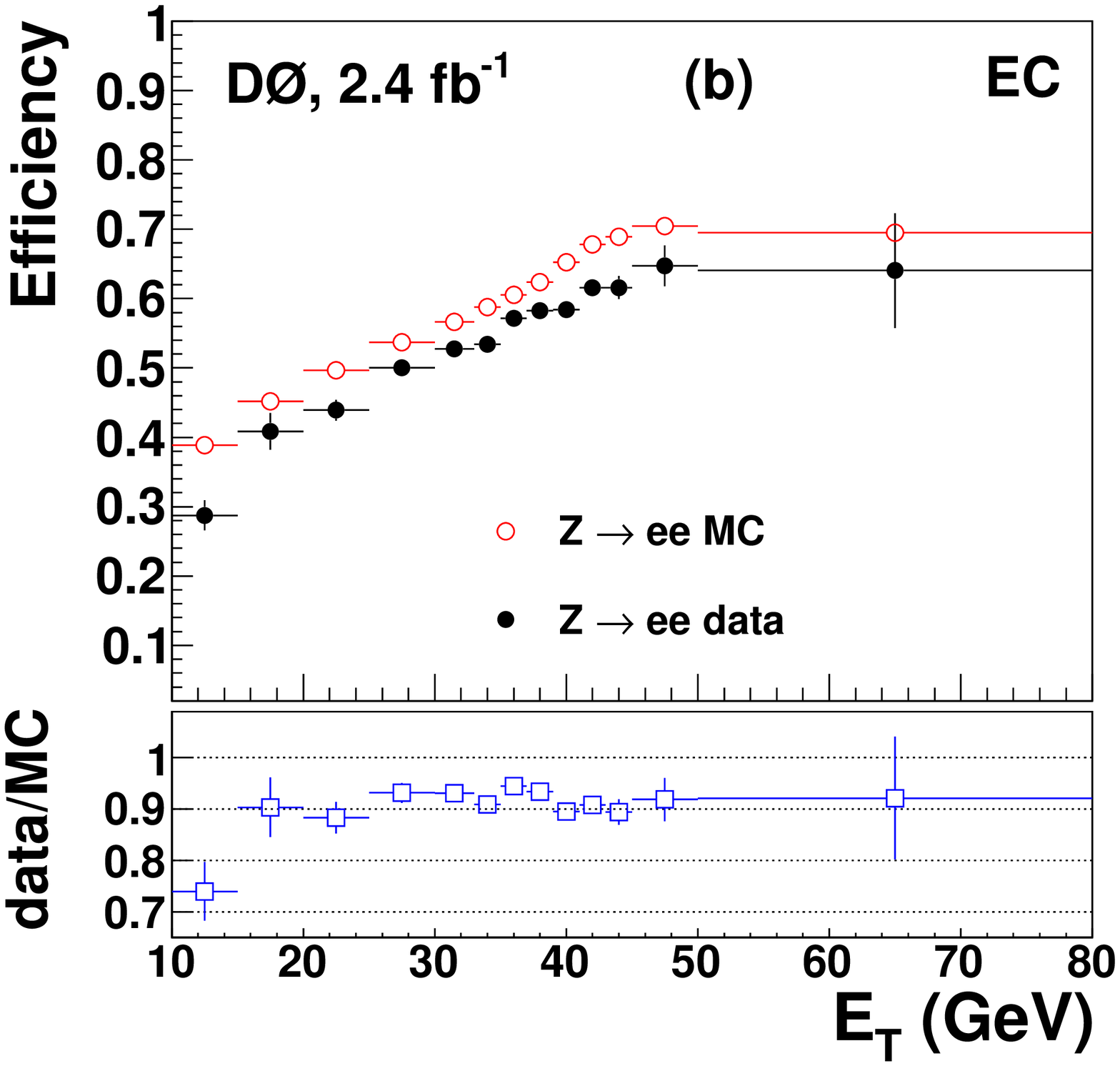}
\includegraphics[width=0.42\textwidth]{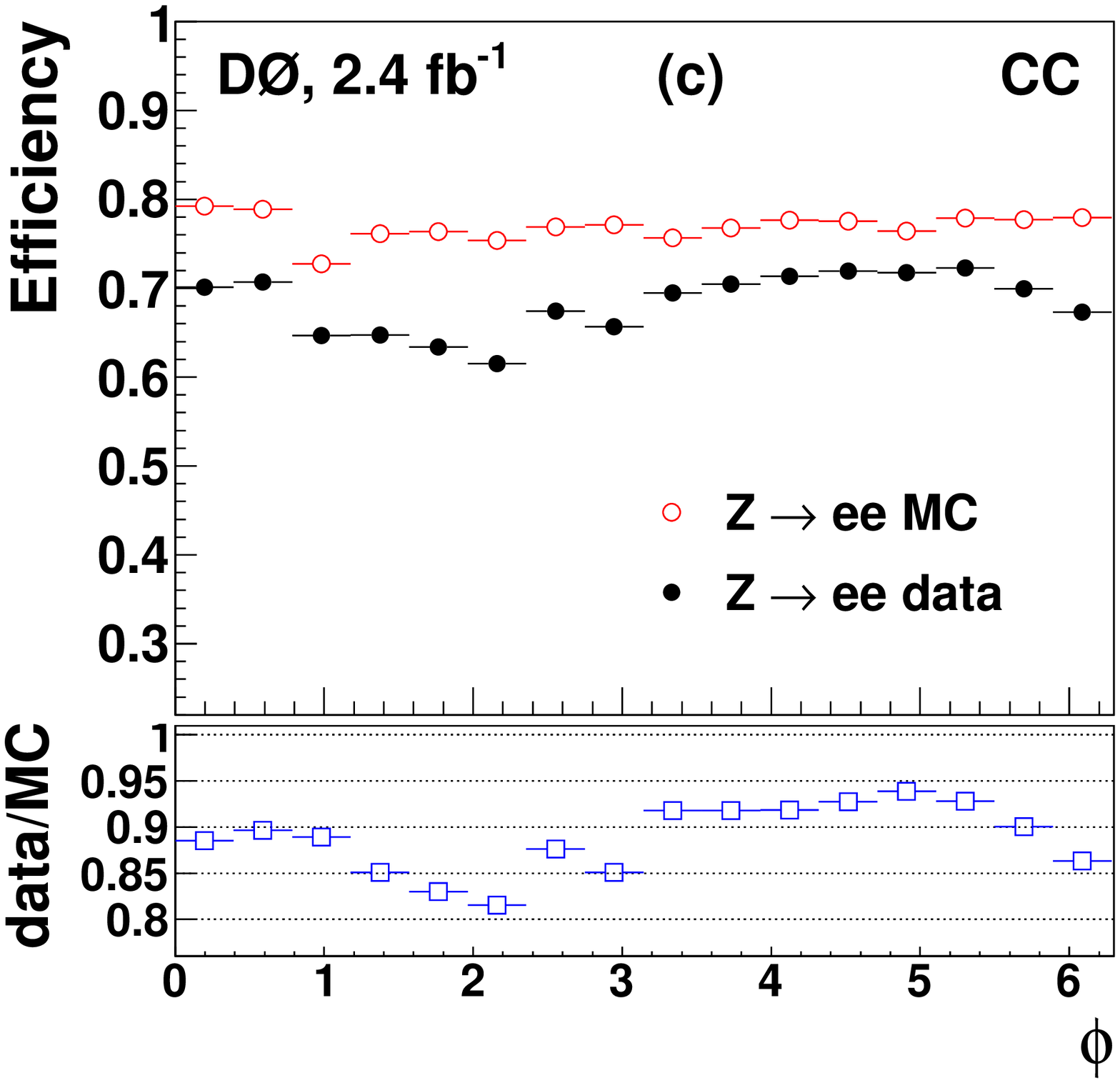}
\includegraphics[width=0.42\textwidth]{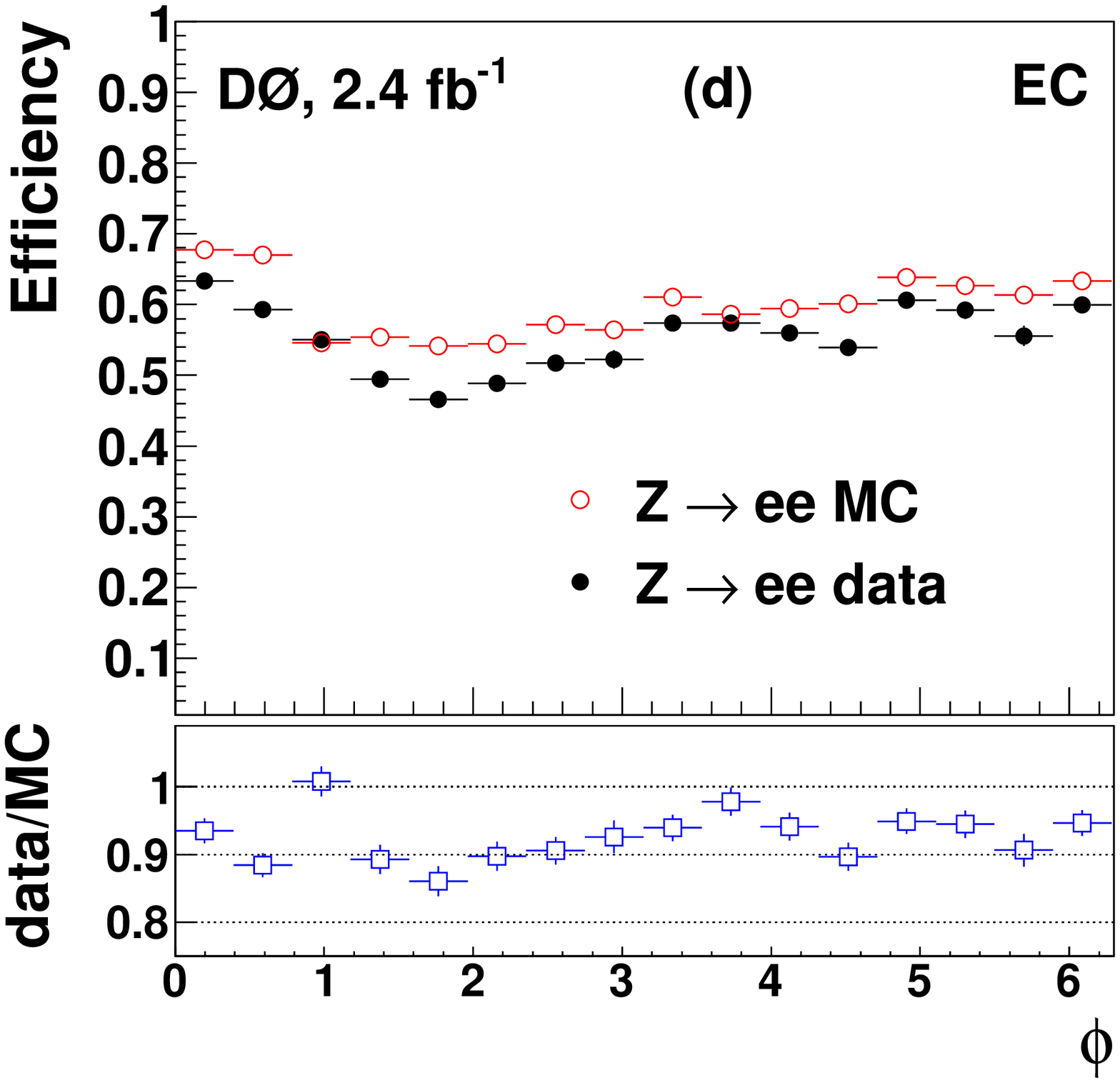}
\includegraphics[width=0.42\textwidth]{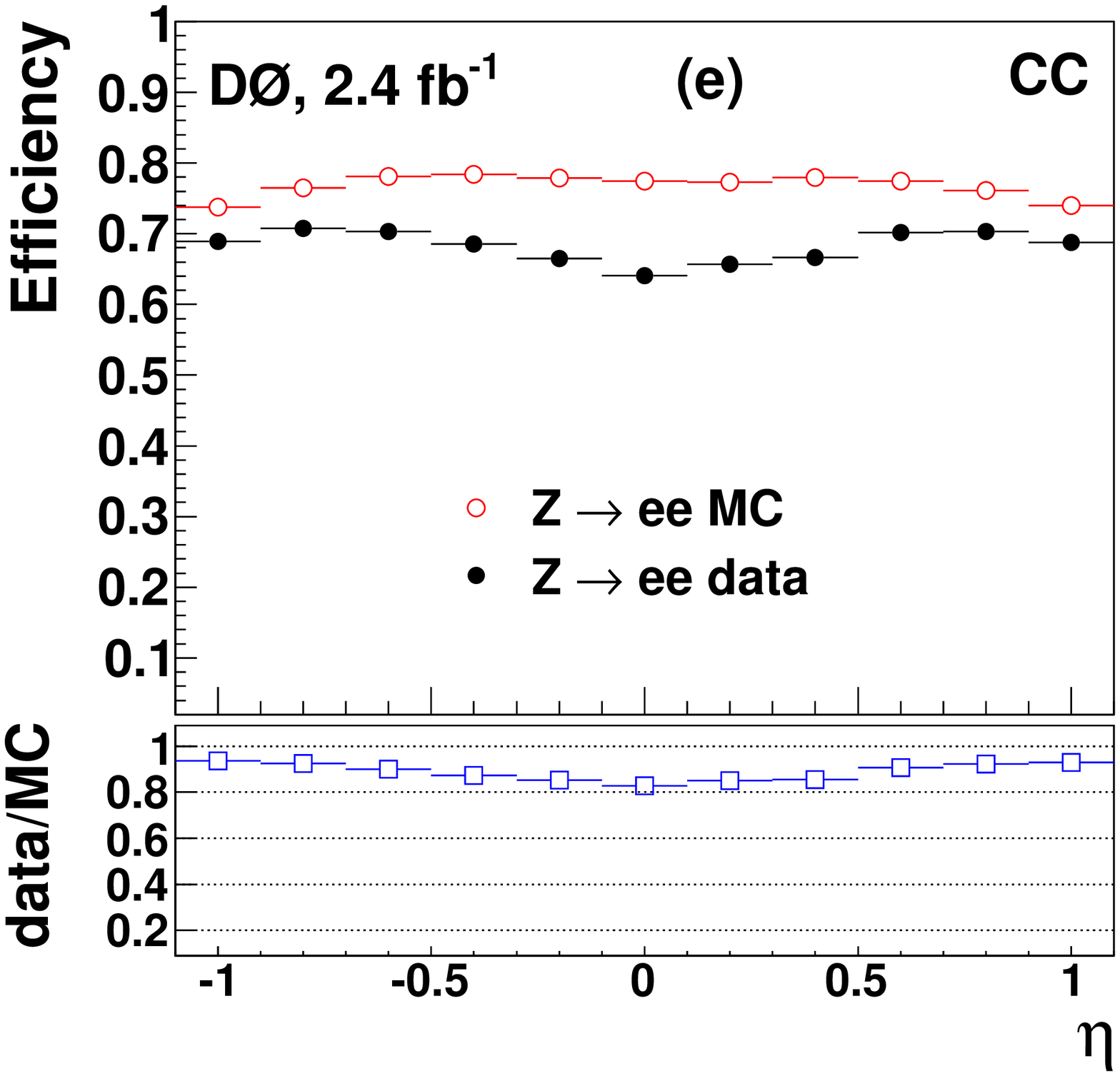}
\includegraphics[width=0.42\textwidth]{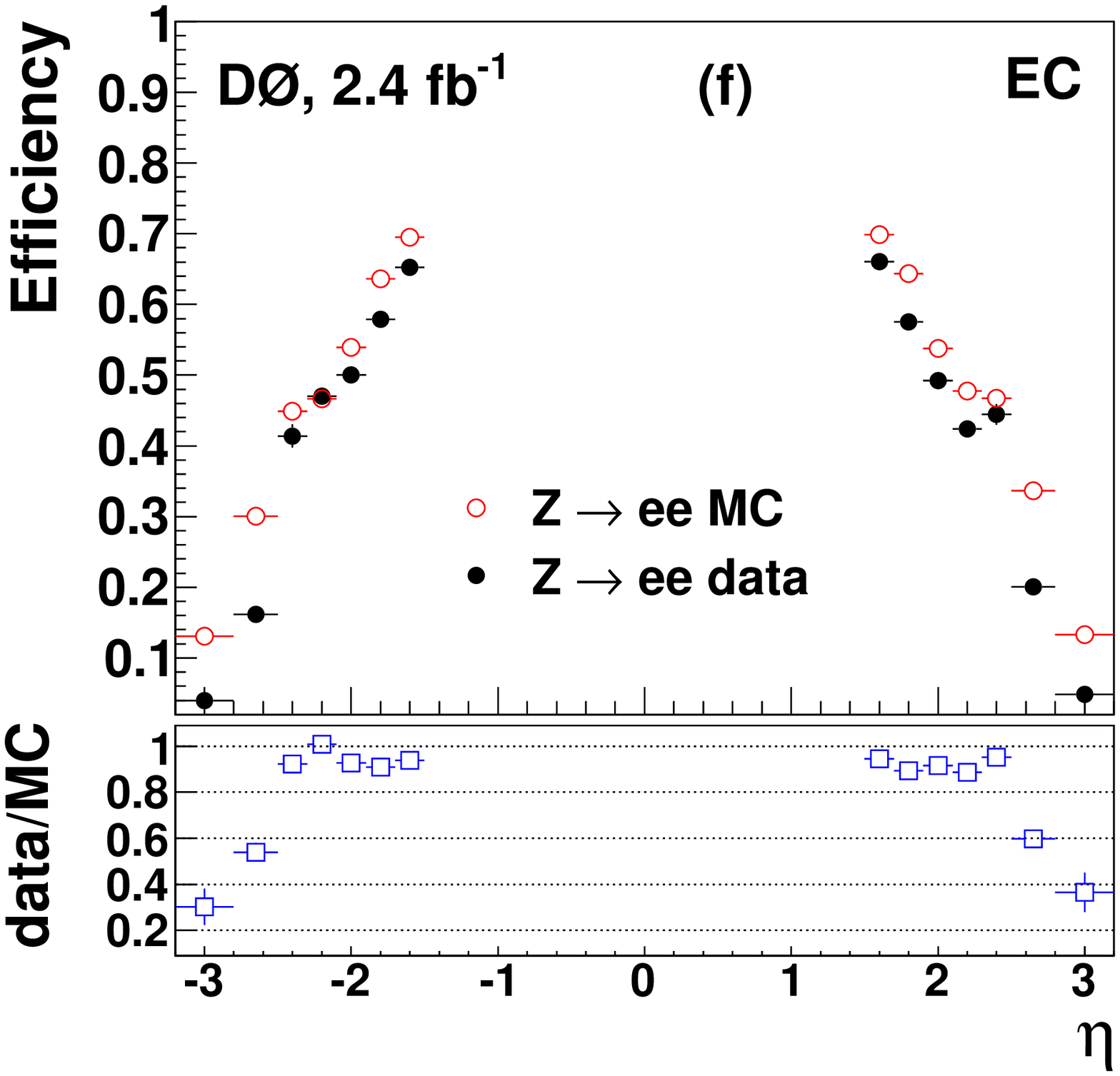}
\caption{
Electron identification efficiencies as a function of (a,b) $E_{T}$,
(c,d) $\phi$ and (e,f) $\eta$ for tight electron requirements in
CC and EC. Displayed are data and MC predictions in $Z
\to ee$ decays and their ratio.}
\label{fig:eTightEff}
\end{figure*}

The tag-and-probe method described in Sect.~\ref{tagandprobe} is used
here with the exception that now the probe electron is required to
fulfill the preselection criteria. 
The identification efficiencies are measured in $\eta-\phi$ phase
space. The resulting efficiencies for electrons in data and MC events
and the ratio of efficiencies in the data and the MC simulation
are shown in Figs.~\ref{fig:eLooseEff}~and~\ref{fig:eTightEff}.
In CC, the efficiencies in the $\eta \approx 0$ region are lower
than in other regions since the light yield of the CFT
is lower due to a shorter path length through the scintillating fiber.
In EC, the efficiencies decrease in high $\eta$ region due to the partial
coverage of the tracking system.
The dependence of the efficiencies on $\phi$ are mainly caused
by the azimuthal variations of the CFT waveguide length
not taken into account in simulation.

To account for deficiencies of the simulation,
the simulation is corrected by applying $\eta$ and $\phi$ dependent correction
factors.
The dependence on instantaneous luminosity for the electron
reconstruction efficiencies is studied and derived
following ($\eta-\phi$)-dependent correction.
Relative to the efficiency at low instantaneous luminosity 
(\ilum~$<$~$1.5 \times 10^{32}$~cm$^{-2}$s$^{-1}$) the efficiency
decreases with increasing \ilum, declining by $\approx$10\% when
\ilum~$=$~$2.5 \times~10^{32}$~cm$^{-2}$s$^{-1}$. 
The ratio of those efficiencies in data and MC simulation
has no dependence on the instantaneous luminosity.

For transverse momenta of $40$~GeV after preselection, loose electrons have a total
identification efficiency of 85\% (95\%) with a fake rate from
misidentified jets
of 5\% (3\%) in the CC (EC). Tight electrons at the same transverse
momentum have an identification efficiency of 72\% (53\%) with a
misidentification rate of 0.2\% (0.1\%) in the CC (EC).

\section{Efficiencies of photon identification}
\label{sec::results_gam}
\subsection{Photon identification efficiencies}
\label{sec:photon_eff}
There are two categories of variables for photon
identification. Variables based mainly on shower information
are used to reject misidentified jets. Tracking-based
variables are used to separate electrons from photons.
There are two main mechanisms by which photons can appear
as electrons. First, if the photon has converted into an electron-positron
pair in the inner tracking system, creating a reconstructed track. 
The probability for conversion is $\approx$6\%,
and we do not reconstruct these converted photons explicitly.
Second, if a track from particles of the underlying event is pointing to the 
EM cluster. In both cases, the matched track information 
for a photon will tend to be different from a real electron.

Because a large sample of pure photons is not available in data,
$Z \to ee$ events are used to derive efficiencies for variables based
mainly on the calorimeter information. For tracking-based variables, the
efficiencies are measured from reconstructed radiated photons in $Z
\to \gamma\ell\ell \ (\ell=e,\mu)$ events in data and MC. In both cases,
differences between data and MC event samples are analyzed to
correct the efficiency in simulation.

Due to different needs in various physics analyses, various sets of
photon identification requirements are developed.
We provide here photon identification efficiencies for two different
sets of photon identification requirements.

The first set of photon identification requirements considered is used
in the search for $H 
\to \gamma\gamma$ decays~\cite{hgg-ref1,hgg-ref2}. The signal is dominated by
high-$p_{T}$ CC photons, and the analysis maximizes the
photon signal acceptance. Photon candidates in the CC are required to
fulfill the preselection requirements as described in
Sect.~\ref{ele_presel}. In addition, it is required that
\begin{itemize}
\item $\Sigma p_{T}^{\rm trk} < 2$ GeV;
\item $\sigma_{\phi}^{2} < 18$ cm$^{2}$;
\item Output of $\gamma$NN5 $> 0.1$.
\end{itemize}
In addition the following requirements are placed on track-based variables:
\begin{itemize}
\item P($\chi^{2}_{\rm spatial}$) $=$ $-1$;
\item $D_{\rm hor} < 0.9$.
\end{itemize}

\begin{figure*}
\centering
\includegraphics[width=0.42\textwidth]{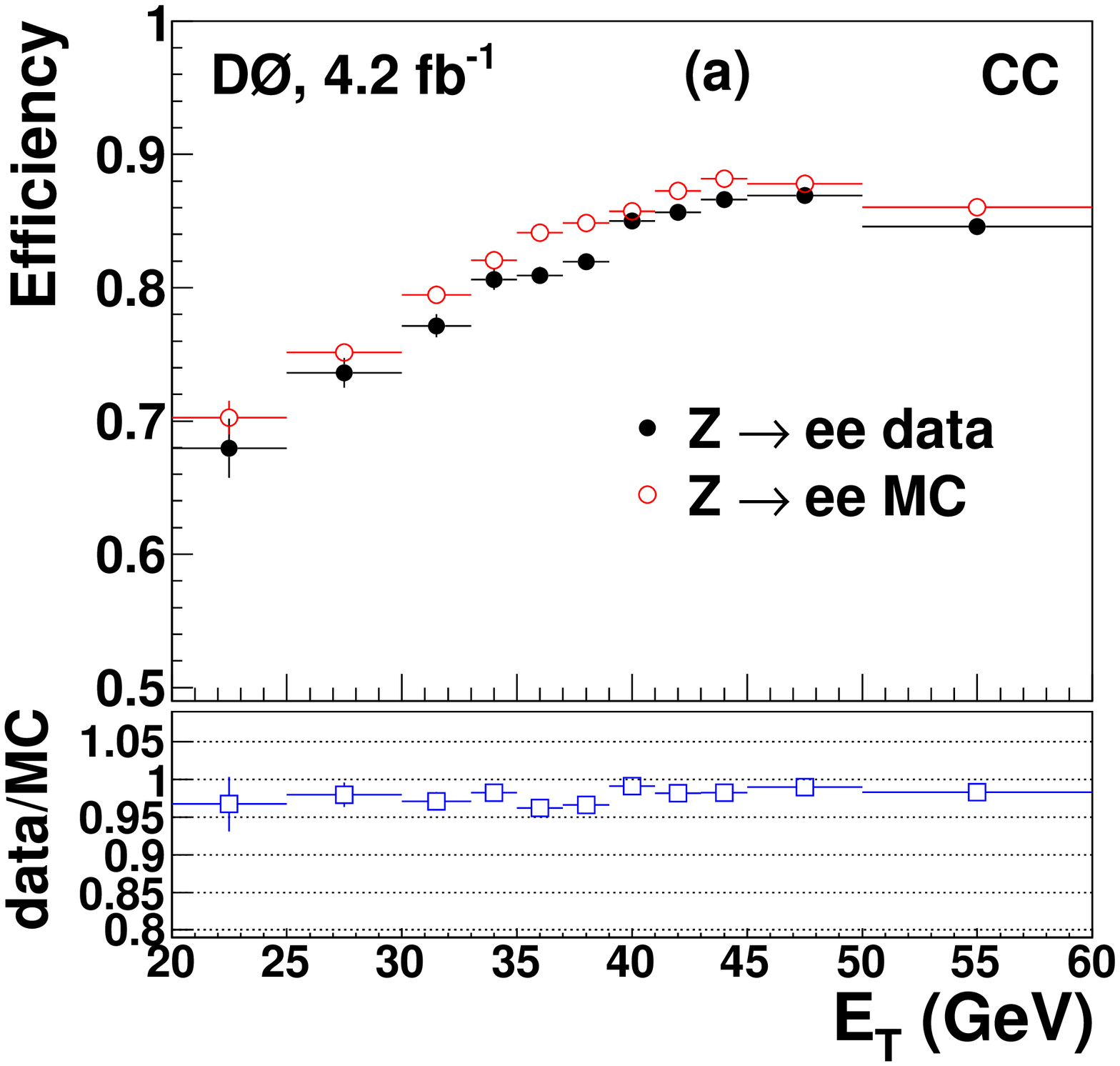}
\includegraphics[width=0.42\textwidth]{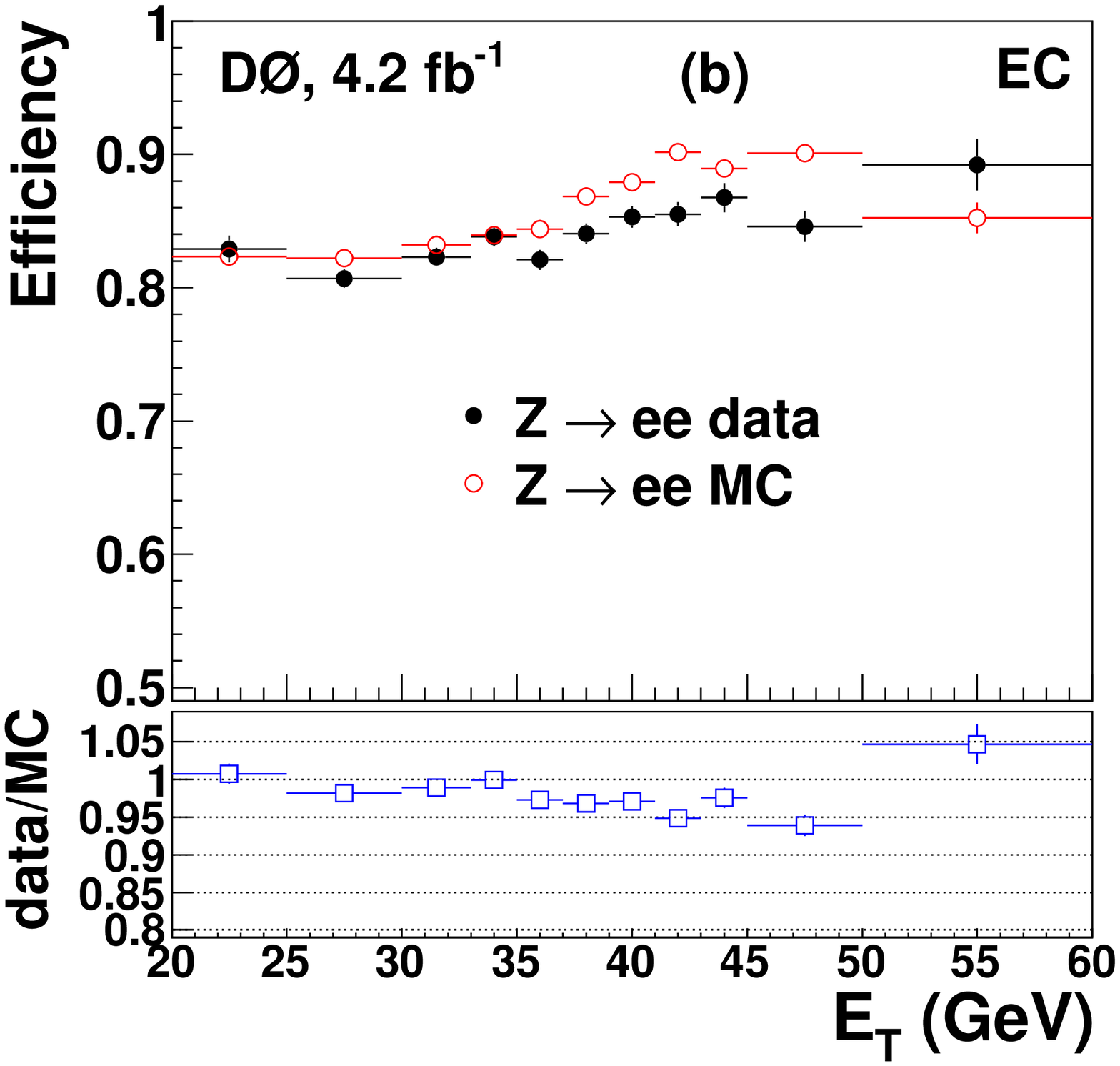}\\
\includegraphics[width=0.42\textwidth]{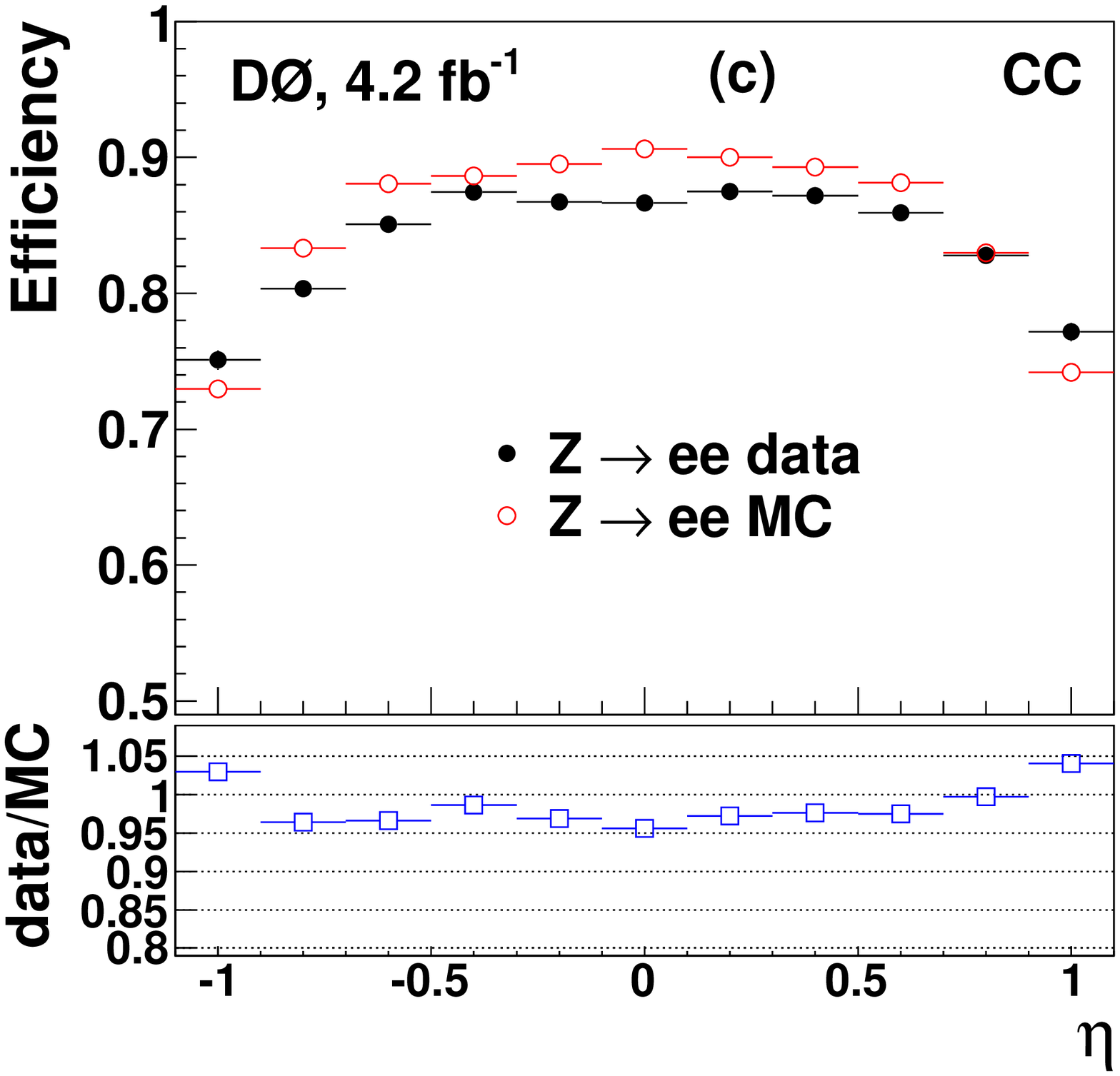}
\includegraphics[width=0.42\textwidth]{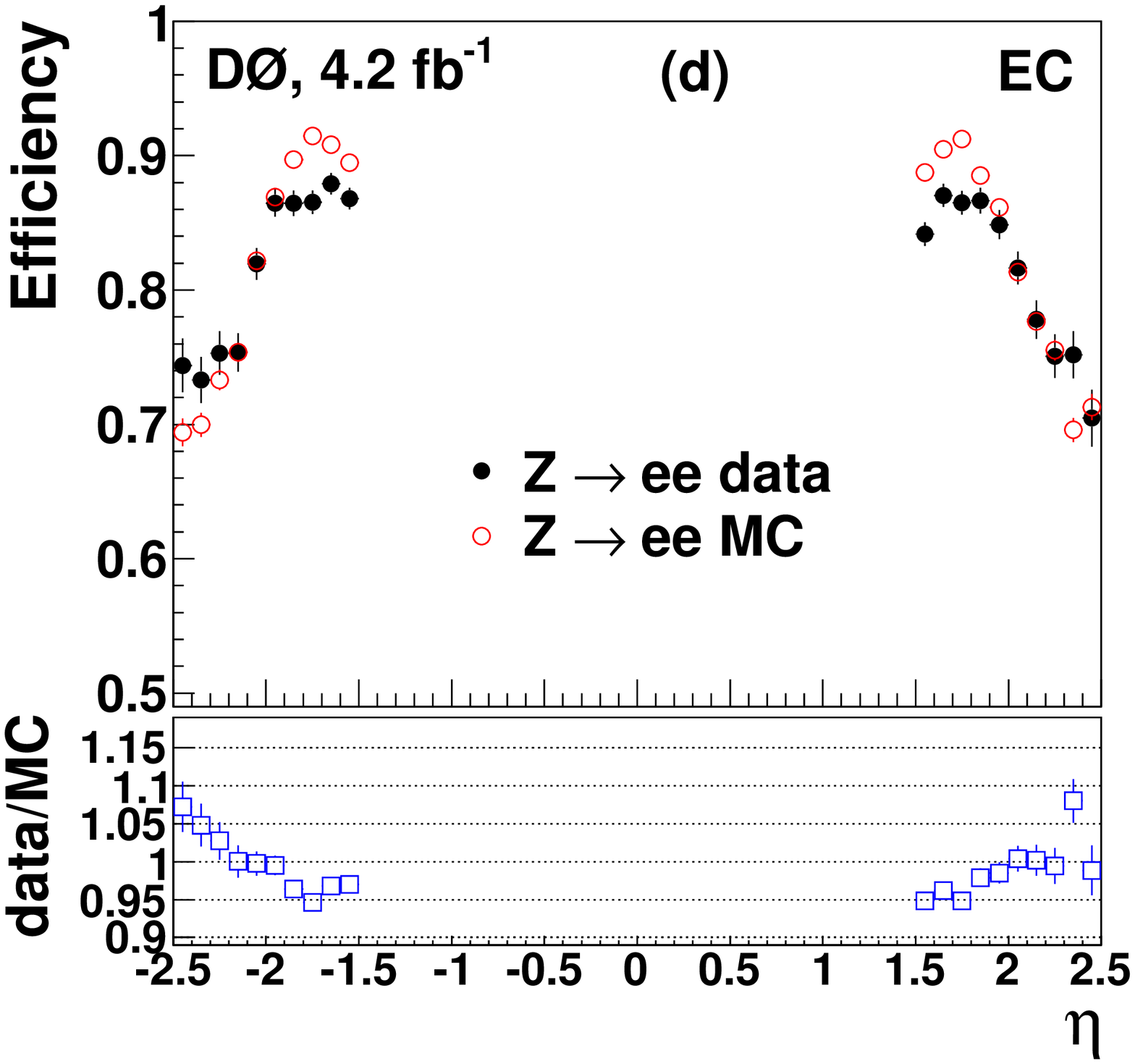}
\includegraphics[width=0.42\textwidth]{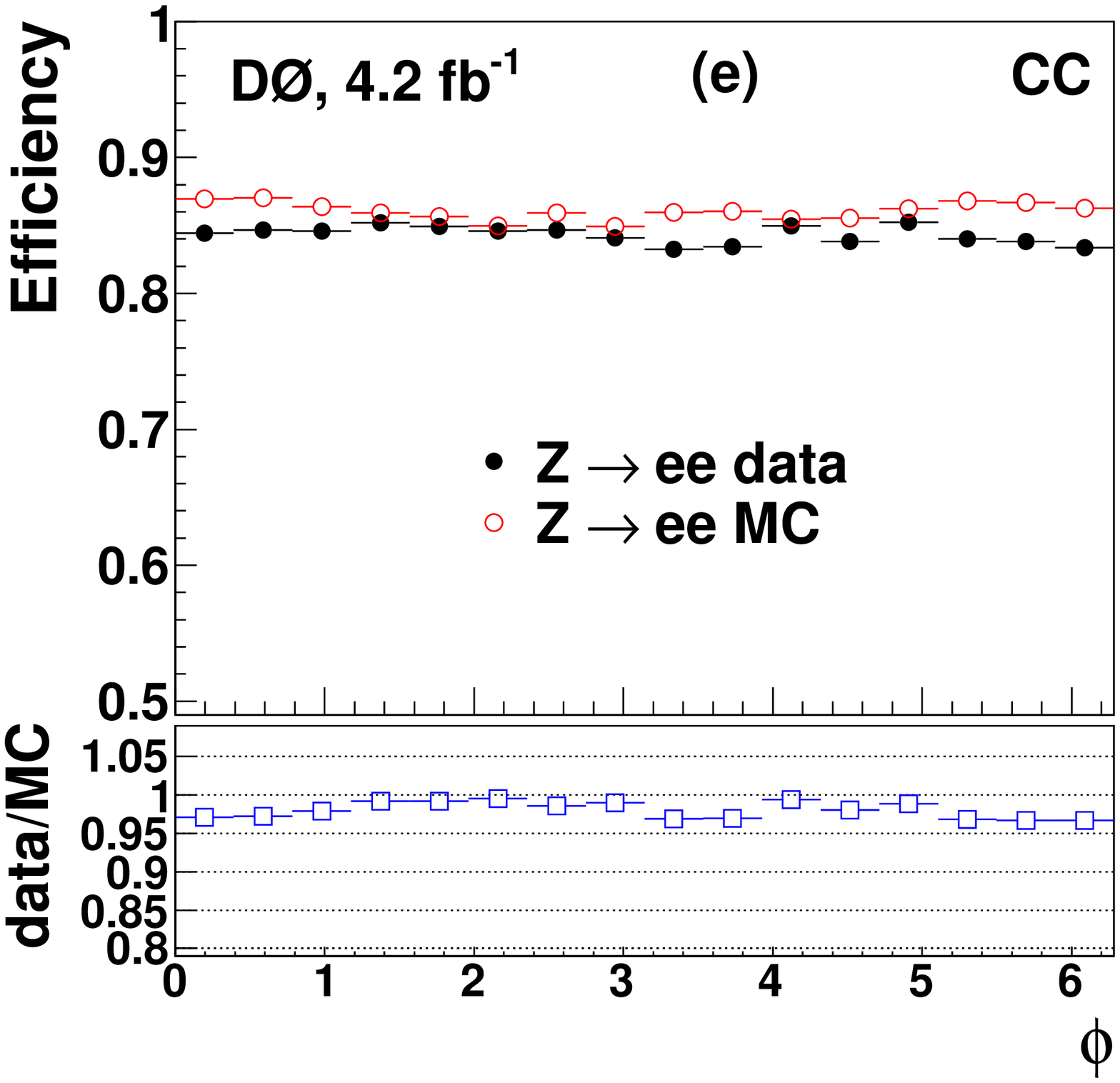}
\includegraphics[width=0.42\textwidth]{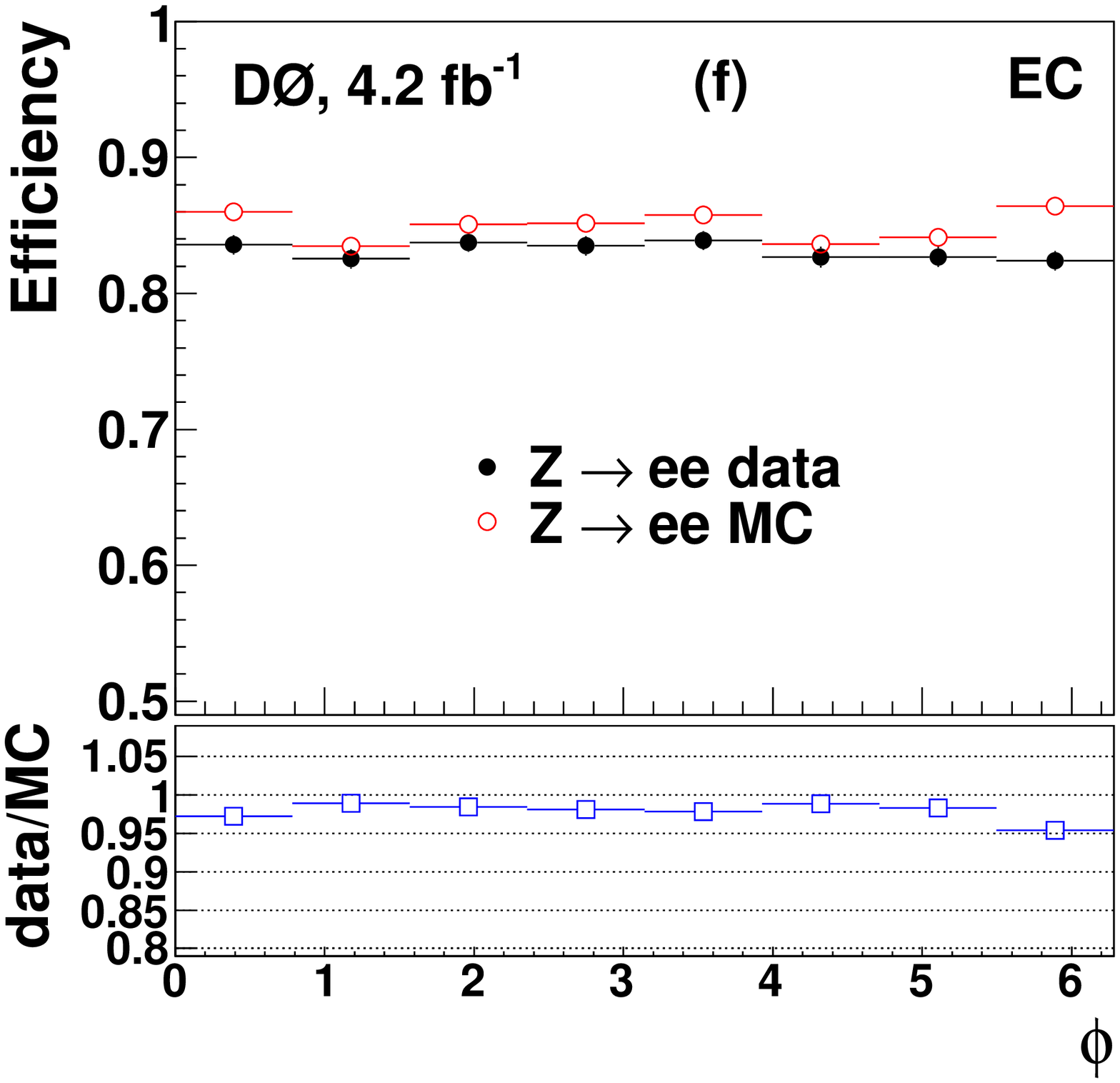}
\caption{
Photon identification efficiencies for identification variables mainly based on calorimeter information 
as derived from $Z \to ee$
decays. Displayed are data and MC predictions and their ratio as a
function of $E_{T}$ (a) (b), $\eta$ (c) (d) and $\phi$ (e) (f) for CC
and EC photons.}
\label{fig:grecoeff}
\end{figure*}

The measured identification efficiencies using the non track-based variables
in this selection are presented in Fig. \ref{fig:grecoeff} (left column) as
a function of $E_{T}$, $\eta$ and $\phi$.
The differences between data and MC are
at the percent level
and are constant in the presented distributions. Therefore,
a single correction factor is applied to MC photon simulation.

The second set of photon identification requirements presented here is used for
measurements of electroweak cross sections, such as the measurement of
the $W\gamma$ production cross section~\cite{wg-ref}. Here, the
photons tend to have low $E_{T}$ and a
high background rejection is required.
The EC photons used are required
to fulfill the preselection criteria of Sect.~\ref{ele_presel} and to
satisfy the following requirements:
\begin{itemize}
\item $\Sigma p_{T}^{\rm trk} < 1.5$ GeV
\item $\sigma_{\phi}^{2} < (7.3\cdot\eta^{2} - 35.9\cdot|\eta|
  + 45.7)$ cm$^{2}$
\item Output of $\gamma$NN4$ > 0.05$
\end{itemize}
In addition, a track-based requirement P($\chi^{2}_{\rm spatial}$) $<$ 0.001 is applied.

Figure \ref{fig:grecoeff} (right column) shows the identification efficiencies using the
non track-based variables in this selection
for data and MC. The difference between data and simulation depends on $\eta$.
To take this into account, the correction to
MC simulation is parametrized as a function of $\eta$.

For both CC and EC photons, exploring the track-based variables as
presented in this section, the efficiencies to identify a photon candidate
are measured.
The $Z\gamma \to \gamma \ell\ell$ $(\ell = e, \mu)$ data and MC comparison justifies that no further
corrections to the photon simulation are required.
The photon identification efficiency for these track-based variables
is 92\% (95\%) in CC (EC) for an electron-to-photon
misidentification rate of 2\% (23\%) in CC (EC) in the selections
described above.
The average photon identification efficiencies for the two sets of requirements described above are
81\% and 83\%
for a rate to misidentify jets as  photons of 4\% and 10\%
for CC and EC photons, respectively.
These identification efficiencies have a similar dependence on the 
instantaneous luminosity as the electron identification, and there 
is no visible dependence on the instantaneous luminosity for 
the ratio of those efficiencies in data and MC simulation.

\subsection{Vertex pointing}

In most physics applications,
it is important to know from which $p\bar{p}$ collision
vertex the photon originated.
Since unconverted photons leave no track,
the default reconstruction vertexing algorithm
does not provide high probability
to find the correct photon origin
if there is no high-$p_{T}$ track in the event.
For events without leptons and with energetic photons,
the most probable photon production vertex can be reconstructed due to
the presence of the underlying event coming
from interactions of spectator quarks, and corresponds to the vertex
with highest track multiplicity \cite{hgg-ref1, hgg-ref2, d0_dpp}.
In such cases, verifying that the true production vertex is found
in data is important, especially
in the high-instantaneous luminosity regime with many $p\bar{p}$ collision vertices.

\begin{figure*}
\includegraphics[scale=0.33]{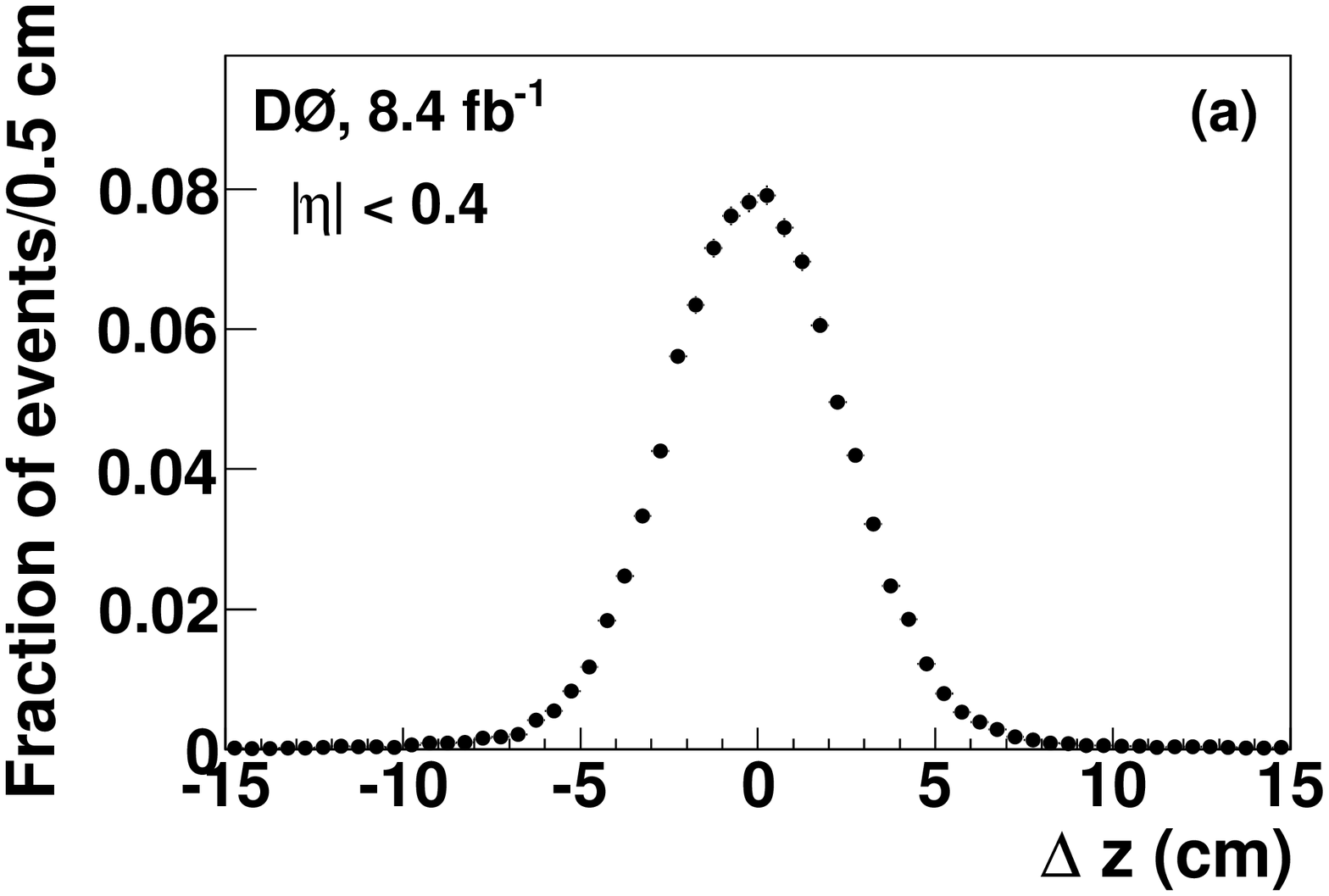} 
\includegraphics[scale=0.33]{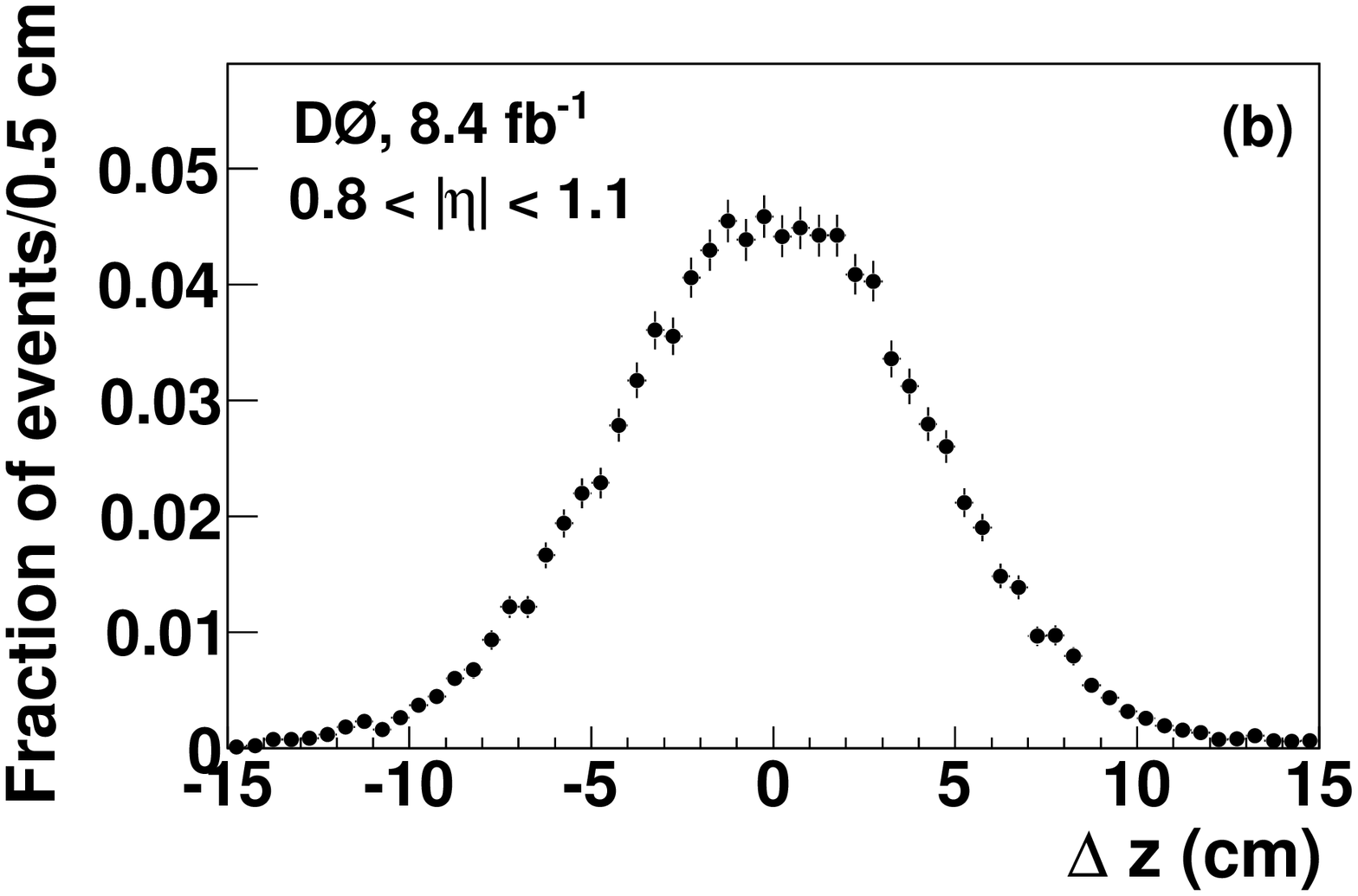}
        \caption{
         Vertex pointing resolution in two rapidity bins:
         $|\eta|<0.4$ (a), and $0.8<|\eta|<1.1$ (b).
         }
        \label{fig:photonPoint}
\end{figure*}

To find the position of the photon origin along the beam line ($z$-axis) 
between $-$60 cm and 60 cm in the CC,
the $(x,y,z)$-coordinates of the EM cluster in the EM1--EM4 layers
and the position of the CPS cluster are used.
Therefore, 5 points are used with radii from about 73~cm to 99~cm.
Using a linear extrapolation to the $z$-axis,
the most probable position of the photon
origin vertex is obtained.
Typical resolution of the algorithm varies between 3 and 4.5~cm.
It becomes larger towards high $\eta$ mainly due to increasing amount of material in
front of the calorimeter (from about $3.4$ to $5X_0$).
The resolution has been tested in data using $Z(\to \ell\ell)+\gamma$ events, in which the ``true'' vertex
($z_{\rm true}$) is reconstructed using the two lepton ($e$ or $\mu$) tracks and
the photon vertex ($z_{\rm point}$) is obtained using the procedure described above.
The distribution of events for $\Delta z=z_{\rm true}-z_{\rm point}$ 
is shown in Fig.~\ref{fig:photonPoint}. 
The resolution is 2.4 cm
for $|\eta| < 0.4$, and 4.3 cm for $0.8 < |\eta| < 1.1$. 

The resolution in MC simulation is a factor of $1.4-1.5$ better
than in data events.
To calibrate the pointing resolution, a smearing procedure
as a function of
photon pseudorapidity is applied.
The $\Delta z$ resolution is almost independent of photon $p_T$.

\section{Conclusions}
\label{sec::conclusions}
The precise and efficient reconstruction and identification of
electrons and photons by the D0 experiment at the 
Tevatron $p\bar{p}$ collider at Fermilab is essential for a broad spectrum of physics
analyses, including high precision SM measurements
and searches for new phenomena.

In this paper, the electron and photon reconstruction and identification algorithms have
been presented using data collected by the D0 detector in
$p\bar{p}$ collisions at a center-of-mass energy of 1.96~TeV.
The separation between electron or photon signal
and multijet background is
considerably improved using multivariate analysis techniques. A
likelihood method for electron identification, a neural network method
for electrons and photons, and a Boosted Decision Tree for electrons
have been developed.  An energy calibration dependent on the azimuthal
angle of the EM cluster, the shower shape and the pseudorapidity has
been performed separately for data and MC, leading to significant
improvements in resolution and uniformity and resulting in a good
agreement between data and MC. 

Single electrons are triggered with an efficiency $\approx$100\%
for transverse momenta above 30~GeV in the fiducial regions of the
calorimeter up to $|\eta| < 2.5$.  For transverse momenta of $E_T =
40$~GeV, in general at D0 electrons can be identified with a total identification
efficiency of 90\% (95\%)
with the rate at which jets are misidentified as electrons being
5\% (3\%) in the CC (EC).
Photons in the CC and EC regions can typically be identified with
efficiencies
varying between 69\%--84\%
with the rate at which electrons or jets are misidentified as photons being
2\%--10\%.

The agreement of electron and photon identification efficiencies
between data and MC
in fiducial regions of the detector is reasonable, with deviations only
at the percent level. Larger correction factors are necessary in
non-fiducial regions close to the boundaries of the calorimeter
modules. These correction factors have been applied to MC events
as a function of kinematic
variables resulting in considerable improvements of the
simulation.

\section*{Acknowledgments}
\input acknowledgement.tex

\begin{comment}

\end{comment}

\end{document}

%% file: author_list.tex
\affiliation{LAFEX, Centro Brasileiro de Pesquisas F\'{i}sicas, Rio de Janeiro, Brazil}
\affiliation{Universidade do Estado do Rio de Janeiro, Rio de Janeiro, Brazil}
\affiliation{Universidade Federal do ABC, Santo Andr\'e, Brazil}
\affiliation{University of Science and Technology of China, Hefei, People's Republic of China}
\affiliation{Universidad de los Andes, Bogot\'a, Colombia}
\affiliation{Charles University, Faculty of Mathematics and Physics, Center for Particle Physics, Prague, Czech Republic}
\affiliation{Czech Technical University in Prague, Prague, Czech Republic}
\affiliation{Institute of Physics, Academy of Sciences of the Czech Republic, Prague, Czech Republic}
\affiliation{Universidad San Francisco de Quito, Quito, Ecuador}
\affiliation{LPC, Universit\'e Blaise Pascal, CNRS/IN2P3, Clermont, France}
\affiliation{LPSC, Universit\'e Joseph Fourier Grenoble 1, CNRS/IN2P3, Institut National Polytechnique de Grenoble, Grenoble, France}
\affiliation{CPPM, Aix-Marseille Universit\'e, CNRS/IN2P3, Marseille, France}
\affiliation{LAL, Universit\'e Paris-Sud, CNRS/IN2P3, Orsay, France}
\affiliation{LPNHE, Universit\'es Paris VI and VII, CNRS/IN2P3, Paris, France}
\affiliation{CEA, Irfu, SPP, Saclay, France}
\affiliation{IPHC, Universit\'e de Strasbourg, CNRS/IN2P3, Strasbourg, France}
\affiliation{IPNL, Universit\'e Lyon 1, CNRS/IN2P3, Villeurbanne, France and Universit\'e de Lyon, Lyon, France}
\affiliation{III. Physikalisches Institut A, RWTH Aachen University, Aachen, Germany}
\affiliation{Physikalisches Institut, Universit\"at Freiburg, Freiburg, Germany}
\affiliation{II. Physikalisches Institut, Georg-August-Universit\"at G\"ottingen, G\"ottingen, Germany}
\affiliation{Institut f\"ur Physik, Universit\"at Mainz, Mainz, Germany}
\affiliation{Ludwig-Maximilians-Universit\"at M\"unchen, M\"unchen, Germany}
\affiliation{Panjab University, Chandigarh, India}
\affiliation{Delhi University, Delhi, India}
\affiliation{Tata Institute of Fundamental Research, Mumbai, India}
\affiliation{University College Dublin, Dublin, Ireland}
\affiliation{Korea Detector Laboratory, Korea University, Seoul, Korea}
\affiliation{CINVESTAV, Mexico City, Mexico}
\affiliation{Nikhef, Science Park, Amsterdam, the Netherlands}
\affiliation{Radboud University Nijmegen, Nijmegen, the Netherlands}
\affiliation{Joint Institute for Nuclear Research, Dubna, Russia}
\affiliation{Institute for Theoretical and Experimental Physics, Moscow, Russia}
\affiliation{Moscow State University, Moscow, Russia}
\affiliation{Institute for High Energy Physics, Protvino, Russia}
\affiliation{Petersburg Nuclear Physics Institute, St. Petersburg, Russia}
\affiliation{Instituci\'{o} Catalana de Recerca i Estudis Avan\c{c}ats (ICREA) and Institut de F\'{i}sica d'Altes Energies (IFAE), Barcelona, Spain}
\affiliation{Uppsala University, Uppsala, Sweden}
\affiliation{Taras Shevchenko National University of Kyiv, Kiev, Ukraine}
\affiliation{Lancaster University, Lancaster LA1 4YB, United Kingdom}
\affiliation{Imperial College London, London SW7 2AZ, United Kingdom}
\affiliation{The University of Manchester, Manchester M13 9PL, United Kingdom}
\affiliation{University of Arizona, Tucson, Arizona 85721, USA}
\affiliation{University of California Riverside, Riverside, California 92521, USA}
\affiliation{Florida State University, Tallahassee, Florida 32306, USA}
\affiliation{Fermi National Accelerator Laboratory, Batavia, Illinois 60510, USA}
\affiliation{University of Illinois at Chicago, Chicago, Illinois 60607, USA}
\affiliation{Northern Illinois University, DeKalb, Illinois 60115, USA}
\affiliation{Northwestern University, Evanston, Illinois 60208, USA}
\affiliation{Indiana University, Bloomington, Indiana 47405, USA}
\affiliation{Purdue University Calumet, Hammond, Indiana 46323, USA}
\affiliation{University of Notre Dame, Notre Dame, Indiana 46556, USA}
\affiliation{Iowa State University, Ames, Iowa 50011, USA}
\affiliation{University of Kansas, Lawrence, Kansas 66045, USA}
\affiliation{Louisiana Tech University, Ruston, Louisiana 71272, USA}
\affiliation{Northeastern University, Boston, Massachusetts 02115, USA}
\affiliation{University of Michigan, Ann Arbor, Michigan 48109, USA}
\affiliation{Michigan State University, East Lansing, Michigan 48824, USA}
\affiliation{University of Mississippi, University, Mississippi 38677, USA}
\affiliation{University of Nebraska, Lincoln, Nebraska 68588, USA}
\affiliation{Rutgers University, Piscataway, New Jersey 08855, USA}
\affiliation{Princeton University, Princeton, New Jersey 08544, USA}
\affiliation{State University of New York, Buffalo, New York 14260, USA}
\affiliation{University of Rochester, Rochester, New York 14627, USA}
\affiliation{State University of New York, Stony Brook, New York 11794, USA}
\affiliation{Brookhaven National Laboratory, Upton, New York 11973, USA}
\affiliation{Langston University, Langston, Oklahoma 73050, USA}
\affiliation{University of Oklahoma, Norman, Oklahoma 73019, USA}
\affiliation{Oklahoma State University, Stillwater, Oklahoma 74078, USA}
\affiliation{Brown University, Providence, Rhode Island 02912, USA}
\affiliation{University of Texas, Arlington, Texas 76019, USA}
\affiliation{Southern Methodist University, Dallas, Texas 75275, USA}
\affiliation{Rice University, Houston, Texas 77005, USA}
\affiliation{University of Virginia, Charlottesville, Virginia 22904, USA}
\affiliation{University of Washington, Seattle, Washington 98195, USA}
\author{V.M.~Abazov} \affiliation{Joint Institute for Nuclear Research, Dubna, Russia}
\author{B.~Abbott} \affiliation{University of Oklahoma, Norman, Oklahoma 73019, USA}
\author{B.S.~Acharya} \affiliation{Tata Institute of Fundamental Research, Mumbai, India}
\author{M.~Adams} \affiliation{University of Illinois at Chicago, Chicago, Illinois 60607, USA}
\author{T.~Adams} \affiliation{Florida State University, Tallahassee, Florida 32306, USA}
\author{J.P.~Agnew} \affiliation{The University of Manchester, Manchester M13 9PL, United Kingdom}
\author{G.D.~Alexeev} \affiliation{Joint Institute for Nuclear Research, Dubna, Russia}
\author{G.~Alkhazov} \affiliation{Petersburg Nuclear Physics Institute, St. Petersburg, Russia}
\author{A.~Alton$^{a}$} \affiliation{University of Michigan, Ann Arbor, Michigan 48109, USA}
\author{A.~Askew} \affiliation{Florida State University, Tallahassee, Florida 32306, USA}
% ...
%
% visitor_addresses.tex                       28 December 2013
%  available symbols are:
%  $\ast, \dag, \ddag, \S, \P, $\|$, $\ast\ast$, \dag\dag, \ddag\ddag ,\#
%
\collaboration{The D0 Collaboration\footnote{with visitors from
%{alton}
$^{a}$Augustana College, Sioux Falls, SD, USA,
%{burdin}
$^{b}$The University of Liverpool, Liverpool, UK,
%{grohsjean}
$^{c}$DESY, Hamburg, Germany,
%{de la cruz-burelo}
$^{d}$Universidad Michoacana de San Nicolas de Hidalgo, Morelia, Mexico
%{partridge}
$^{e}$SLAC, Menlo Park, CA, USA,
%{hesketh}
$^{f}$University College London, London, UK,
%{luna-garcia}
$^{g}$Centro de Investigacion en Computacion - IPN, Mexico City, Mexico,
%{santos}
$^{h}$Universidade Estadual Paulista, S\~ao Paulo, Brazil,
%{meyer}
$^{i}$Karlsruher Institut f\"ur Technologie (KIT) - Steinbuch Centre for Computing (SCC),
D-76128 Karlsrue, Germany,
%{patwa}
$^{j}$Office of Science, U.S. Department of Energy, Washington, D.C. 20585, USA,
%{cooke}
$^{k}$American Association for the Advancement of Science, Washington, D.C. 20005, USA
and
%{borysova}
$^{l}$Kiev Institute for Nuclear Research, Kiev, Ukraine
%{montgomery}
%$^{?}$Thomas Jefferson National Accelerator Facility, Newport News, VA 23606, USA,
%{falkowski}
%$^{?}$Laboratoire de Physique Theorique, Orsay, FR,
%{hooper,kozminski}
%$^{?}$}Visitor from Lewis University, Romeoville, IL, USA.
%{weber}
%$^{?}$Universit{\"a}t Bern, Bern, Switzerland.
%{deceased}
%{zanabria}
%$^{?}$City Colleges of Chicago, Chicago, IL, USA}
%$^{\ddag}$Deceased.
}} \noaffiliation
\vskip 0.25cm

%% file: acknowledgement.tex
% acknowledgement.tex                            28 December 2013
%
We thank the staffs at Fermilab and collaborating institutions,
and acknowledge support from the
DOE and NSF (USA);
CEA and CNRS/IN2P3 (France);
MON, NRC KI and RFBR (Russia);
CNPq, FAPERJ, FAPESP and FUNDUNESP (Brazil);
DAE and DST (India);
Colciencias (Colombia);
CONACyT (Mexico);
NRF (Korea);
FOM (The Netherlands);
STFC and the Royal Society (United Kingdom);
MSMT and GACR (Czech Republic);
BMBF and DFG (Germany);
SFI (Ireland);
The Swedish Research Council (Sweden);
and
CAS and CNSF (China).
%